\documentclass[11pt]{article}%
\usepackage{amsmath}
\usepackage{graphicx}%
\usepackage{amsfonts}%
\usepackage{amssymb}
\newtheorem{theorem}{Theorem}

\newtheorem{proposition}[theorem]{Proposition}

\newenvironment{proof}[1][Proof]{\textbf{#1.} }{\ \rule{0.5em}{0.5em}}
\begin{document}

\title{The Liar and Related Paradoxes:\\Fuzzy Truth Value Assignment \\for Collections of Self-Referential Sentences}
\author{K. Vezerides\\vezerid@ac.anatolia.edu.gr
\and and
\and Ath. Kehagias\\kehagiat@gen.auth.gr\\http://users.auth.gr/ \symbol{126}kehagiat}
\maketitle

\begin{abstract}
We study self-referential sentences of the type related to the \emph{Liar
paradox}. In particular, we consider the problem of assigning consistent fuzzy
truth values to \emph{collections of self-referential sentences}. We show that
the problem can be reduced to the solution of a system of nonlinear equations.
Furthermore, we prove that, under mild conditions, such a system always has a
solution (i.e. a consistent truth value assignment) and that, for a particular
implementation of logical ``and'', ``or'' and ``negation'', the ``mid-point''
solution is always consistent. Next we turn to computational issues and
present several truth-value assignment algorithms; we argue that these
algorithms can be understood as generalized sequential reasoning. In an
Appendix we present a large number of examples of self-referential collections
(including the Liar and the Strengthened Liar), we formulate the corresponding
truth value equations and solve them analytically and/ or numerically.

\end{abstract}


\section{Introduction}

\label{sec01}

\emph{Self referential sentences }are sentences which talk about themselves.
Sentences of this type often generate logical \emph{paradoxes}, the study of
which goes back to the ancient Greeks. Many approaches have been proposed to
neutralize the paradoxes. In this paper we study the problem from the point of
fuzzy logic.

The prototypical example of a self-referential sentence is the \emph{Liar
Sentence}: Epimenides the Cretan has supposedly uttered the following
sentence:
\[
\text{``All Cretans are liars.''}%
\]
The Liar sentence generates a paradox. To see this consider whether the
sentence is true or false. Following a somewhat loose chain of reasoning, let
us assume ``All Cretans are liars'' to mean that everything a Cretan says is
not true. Since Epimenides is a Cretan, his statement that ``All Cretans are
liars'' is not true; if we take ``not true'' to mean ``false'', then the
\emph{opposite }of ``All Cretans are liars'' must be true; and if we take this
opposite to be ``All Cretans are truth-tellers'', then what Epimenides says
must be true, i.e. it is true that ``All Cretans are liars''; but then it must
be the case that what Epimenides says is false. It seems that we have entered
a vicious circle, concluding first that Epimenides' statement is false, then
that it is true, then again that it is false and so on ad infinitum.

The above reasoning is not rigorous, since, for example, ``All Cretans are
liars'' does not necessarily mean that all Cretans \emph{always }utter false
statements; similarly the negation of ``All Cretans are liars'' is
``\emph{Some} Cretans are truth-tellers'' and so on.

But the paradox also appears under more exact reasoning. Consider the
following sentence:
\[
\text{``This sentence is false.''}%
\]
In this case, if we assume the sentence to be true, then what it says must
hold, i.e. the sentence must be false. But then its opposite must be true,
i.e. it must be true that ``This sentence is true''. But then it is true that
``This sentence is false'' and we have again entered an oscillation between
two conclusions: first that the sentence is false, then that it is true.
Notice that we would enter a similar oscillation if we started by assuming the
sentence to be false; then we would conclude that the sentence is true, which
would mean that the sentence is false etc. A similar effect can be obtained
using two sentences. The following pair is the so-called \emph{inconsistent
dualist}:
\begin{align*}
A_{1}  &  =\text{``Sentence }A_{2}\text{ is true''}\\
A_{2}  &  =\text{``Sentence }A_{1}\text{ is false''.}%
\end{align*}
If $A_{1}$ is true, then $A_{2}$ is also true; but then $A_{1}$ must be false
and so $A_{2}$ must be false and so on ad infinitum.

Note that self-reference does not \emph{necessarily }lead to paradox. The
following pair is the so-called \emph{consistent dualist}:
\begin{align*}
A_{1} &  =\text{``Sentence }A_{2}\text{ is true''}\\
A_{2} &  =\text{``Sentence }A_{1}\text{ is true''.}%
\end{align*}
If $A_{1}$ is true, then $A_{2}$ is also true, which confirms that $A_{1}$ is
true. In short, accepting that $A_{1}$ and $A_{2}$ are both true is perfectly
consistent. However, note that we could equally well assume that $A_{1}$ is
false, which would mean that $A_{2}$ is also false, which confirms that
$A_{1}$ is false. In short, we can also accept that $A_{1}$ and $A_{2}$ are
both false. Is this a problem? We will consider some possible answers in the sequel.

We have already mentioned that self-referential sentences have been studied
extensively, and from several different points of view. In this section we
will only discuss some work which is directly related to the current paper; a
more extensive discussion of the literature will be presented in Section
\ref{sec06}.

The application of fuzzy logic to the Liar paradox goes back to a paper by
Zadeh \cite{Zadeh}; in summary, he resolves the paradox by assigning to the
Liar sentence a truth value of 1/2. Following Zadeh's paper, several authors
have analyzed self-reference using fuzzy logic (for more details see Section
\ref{sec06}).

The current paper has been heavily motivated by the work of Grim and his
collaborators \cite{Grim1,Grim4,Grim3,Grim2}. Grim considers collections of
self-referential sentences and models the \emph{fuzzy reasoning process }as a
\emph{dynamical system}. A method is presented to map each self-referential
collection to a dynamical system which represents the reasoning process. Each
sentence of the self-referential collection has a \emph{time-evolving fuzzy
truth value }which corresponds to a \emph{state variable} of the dynamical
system. Grim presents several examples of self-referential collections and
studies the properties of the corresponding dynamical systems. One of the main
points of \cite{Grim1} is that self-referential collections can generate
\emph{oscillating} or \emph{chaotic} dynamical behavior.

Grim's formulation is an essential starting point for the current paper; but
the issues we address are rather different. Our main interest is in obtaining
\emph{consistent truth value assignments}, similarly to \cite{Zadeh} and
unlike Grim who concentrates on oscillatory behavior. A more detailed
understanding of our approach can be obtained by the following outline of the paper.

In Section \ref{sec02} we present the logical framework which will be used for
the study of self-referential sentences. In this we follow very closely Grim's
formulation. In Section \ref{sec03}, we reduce the problem of consistent truth
value assignment to the solution of a system of \emph{algebraic}\footnote{I.e.
static, \emph{not} time-evolving.} nonlinear equations. Such a system will, in
general, possess more than one solution. It is a rather remarkable fact that,
for a very broad family of self-referential collections, the corresponding
equations possess \emph{at least }one solution, i.e. a consistent fuzzy truth
value assignment is \emph{always} possible, under some mild continuity
conditions; this is the subject of our Proposition \ref{cnt0303}. In Section
\ref{sec04} we turn to the \emph{computation }of the solutions. We consider
this as a separate, \emph{algorithm-dependent} sub-problem. In other words,
the same system of algebraic equations can be solved by many different
algorithms. We consider several such algorithms and, returning to Grim's point
of view, we study the dynamics of each algorithm. In a certain sense, each
such algorithm can be understood as a particular reasoning style, and
``human-style'', sequential reasoning of the form ``if $A_{1}$ then $A_{2}$,
if $A_{2}$ then not-$A_{1}$ etc.'' \ is one among many options. In Section
\ref{sec06} we take a brief look at the literature on self-referential
sentences, paradoxes and related topics and relate it to our approach. In
Section \ref{sec07} we summarize and discuss our results. Finally, in the
Appendix we present several specific examples of self-referential collections,
formulate the corresponding equations and solve them analytically and/or
numerically, by using the previously presented algorithms; we also compare the
behavior of the algorithms.

\section{The Logic Framework}

\label{sec02}

The general object of our interest is a finite collection of $M$ sentences
which talk about each other's truth value (i.e. a self-referential
collection); in particular we explore the extend to which the internal
structure of such a system determines the truth values of the sentences. This
question will be answered mainly in Section \ref{sec0302}. In this section we
introduce the logical framework which is necessary to address the problem.
This framework is quite similar to the one used in \cite{Grim1}.

We start with a finite set $\mathbf{V}_{1}$, the set of \emph{1st-level
elementary sentences }(also called \emph{variables}):
\[
\mathbf{V}_{1}=\left\{  A_{1},A_{2},...,A_{M}\right\}  .
\]
From the 1st level variables we \emph{recursively }build $\mathbf{S}_{1}$, the
set of \emph{1st level sentences} (also called logical \emph{formulas}):
\begin{align}
\text{If }A_{m} &  \in\mathbf{V}_{1}\text{ then }A_{m}\in\mathbf{S}%
_{1}\nonumber\\
\text{If }B_{1},B_{2} &  \in\mathbf{S}_{1}\text{ then }B_{1}\vee B_{2}%
,B_{1}\wedge B_{2},B_{1}^{\prime}\in\mathbf{S}_{1,}\label{eq001}%
\end{align}
where $\vee,\wedge,^{\prime}$ are the logical operators \emph{or, and,
negation} \footnote{We take the informal point of view that precedence of
operators, grouping of terms etc. are well understood from the context and do
not require special explanation. Similalrly, we treat the use of parentheses
in a completely informal manner.}. Note that $\mathbf{V}_{1}\subseteq
\mathbf{S}_{1}$.

Next we build $\mathbf{V}_{2}$, the set of \emph{2nd level elementary
sentences}:
\[
\mathbf{V}_{2}=\left\{  \text{``Tr}\left(  B\right)  =b\text{''}%
:B\in\mathbf{S}_{1},b\in\left[  0,1\right]  \right\}
\]
where Tr$\left(  ...\right)  $ is shorthand for ``The truth value of ...''; in
other words, ``Tr$\left(  B\right)  =b$'' means ``The truth value of $B$ is
$b$''. Finally, we recursively build $\mathbf{S}_{2}$, the set \emph{2nd level
sentences}:
\begin{align}
\text{If }C  &  \in\mathbf{V}_{2}\text{ then }C\in\mathbf{S}_{2}\nonumber\\
\text{If }D_{1},D_{2}  &  \in\mathbf{S}_{2}\text{ then }D_{1}\vee D_{2}%
,D_{1}\wedge D_{2},D_{1}^{\prime}\in\mathbf{S}_{2.} \label{eq002}%
\end{align}
where $\vee,\wedge,^{\prime}$ again stand for or, and, negation \footnote{From
the mathematical point of view, the symbols $\vee,\wedge,^{\prime}$ denote
different operators in (\ref{eq001}) and in (\ref{eq002}): in the first case
they operate on elements of $\mathbf{S}_{1}$ while in the second case they
operate on elements of $\mathbf{S}_{2}$; but their logical interpretation is
the same in both cases.}. Note that $\mathbf{V}_{2}\subseteq\mathbf{S}_{2}$.

We will occasionally use the terms \emph{elementary truth value assessments
}for the elements of $\mathbf{V}_{2}$ and \emph{truth value assessments }for
the elements of $\mathbf{S}_{2}$. Also, we define a special subset of
\ $\mathbf{V}_{2}$ and the corresponding subset of $\mathbf{S}_{2}$ as
follows. The elementary \emph{Boolean }truth value assessments are denoted by
$\widetilde{\mathbf{V}}_{2}$ and defined by
\[
\widetilde{\mathbf{V}}_{2}=\left\{  \text{``Tr}\left(  B\right)
=b\text{''}:B\in\mathbf{S}_{1},b\in\left\{  0,1\right\}  \right\}  .
\]
The \emph{Boolean }truth value assessments are denoted by $\widetilde
{\mathbf{S}}_{2}$ and defined recursively as follows
\begin{align*}
\text{If }C  &  \in\widetilde{\mathbf{V}}_{2}\text{ then }C\in\widetilde
{\mathbf{S}}_{2}\\
\text{If }D_{1},D_{2}  &  \in\widetilde{\mathbf{S}}_{2}\text{ then }D_{1}\vee
D_{2},D_{1}\wedge D_{2},D_{1}^{\prime}\in\widetilde{\mathbf{S}}_{2.}%
\end{align*}
Again we have $\widetilde{\mathbf{V}}_{2}\subseteq\widetilde{\mathbf{S}}_{2}$.

Obviously, we can keep building up the hierarchy of sentences, defining
$\mathbf{V}_{n}$ in terms of $\mathbf{S}_{n-1}$ and $\mathbf{S}_{n}$ in terms
of $\mathbf{V}_{n}$; but going up to $\mathbf{V}_{2},\mathbf{S}_{2}$ will be
sufficient for the purposes of this paper. Let us conclude this section by
giving some examples of elements from $\mathbf{V}_{1},\mathbf{S}%
_{1},\mathbf{V}_{2},\mathbf{S}_{2}$.
\begin{align*}
\mathbf{V}_{1}  &  :A_{1},A_{2},...,A_{M}.\\
\mathbf{S}_{1}  &  :A_{1}\vee A_{3},A_{2}^{\prime},\left(  A_{2}\wedge
A_{4}\right)  \vee A_{5}^{\prime}\text{ ... etc.}\\
\mathbf{V}_{2}  &  :\text{``Tr}\left(  A_{1}\right)  =1\text{'', ``Tr}\left(
A_{7}^{\prime}\right)  =0\text{'', ``Tr}\left(  \left(  A_{2}\vee
A_{3}\right)  \wedge A_{1}\right)  =0.3\text{'' ... etc.}\\
\mathbf{S}_{2}  &  :\left[  \text{``Tr}\left(  A_{1}^{\prime}\right)
=0\text{''}\wedge\text{ ``Tr}\left(  \left(  A_{1}\vee A_{4}\right)  \wedge
A_{2}\right)  =0.3\text{''}\right]  \vee\text{``Tr}\left(  A_{3}\right)
=0.8\text{'' ... etc.}%
\end{align*}

\section{Truth Value Assignment}

\label{sec03}

Our main goal is to assign truth values to sentences of the 2nd level (i.e. to
elements of $\mathbf{S}_{2}$). As it turns out we will also need, as an
intermediate step, to assign truth values to sentences of the 1st level (i.e.
to elements of $\mathbf{S}_{1}$). In this section we present two ways of
achieving this. Our main interest is in the second method, presented in
Section \ref{sec0302}, the so-called \emph{implicit truth }value assignment,
which makes use of the self-referential nature of a particular collection.
However, for purposes of comparison, in Section \ref{sec0301} we will review
the ``classical'', \emph{explicit }method for truth value assignment to
\emph{non}-self-referential sentences.

\subsection{Explicit Truth Value Assignment}

\label{sec0301}

The following explicit method of assigning truth values to elements of
$\mathbf{S}_{1}$ is well known. We start by assigning an arbitrary truth value
to every element of $\mathbf{V}_{1}$ (1st level variable). This is equivalent
to selecting a function $x:\mathbf{V}_{1}\rightarrow\left[  0,1\right]  $,
i.e. $\forall A_{m}\in\mathbf{V}_{1}$ we have Tr$\left(  A_{m}\right)
=x\left(  A_{m}\right)  $; for the sake of brevity we will henceforth use the
simpler notation
\[
\forall A_{m}\in\mathbf{V}_{1}:\text{Tr}\left(  A_{m}\right)  =x_{m}.
\]
Next, take any $B\in\mathbf{S}_{1}$; it is a logical formula with variables
$A_{1},...,A_{M}$. If we replace every occurrence of $A_{m}$ with $x_{m}$ then
we obtain a numerical formula containing the variables $x_{1},...,x_{M}$ and
the operators $\vee,\wedge,^{\prime}$. To obtain the truth value of $B$ we
choose a particular \emph{numerical implementation of }$\vee,\wedge,^{\prime}$
and perform the numerical calculations . I.e.
\begin{align*}
B &  =F_{B}\left(  A_{1},...,A_{m}\right)  \\
\text{Tr}\left(  B\right)   &  =f_{B}\left(  \text{Tr}\left(  A_{1}\right)
,...,\text{Tr}\left(  A_{M}\right)  \right)
\end{align*}
or, more concisely,%
\[
\text{Tr}\left(  B\right)  =f_{B}\left(  x_{1},...,x_{M}\right)  .
\]
Here $F_{B}$ is a \emph{logical} formula ($F_{B}:\mathbf{V}_{1}\rightarrow
\mathbf{S}_{1}$) and $f_{B}$ is the corresponding \emph{numerical }formula
($f_{B}:\left[  0,1\right]  ^{M}\rightarrow\left[  0,1\right]  $) obtained by
replacing $A_{m}$ with $x_{m}$ and now understanding the symbols $\vee
,\wedge,^{\prime}$ as \emph{numerical }operators (specific examples of the
procedure appear in the Appendix). The implementation of the logical operators
$\vee,\wedge,^{\prime}$ by numerical operators, namely \emph{t-conorms},
\emph{t-norms} and \emph{negations},\emph{\ }has been studied extensively by
fuzzy logicians \cite{Klir,Nguyen}. Several typical implementations are
presented in Table 1.
\[%
\begin{tabular}
[c]{|c|c|c|c|}\hline
\textbf{Family} & $x\wedge y$ & $x\vee y$ & $x^{\prime}$\\\hline\hline
Standard & min($x,y$) & max($x,y$) & $1-x$\\\hline
Algebraic & $xy$ & $x+y-xy$ & $1-x$\\\hline
Bounded & max($0,x+y-1$) & min($1,x+y$) & $1-x$\\\hline
Drastic & $\left(
\begin{array}
[c]{cl}%
x & \text{when }y=1\\
y & \text{when }x=1\\
0 & \text{else}%
\end{array}
\right)  $ & $\left(
\begin{array}
[c]{cl}%
x & \text{when }y=0\\
y & \text{when }x=0\\
1 & \text{else}%
\end{array}
\right)  $ & $1-x$\\\hline
\end{tabular}
\ \ \ \ \ \ \ \
\]

\begin{center}
\textbf{Table 1}
\end{center}

In the above manner, the truth value assignment originally defined on
$\mathbf{V}_{1}$ (i.e. Tr$\left(  A_{m}\right)  =x_{m}$, $m=1,2,...,M$)\ has
been extended to $\mathbf{S}_{1}$. Now we can use the truth values of 1st
level sentences to assign truth values to 2nd level elementary sentences as
follows. Given a $C\in\mathbf{V}_{2}$, which has the form
\[
C=\text{``Tr}\left(  B\right)  =b\text{''}%
\]
with $B\in\mathbf{S}_{1}$, $b\in\left[  0,1\right]  $, we \emph{define}
\begin{equation}
\text{Tr}\left(  C\right)  =1-\left|  \text{Tr}\left(  B\right)  -b\right|  .
\label{eq0203}%
\end{equation}
Tr$\left(  B\right)  $ in (\ref{eq0203}) has already been defined, since
$B\in\mathbf{S}_{1}$. Note that, according to (\ref{eq0203}), the maximum
truth value of $C$ is 1 and it is achieved when Tr$\left(  B\right)  =b$; the
latter is exactly what $C$ says. More generally, the truth value of $C$ is a
decreasing function of the absolute difference between Tr$\left(  B\right)  $
and $b$. This certainly appears reasonable\footnote{For further justification
of (\ref{eq0203}) see \cite{Grim1}. Note however, that a number of other
functions could be used; a simple example is Tr$\left(  C\right)  =1-\left(
\text{Tr}\left(  B\right)  -b\right)  ^{2}$.}. In this manner we can compute
the truth value of every $C\in\mathbf{V}_{2}$. Finally, we can extend truth
values from $\mathbf{V}_{2}$ to $\mathbf{S}_{2}$ in exactly the same manner as
we extended truth values from $\mathbf{V}_{1}$ to $\mathbf{S}_{1}$.

Hence starting with a truth value assignment $x_{1},...,x_{M}$ on
$\mathbf{V}_{1}$ we have obtained a truth value assignment for every
$D\in\mathbf{S}_{2}$, namely
\[
\text{Tr}\left(  D\right)  =f_{D}\left(  x_{1},...,x_{M}\right)
\]
where $f_{D}:\left[  0,1\right]  ^{M}\rightarrow\left[  0,1\right]  $. This
\emph{explicit }assignment of truth values does not involve any self-reference
or circularity: starting with the initial specification of the truth values of
$A_{1},...,A_{M}$ as $x_{1},...,x_{M}$, the truth values of all (1st and 2nd
level)\ sentences are obtained as functions of $x_{1},...,x_{M}$.

\subsection{Implicit Truth Value Assignment}

\label{sec0302}

Now suppose that we have a collection of $M$ sentences, which talk about the
truth values of each other. We will show a method by which the information
each sentence conveys about the truth value of the other sentences (possibly
also about its own truth value)\ can be used to determine (up to a point) the
truth values in a consistent manner.

The previously presented framework gives a formal description of sentences
which talk about the truth values of other sentences. In particular, members
of $\mathbf{S}_{2}$ talk about the truth values of members of $\mathbf{S}%
_{1}\supseteq\mathbf{V}_{1}$. Our approach can be summarized as follows.

Suppose we are given $M$ self-referential sentences and consider the set
$\mathbf{S}_{2}$ which is generated from $M$ elementary sentences. We can pick
sentences $D_{1},$..., $D_{M}$ $\in\mathbf{S}_{2}$ which have the same
structure as the original self-referential sentences (examples of the
procedure appear in the Appendix). The only difference is that the $M$
self-referential sentences talk about each other, while $D_{1},$..., $D_{M}$
talk about some elementary, unspecified sentences $A_{1}$, ..., $A_{M}$.
However, since $A_{1}$, ..., $A_{M}$ are unspecified, we can identify $A_{m}$
with $D_{m}$ (for $m\in\left\{  1,...,M\right\}  $). Intuitively, this means
that $D_{m}$ says something about the truth values of $A_{1},$..., $A_{M}$,
i.e. about the truth values of $D_{1},$..., $D_{M}$. This is exactly the
situation which we were trying to model in the first place. While
philosophical objections can be raised about this type of self-reference, the
situation is quite straightforward from the computational point of view (as
will be seen in the following) and allows the determination of consistent
truth values.

Let us now give the mathematical details. As mentioned at the end of Section
\ref{sec0301}, the truth value of every 2nd level sentence $D\in\mathbf{S}%
_{2}$ (for fixed $M$ and a specific choice of t-norm, t-conorm and negation)
is a numerical function $f_{D}\left(  x_{1},...,x_{M}\right)  $, the
independent variables $x_{1},...,x_{M}$ being the truth values of
$A_{1},...,A_{M}$. To obtain specific truth values by the procedure of Section
\ref{sec0301}, it is necessary to specify $x_{1},...,x_{M}$. Choose a function
$\Phi:\left\{  1,2,...,M\right\}  \rightarrow\mathbf{S}_{2}$. $\Phi\left(
1\right)  $, $\Phi\left(  2\right)  $, ..., $\Phi\left(  M\right)  $ are 2nd
level sentences\footnote{Obviously, in a particular situation we will choose
these sentences which have the same structure as the self-referntial sentences
in which we are interested -- see the Appendix.} which can also be denoted as
$D_{1},$..., $D_{M}$. Now (for $m\in\left\{  1,...,M\right\}  $) $D_{m}$ is a
logical formula $F_{m}\left(  A_{1},...,A_{M}\right)  $. In other words, we
have (for $m=1,2,...,M$):
\begin{align*}
D_{1}  &  =F_{1}\left(  A_{1},...,A_{M}\right) \\
D_{2}  &  =F_{2}\left(  A_{1},...,A_{M}\right) \\
&  ...\\
D_{M}  &  =F_{M}\left(  A_{1},...,A_{M}\right)  .
\end{align*}
Let us form the system of logical equations
\begin{align}
A_{1}  &  =D_{1}=F_{1}\left(  A_{1},...,A_{M}\right) \nonumber\\
A_{2}  &  =D_{2}=F_{2}\left(  A_{1},...,A_{M}\right) \nonumber\\
&  ...\label{eq0205}\\
A_{M}  &  =D_{M}=F_{M}\left(  A_{1},...,A_{M}\right)  .\nonumber
\end{align}
The ``natural'' interpretation of (\ref{eq0205}) is that $A_{m}$ \emph{says }
(or is, or means) $D_{m}$. (\ref{eq0205}) implies that:
\begin{align}
\text{Tr}\left(  A_{1}\right)   &  =\text{Tr}\left(  D_{1}\right)
=f_{1}\left(  \text{Tr}\left(  A_{1}\right)  ,...,\text{Tr}\left(
A_{M}\right)  \right) \nonumber\\
\text{Tr}\left(  A_{2}\right)   &  =\text{Tr}\left(  D_{2}\right)
=f_{2}\left(  \text{Tr}\left(  A_{1}\right)  ,...,\text{Tr}\left(
A_{M}\right)  \right) \nonumber\\
&  ...\label{eq0206}\\
\text{Tr}\left(  A_{M}\right)   &  =\text{Tr}\left(  D_{M}\right)
=f_{M}\left(  \text{Tr}\left(  A_{1}\right)  ,...,\text{Tr}\left(
A_{M}\right)  \right) \nonumber
\end{align}
where $f_{m}:\left[  0,1\right]  ^{M}\rightarrow\left[  0,1\right]  $ is the
numerical formula obtained (by the procedure of Section \ref{sec0301}) from
$F_{m}$. A simpler way to write (\ref{eq0206}) is
\begin{align}
x_{1}  &  =f_{1}\left(  x_{1},...,x_{M}\right) \nonumber\\
x_{2}  &  =f_{2}\left(  x_{1},...,x_{M}\right) \nonumber\\
&  ...\label{eq0207}\\
x_{M}  &  =f_{M}\left(  x_{1},...,x_{M}\right)  .\nonumber
\end{align}
(\ref{eq0207}) is a system of $M$ numerical equations in $M$ unknowns; we will
refer to it as the system of \emph{truth value equations}. Note that in
general the truth value equations will be nonlinear.

Depending on the particular $\Phi$ used, (\ref{eq0207})\ may have none, one or
more than one solutions in $\left[  0,1\right]  ^{M}.$ Hence, by specifying a
particular $\Phi$, we obtain a \emph{set} of possible \emph{consistent }truth
value assignments for the 1st level elementary sentences. In other words,
every solution of (\ref{eq0207})\ is a consistent truth value assignment. At
this point we must consider the possibility that the set of solutions is
empty, i.e. that there is no consistent truth value assignment. However, as we
will soon see, under mild conditions there always exists at least one
consistent assignment. Assuming that (\ref{eq0207})\ has one or more
solutions, we can choose one of these to assign truth values to the 1st level
elementary sentences; next, using exactly the same construction as in Section
\ref{sec0301}, we can assign truth values to 1st level sentences, then to 2nd
level elementary sentences and finally to 2nd level sentences. In particular,
it is easy to check that at the end of the procedure the 2nd level sentences
$D_{1},...,D_{M}$ will receive the truth values originally specified by the
solution of (\ref{eq0207}) -- hence the truth value assignment is, indeed, consistent.

Let us now show that, under mild conditions, every $\Phi$ function specifies
\emph{at least} one consistent truth value assignment. This is the subject of
Proposition \ref{cnt0303}. However, we first need two auxiliary propositions.

\begin{proposition}
\label{cnt0301}If the implementations of $\vee,\wedge,^{\prime}$ are,
respectively, a continuous t-conorm, a continuous t-norm and a continuous
negation, then $f_{1},f_{2},...,f_{M}$ in (\ref{eq0207})\ are continuous
functions of $\left(  x_{1},x_{2},...,x_{M}\right)  $.
\end{proposition}

\begin{proof}
We give a sketch of the proof (we omit the complete proof for the sake of
brevity; the basic idea is clear enough.). Suppose that $\vee,\wedge,^{\prime
}$ are a continuous t-conorm, t-norm and negation. Take some $m\in\left\{
1,2,...,M\right\}  $. Recall that $f_{m}\left(  x_{1},x_{2},...,x_{M}\right)
=$Tr$\left(  D_{m}\right)  $, where $D_{m}\in\mathbf{S}_{2}$. Now, take any
$B\in\mathbf{S}_{1}$; then Tr$\left(  B\right)  =y\left(  x_{1},x_{2}%
,...,x_{M}\right)  $ and $y$ is a finite combination of $\vee,\wedge,^{\prime
}$ and $x_{1},...,x_{M}$, which is clearly a continuous function of the
vector\footnote{We use column vectors; the superscript $T$ indicates the
transpose.} $\left(  x_{1},x_{2},...,x_{M}\right)  ^{T}$. Furthermore, let
$C=$``Tr$\left(  B\right)  =b$''; then
\[
\text{Tr}\left(  C\right)  =1-\left|  \text{Tr}\left(  B\right)  -b\right|
=1-\left|  y\left(  x_{1},x_{2},...,x_{M}\right)  -b\right|
\]
which is also a continuous function of $\left(  x_{1},x_{2},...,x_{M}\right)
^{T}$. Since this is true for every $B\in\mathbf{S}_{1}$ and every
$b\in\left[  0,1\right]  $, we conclude that Tr$\left(  C\right)  $ is a
continuous function of $x=\left(  x_{1},x_{2},...,x_{M}\right)  ^{T}$ for
\emph{every} $C\in\mathbf{V}_{2}$. Finally, since $D_{m}\in\mathbf{S}_{2}$,
and Tr$\left(  D_{m}\right)  $ is a finite combination of $\vee,\wedge
,^{\prime}$ and a finite number of terms Tr$\left(  C_{1}\right)  $,
Tr$\left(  C_{2}\right)  $, ..., Tr$\left(  C_{L}\right)  $ (where
$C_{1},C_{2},...,C_{L}\in\mathbf{V}_{2}$) it follows that Tr$\left(
D_{m}\right)  $ is a continuous function of $\left(  x_{1},x_{2}%
,...,x_{M}\right)  ^{T}$.
\end{proof}

\begin{proposition}
\label{cnt0302}Suppose that $X$ is a nonempty, compact, convex set in $R^{M}$.
If the function $f:X\rightarrow X$ is continuous, then there exists at least
one fixed point $\overline{x}\in X$ satisfying
\[
\overline{x}=f\left(  \overline{x}\right)  .
\]
\end{proposition}

\begin{proof}
This the well-known \emph{Brouwer's Fixed Point Theorem}. Its proof can be
found in a number of standard texts, for instance in \cite[pp.323-329]%
{Binmore}.
\end{proof}

Now we can easily prove the existence of consistent truth value assignments.

\begin{proposition}
\label{cnt0303}If the implementations of $\vee,\wedge,^{\prime}$ are,
respectively, a continuous t-conorm, a continuous t-norm and a continuous
negation, then (\ref{eq0207})\ has at least one solution $\overline{x}=\left(
\overline{x}_{1},\overline{x}_{2},...,\overline{x}_{M}\right)  ^{T}\in\left[
0,1\right]  ^{M}$.
\end{proposition}

\begin{proof}
We define the vector function $f\left(  x_{1},x_{2},...,x_{M}\right)  \ $as
follows:
\[
f\left(  x_{1},x_{2},...,x_{M}\right)  =\left(  f_{1}\left(  x_{1}%
,x_{2},...,x_{M}\right)  ,f_{2}\left(  x_{1},x_{2},...,x_{M}\right)
,...,f_{M}\left(  x_{1},x_{2},...,x_{M}\right)  \right)  ^{T}%
\]
where (for $m\in\left\{  1,2,...,M\right\}  $) $f_{m}\left(  x_{1}%
,x_{2},...,x_{M}\right)  $ is the function appearing in (\ref{eq0207}). Since
$f_{m}\left(  x_{1},x_{2},...,x_{M}\right)  $ computes a truth value, we have
$f_{m}:\left[  0,1\right]  ^{M}\rightarrow\left[  0,1\right]  $ and hence
$f:\left[  0,1\right]  ^{M}\rightarrow\left[  0,1\right]  ^{M}$ . Furthermore,
by Proposition \ref{cnt0301} each $f_{m}$ is a continuous function and so $f$
is also a continuous function. Now we can apply Proposition \ref{cnt0302} with
$X=\left[  0,1\right]  ^{M}$.
\end{proof}

When we use Boolean truth value assessments, we can prove an additional result
about consistent truth value assignments.

\begin{proposition}
\label{cnt0304}Suppose that in (\ref{eq0205}) $D_{1},D_{2},...,D_{M}%
\in\widetilde{\mathbf{S}}_{2}$ and the implementations of $\vee,\wedge
,^{\prime}$ are, respectively, max, min and the standard negation. Then
(\ref{eq0207}) admits the solution $\left(  1/2,1/2,...,1/2\right)  ^{T}$.
\end{proposition}

\begin{proof}
Take any $m\in\left\{  1,2,...,M\right\}  $; then $F_{m}\left(  A_{1}%
,A_{2},...,A_{M}\right)  $ is a combination (through $\vee,\wedge,^{\prime}$)
of a finite number of elements $C_{1},C_{2},...,C_{L}\in\widetilde{\mathbf{V}%
}_{2}$. Take any $C_{l}$ (with $l\in\left\{  1,2,...,L\right\}  $); it has the
form
\begin{equation}
C_{l}=\text{``Tr}\left(  B_{l}\right)  =b_{l}\text{''} \label{eaq0301}%
\end{equation}
where $B_{l}\in\mathbf{S}_{2}$ and $b_{l}\in\left\{  0,1\right\}  $. The
corresponding numerical term will have the form
\begin{equation}
\text{Tr}\left(  C_{l}\right)  =1-\left|  \text{Tr}\left(  B_{l}\right)
-b_{l}\right|  . \label{eq0302}%
\end{equation}
or
\begin{equation}
z_{l}=1-\left|  y_{l}-b_{l}\right|  \label{eq0303}%
\end{equation}
where $y_{l}=$Tr$\left(  B_{l}\right)  $ and $z_{l}=$Tr$\left(  C_{l}\right)
$. Now, $y_{l}$ will be a finite combination of $x_{1}$, ..., $x_{M}$ through
max, min and negation operators, hence when $x_{1}$= $x_{2}$=$...$= $x_{M}$=
$1/2$ we also get $y_{l}=1/2$. Then, for $b_{l}\in\left\{  0,1\right\}  $ we
also get from (\ref{eq0303}) that $z_{l}=1/2.$

Hence every term appearing in $f_{m}\left(  1/2,1/2,...,1/2\right)  $ (the
numerical version of $F_{m}\left(  A_{1},A_{2},...,A_{M}\right)  $) will be
equal to 1/2. Since these terms will be combined with max, min and negation
operators it follows that $f_{m}\left(  1/2,1/2,...,1/2\right)  =1/2$ and this
satisfies the $m$-th truth value equation:
\begin{equation}
x_{m}=\frac{1}{2}=f_{m}\left(  1/2,1/2,...,1/2\right)  . \label{eq0304}%
\end{equation}
Since (\ref{eq0304})\ holds for every $m\in\left\{  1,2,..,M\right\}  $, it
follows that (\ref{eq0207}) admits the solution $\left(
1/2,1/2,...,1/2\right)  ^{T}$.
\end{proof}

\subsection{Discussion}

\label{sec0303}

We have seen that the assignment of consistent truth values to a collection of
self-referential propositions can be reduced to solving the system of
(algebraic) truth value equations. Furthermore, Proposition \ref{cnt0303}
shows that, subject to some mild continuity conditions, the truth value
equations admit at least one solution.

Proposition \ref{cnt0303} indicates that the Liar and related self referential
collections cease to be paradoxical in the context of fuzzy logic. From the
mathematical point of view the situation is rather straightforward. A system
of truth value equations may have no solution in $\left\{  0,1\right\}  ^{M}$;
by expanding to $\left[  0,1\right]  ^{M}$ the set in which we seek solutions
we can guarantee that the system \emph{always} has at least one solution. Of
course, this result is achieved at the price of admitting fuzzy solutions,
i.e. truth values which fall short of certainty.

Next, let us briefly compare explicit and implicit truth value assignment.
Explicit assignment is straightforward: starting with the initial (and more or
less arbitrary) choice of truth values for the 1st level variables, the truth
values of 1st and 2nd level sentences are uniquely determined. If initial
truth values were restricted to be either 0 or 1, then the initial choice
would essentially be a choice of axioms, i.e. the assertion of certain
propositions (or their negations); choosing truth values in $\left[
0,1\right]  $ can be seen as a ``\emph{generalized axiomatization}''.
Generally, in implicit truth value assignment the truth values of 1st and 2nd
level sentences are not uniquely determined; rather a number of possible
consistent truth values are obtained. There are of course cases (examples will
be presented in the Appendix) where there is a single consistent truth value
assignment. One is tempted to remark that in such cases the collection of
self-referential sentences is equipped with an \emph{implicit axiomatization}.

Let us now return to the system of truth value equations. We have established
that it always has a solution; but many additional questions can be asked
about it. Let us list a few such questions\footnote{We make no attempt to
\emph{answer }these questions in the current paper; they will be the subject
of future research.}.

\begin{enumerate}
\item Under what conditions does (\ref{eq0207}) possess a unique solution?

\item Are some solutions ``better'' than other? For instance, solutions in the
interior of the hypercube $\left[  0,1\right]  ^{M}$ are ``fuzzier'' than
solutions on the boundary, which in turn are fuzzier than solutions on the
vertices. What are existence and uniqueness conditions for vertex or boundary solutions?

\item What is the structure of the set of solutions? For instance:

\begin{enumerate}
\item are there conditions under which it is a vector space?\ 

\item what is its dimension (vector space dimension, Hausdorff dimension etc.)?

\item is it equipped with a ``natural'' order? is it a lattice?
\end{enumerate}

\item Assuming that some of the above properties are established, are they
invariant under different implementations of $\vee,\wedge,^{\prime}$ and/ or
the function Tr$\left(  \cdot\right)  $?
\end{enumerate}

One can also ask computationally oriented questions; the most obvious one is:
``how to solve the truth value equations?''. There is a large number of root
finding algorithms that can be used to this end and we will discuss some of
them in the next section. However, we believe it is useful to distinguish
between properties of the truth value equations (such as the properties listed
above)\ and algorithm properties. We will further discuss this distinction in
Section \ref{sec0404} \footnote{Note that in our formulation the numbers of
unknowns and equations are both equal to $M$. This is a natural consequence of
the formulation, since both the unknowns and equations are corresponded to the
original $M$ self-referential sentences. From the mathematical point of view,
since every variable is associated to an equation of the form \ $x_{m}%
=f_{m}\left(  x_{1},...,x_{M}\right)  $, the number of variables \emph{must}
equal the number of equations. Of course, the $m$-th function $f_{m}$ will
generally \emph{not} depend on all of $x_{1},...,x_{M}$.
\par
One could also consider more complicated situations, for instance a collection
of $M$ sentences which can be separated into two groups of sizes $M_{1}$ and
$M_{2}$(with $M_{1}+M_{2}=M$) having the following property: the sentences of
the first group talk only about this group, while the sentences of the second
group talk about both groups. This is a self-referential collection which
contains a \emph{closed} self-referential sub-collection (namely the first
group). The sub-collection corresponds to a smaller system of truth value
equations; the solutions obtained from this sub-system of $M_{1}$ equations
will in general be different from the ones obtained from the original system
of $M$ equations.
\par
Another possibility is to consider a collection of $N$ sentences for which
truth values are given and fixed and then a second collection of $M$ sentences
which talk about themselves and the first collection of sentences. This still
results in a set of $M$ equations which involve $M$ unknown truth values
$x_{1},x_{2},...,x_{M}$ and $N$ fixed truth values (call them
\emph{parameters}) $z_{1},z_{2},...,z_{N}$. Our Proposition \ref{cnt0303}
still applies and guarantees the existence of a consistent truth value
solution $(x_{1},x_{2},...,x_{M})^{T}$ for every choice of $z_{1}%
,z_{2},...,z_{N}$.}.

\section{Computational Issues}

\label{sec04}

We now discuss a number of algorithms which can be used to solve the truth
value equations.

\subsection{Root Finding}

\label{sec0401}

Probably the most popular method for solving systems of nonlinear equations is
the iterative Newton-Raphson algorithm. A description and analysis can be
found in many numerical analysis textbooks (e.g. \cite{Numerit}). Using
\[
x=\left(  x_{1},...,x_{M}\right)  ^{T},
\]
let us rewrite the truth value equations (\ref{eq0207}) as follows. Define
(for $m\in\left\{  1,2,...,M\right\}  $):
\[
h_{m}\left(  x\right)  =x_{m}-f_{m}\left(  x_{1},...,x_{M}\right)
\]
and
\[
h\left(  x\right)  =\left(  h_{1}\left(  x\right)  ,...,h_{M}\left(  x\right)
\right)  ^{T}.
\]
Then the truth value equations (\ref{eq0207}) are equivalent to%

\[
h\left(  x\right)  =0.
\]
Let us define a time varying vector
\[
x\left(  t\right)  =\left(  x_{1}\left(  t\right)  ,...,x_{M}\left(  t\right)
\right)  ^{T};
\]
then the Newton-Raphson method consists in the iteration ($t=0,1,...$):
\begin{equation}
x\left(  t+1\right)  =x\left(  t\right)  -\left[  G\left(  x\left(  t\right)
\right)  \right]  ^{-1}\cdot h\left(  x\left(  t\right)  \right)
\label{eq0406}%
\end{equation}
with
\[
G\left(  x\right)  =\left(
\begin{array}
[c]{cccc}%
\frac{\partial h_{1}}{\partial x_{1}} & \frac{\partial h_{1}}{\partial x_{2}}
& .. & \frac{\partial h_{1}}{\partial x_{M}}\\
\frac{\partial h_{2}}{\partial x_{1}} & \frac{\partial h_{2}}{\partial x_{2}}
& ... & \frac{\partial h_{2}}{\partial x_{M}}\\
... & ... & ... & ...\\
\frac{\partial h_{M}}{\partial x_{1}} & \frac{\partial h_{M}}{\partial x_{2}}
& ... & \frac{\partial h_{M}}{\partial x_{M}}%
\end{array}
\right)  .
\]
Depending on the choice of t-norms and t-conorms, the partial derivatives of
$h_{1}$,..., $h_{M}$ may be undefined at some points of $\left[  0,1\right]
^{M}$; however at such points $G\left(  x\right)  $ can be approximated by a
differentiable function.

It is usually stated in textbooks that the Newton-Raphson algorithm will
converge to a solution $\overline{x}$ if it starts with $x\left(  0\right)  $
``suficiently close to $\overline{x}$''. But this statement needs some clarifications.

A \emph{fixed point }of (\ref{eq0406}) is a point $\overline{x}$ which has the
following property: if for some $t$ we have $x\left(  t\right)  =\overline{x}%
$, then for $s=1,2,...$ we have $x\left(  t+s\right)  =x\left(  t\right)  $.
For this to hold it is necessary and sufficient that
\begin{equation}
\left[  G\left(  \overline{x}\right)  \right]  ^{-1}\cdot h\left(
\overline{x}\right)  =0. \label{eq0409}%
\end{equation}
It is clear from (\ref{eq0409})\ that every solution $\overline{x}$ of
(\ref{eq0207}) is a fixed point of (\ref{eq0406}) but the converse is not
necessarily true, i.e. (\ref{eq0406}) may have fixed points which are not
solutions of (\ref{eq0207}). Furthermore, a fixed point is called \emph{stable
}if
\[
\lim_{t\rightarrow\infty}x\left(  t\right)  =\overline{x}.
\]
for every choice of $x\left(  0\right)  $ ``sufficiently close'' to
$\overline{x}$; otherwise it is called \emph{unstable}. The iteration of
(\ref{eq0406}) can, by definition, converge only to stable fixed points, i.e.
some solutions of (\ref{eq0207}) cannot be obtained because they are unstable
fixed points.

There is an additional possibility: (\ref{eq0406}) may converge to a solution
of (\ref{eq0207}) which lies outside of $\left[  0,1\right]  ^{M}$. This
problem can be addressed by the following modification of (\ref{eq0406}):

\begin{enumerate}
\item if, for some $t$ and $m,$ (\ref{eq0406}) gives $x_{m}\left(  t\right)
>1$, then set $x_{m}\left(  t\right)  =1$;

\item if, for some $t$ and $m,$ (\ref{eq0406}) gives $x_{m}\left(  t\right)
<0$, then set $x_{m}\left(  t\right)  =0$.
\end{enumerate}

In conclusion, the Newton-Raphson algorithm may fail to find a consistent
truth value assignment for several reasons:

\begin{enumerate}
\item the algorithm may fail to converge;

\item it may converge to a value which does not solve (\ref{eq0207}) or lies
outside of $\left[  0,1\right]  ^{M}$;

\item it may miss a solution which is an unstable fixed point.
\end{enumerate}

Furthermore, there is no obvious way in which to obtain \emph{all} the
solutions of (\ref{eq0207}) using (\ref{eq0406})\footnote{In fact this remark
holds not only for Newton-Raphson, but for all numerical root finding
algorithms of which we are aware.}.

However, as a practical matter, the numerical experiments of the Appendix
indicate that (\ref{eq0406}) generally converges to a consistent truth value
assignment, i.e. a solution of (\ref{eq0207}) belonging to $\left[
0,1\right]  ^{M}$.

\subsection{Inconsistency Minimization}

\label{sec0403}

Root finding can also be formulated (in a standard manner) as a minimization
problem. Let us apply this approach to truth value assignment. We write the
following \emph{inconsistency }function
\begin{equation}
J\left(  x_{1},...,x_{M}\right)  =\sum_{m=1}^{M}\left(  x_{m}-f_{m}\left(
x_{1},...,x_{M}\right)  \right)  ^{2}. \label{eq0421}%
\end{equation}
Note that $J\left(  x_{1},...,x_{M}\right)  $ is a reasonable measure of the
\emph{inaccuracy} in satisfying the truth value equations. For every choice of
$\left(  x_{1},...,x_{M}\right)  ^{T}$, $J$ takes a nonnegative value; every
solution of (\ref{eq0207}) yields a global minimum of $J$. An intuitive
interpretation of (\ref{eq0421}) goes as follows: $\left(  x_{m}-f_{m}\left(
x_{1},...,x_{M}\right)  \right)  ^{2}$ is the inconsistency of the $m$-th
sentence and $\sum_{m=1}^{M}\left(  x_{m}-f_{m}\left(  x_{1},...,x_{M}\right)
\right)  ^{2}$ is the \emph{total }inconsistency of the self-referential
collection. From Proposition \ref{cnt0303} we know that there is at least one
consistent truth value assignment, i.e. a truth value assignment of zero inconsistency.

There is a vast number of minimization algorithms; indeed the ``standard''
sequential reasoning of classical Aristotelian logic can be understood as a
simplistic inconsistency minimization algorithm which attempts to select the
optimizing value of one variable at a time.

In the experiments reported in the Appendix we will use another simple
algorithm, namely \emph{steepest descent minimization}. This consists of the
following iteration (for $t=1,2,...$):
\begin{equation}
x\left(  t+1\right)  =x\left(  t\right)  -k\cdot\frac{\partial J}{\partial x}.
\label{eq0422}%
\end{equation}
The term $\frac{\partial J}{\partial x}$ in (\ref{eq0422}) is the gradient of
$J$ with respect to $x$; at points where $J$ is non-differentiable, similarly
to the Newton-Raphson case, we can use an appropriate approximation by a
differentiable function.

For small enough $k$ every step of (\ref{eq0422})\ yields a decrease of $J$.
Hence (\ref{eq0422})\ will at least converge to the neighborhood of a local
minimum of $J$; however this is not enough to guarantee that (\ref{eq0207}%
)\ is satisfied. Furthermore, there is no guarantee that $x\left(  t\right)  $
as given by (\ref{eq0422}) will always belong to $\left[  0,1\right]  ^{M}$.
However, note that in place of steepest descent one could use a more
sophisticated, \emph{constrained} optimization algorithm to minimize some
appropriate function of $\left(  x_{1},...,x_{M}\right)  ^{T}$ under the
constraints of zero inconsistency and staying in $\left[  0,1\right]  ^{M}$.

Let us also note that the ``inconsistency minimization'' approach is somewhat
related to connectionist cognitive models and could conceivably be used by
actual human reasoners. However, we are not particularly concerned about the
``psychological plausibility'' of our approach, neither do we propose it as a
description of actual human reasoning.

\subsection{A Control-Theoretic Algorithm}

\label{sec0402}

Finally, inspired from control theory, we propose a root-finding algorithm
which can be summarized by the following equation:
\begin{equation}
x\left(  t+1\right)  =x\left(  t\right)  -k\cdot\left(  x\left(  t\right)
-f\left(  x_{1}\left(  t\right)  ,...,x_{M}\left(  t\right)  \right)  \right)
\label{eq0411}%
\end{equation}
The motivation for (\ref{eq0411}) can be explained as follows. Choose any
$m\in\left\{  1,2,...,M\right\}  $. From (\ref{eq0411}) we see that the
difference $x_{m}\left(  t+1\right)  -x_{m}\left(  t\right)  $ is equal to
$-k\cdot\left(  x_{m}\left(  t\right)  -f_{m}\left(  x_{1}\left(  t\right)
,...,x_{M}\left(  t\right)  \right)  \right)  $. The following informal
convergence argument holds provided that $k$ is small and, consequently, the
change between $x_{m}\left(  t\right)  $ and $x_{m}\left(  t+1\right)  $ is
also small. In that case we can simply write $-k\cdot\left(  x_{m}%
-f_{m}\left(  x_{1},...,x_{M}\right)  \right)  $ and argue as follows. If
$x_{m}\left(  t\right)  $ is greater than $f_{m}\left(  x_{1}\left(  t\right)
,...,x_{M}\left(  t\right)  \right)  $, then $x_{m}\left(  t+1\right)  $
becomes smaller than $x_{m}\left(  t\right)  $ and (assuming $k$ to be
sufficiently small) the difference $x_{m}-f_{m}\left(  x_{1},...,x_{M}\right)
$ gets closer to zero. Similarly, if $x_{m}\left(  t\right)  $ is smaller than
$f_{m}\left(  x_{1}\left(  t\right)  ,...,x_{M}\left(  t\right)  \right)  $,
then $x_{m}\left(  t+1\right)  $ becomes larger than $x_{m}\left(  t\right)  $
increases and the difference $x_{m}-f_{m}\left(  x_{1},...,x_{M}\right)  $
again gets closer to zero. Finally, if $\left(  x_{m}-f_{m}\left(
x_{1},...,x_{M}\right)  \right)  $ equals zero, then $x_{m}$ remains
unchanged. Hence, unlike Newton-Raphson and Steepest Descent, the fixed points
of (\ref{eq0411}) are exactly the solutions of the truth value equations.
However, it is not guaranteed that every fixed point is stable.

Eq.(\ref{eq0411}) is very similar to schemes used for the control of
\emph{servomechanisms}; standard control theoretic methods can be used to
investigate the behavior of $x\left(  t\right)  $, i.e. prove rigorously
existence of stable fixed points, convergence, boundedness etc. (for example,
these issues can be investigated by constructing a \emph{Lyapunov function} of
(\ref{eq0411})).

Similarly to steepest descent, the control theoretic algorithm has a certain
degreee of psychological plausibility. Namely, we can imagine a human reasoner
who adjusts truth values in small increments, according to the discrepancy
from the values predicted by the truth value equations.

\subsection{Dynamical Systems and Reasoning}

\label{sec0404}

As is the case for every iterative algorithm, the three algorithms presented
above can be viewed as \emph{dynamical systems}. Hence our approach is
somewhat similar to Grim's, since he also has modelled reasoning about
self-referential sentences as a dynamical system. There is of course an
important difference: Grim appears to be more interested in oscillatory and
chaotic behavior than in convergence to consistent truth values.

While we find the chaotic behavior of Grim's dynamical systems quite
interesting, we believe it is a rather secondary aspect of self-referential
collections. As we have already mentioned, we believe it is a property of a
particular algorithm rather than of the self-referential collection itself. Of
course, the algorithms proposed by Grim are somewhat special, in the sense
that they are inspired by human reasoning. However, the analogy should not be
drawn too far. It is rather unlikely that a human, when reasoning about
self-referential sentences, simulates chaotic dynamical systems with infinite
precision arithmetic. We do not bring this issue up to criticize Grim's
approach\footnote{Neither has Grim claimed that his model describes actual
human reasoning.}. Our point is that both Grim's dynamical systems and the
algorithms we have presented in Sections \ref{sec0401} -- \ref{sec0402} can be
seen as ``generalized reasoning systems''. In other words, despite our
previous remarks about psychological plausibility, the connection of such
systems to human reasoning is rather slender; however, they may reveal
interesting aspects of the \emph{mathematical }structure of self-referential collections.

\section{Bibliographic Remarks}

\label{sec06}

The literature on self-referential sentences is extensive and originates from
many disciplines and points of view. Here we only present a small part of this
literature: papers which we have found related to our investigation.

There is a vast philosophical literature on self-referential sentences and the
related topic of \emph{truth}. This literature goes back to the ancient
Greeks. Some pointers to the early literature can be found in the books
\cite{Etchemendy,Martin1,Martin2,McGee}. These books also cover recent
developments. Regarding the modern literature, we must list Tarski's
fundamental paper on the \emph{truth predicate} \cite{Tarski}. Two other
interesting early papers are \cite{Prior1,Prior2}. Important developments in
the 70's were the concept of \emph{truth gaps} \cite{VanFras1,VanFras2} and
Kripke's theory of truth \cite{Kripke}. Some important papers from the 80's
and 90's are \cite{Goldstein,Gupta,Priest,Tappenden}.

Preliminary concepts of \emph{multi-valued logic} \cite{Rescher} have been
used in connection to the Liar paradox already in the Middle Ages. The notion
of truth gaps is also closely related to multi-valued logic, as is Skyrms'
resolution of the Liar \cite{Skyrms}. The treatment of the Liar by the methods
of fuzzy logic was a natural development. We have already mentioned Zadeh's
paper \cite{Zadeh} which is, as far as we know, the first work dealing with
the Liar in the context of fuzzy logic. We have also mentioned that Grim's
work \cite{Grim1,Grim4,Grim3,Grim2} has been a major motivation for the
current paper. In the fuzzy logic literature there is a considerable amount of
work on self-referential sentences and the concept of truth (for example
\cite{Hajek1,Hajek2}). These approaches fall within the mathematical logic
tradition and are quite different from ours. But some logicians have also
addressed computational issues which are related to our concerns. For example
the concept of \emph{fuzzy} \emph{satisfiability} is addressed in
\cite{Klement1,Klement2,Navara,Wang1}; on the same topic see \cite{Sudarsky}.

We believe that ideas from the areas of \emph{nonmonotonic reasoning }and
\emph{belief revision} could yield fruitful insights on the Liar and related
paradoxes, but we are not aware of any work in this direction. However the
AGU\ axioms for belief revision (see for example \cite{AGU,Gardenfors1} and
the book \cite{Gardenfors2}) appear to us very relevant to the study of
logical paradoxes. Many of these issues are explored in \cite{GuptaBelnap};
further interesting recent work includes \cite{Fuhrmann,Kyburg}. Also,
self-reference is explicitly treated in \cite{Perlis1,Perlis2}.

We consider our approach to self-referential sentences only marginally related
to psychology and cognitive science. However the so-called \emph{coherence}
theory (or theories) of truth, a topic at the interface between psychology and
philosophy, appears related to self-reference. Probably the best known
coherence theory of truth is the one expounded by Thagard in \cite{Thagard1}
and further elaborated in \cite{Thagard3,Thagard4} and several other
publications. Thagard has also applied his theory to reasoning about a set of
sentences, each pair of which may \emph{cohere} or \emph{incohere}. This
theory can also be applied to self-referential sentences (though we are not
aware of work exploring this connection). Thagard has given a computational
formulation of his theory as a \emph{constraint satisfaction} problem
\cite{Thagard3}; this formulation is rather similar to the inconsistency
minimization approach we have presented in Section \ref{sec0403}. The
relationship of coherence to \emph{belief networks }is discussed
\cite{Thagard4}. It is interesting to note that Thagard has also presented a
\emph{connectionist }formulation of the constraint satisfaction problem, which
is related to several connectionist models of \emph{nonmonotonic reasoning
}\cite{Ralescu,Pinkas,Raghu}. Finally, from our point of view, a particularly
interesting paper is \cite{Schoch}, which proposes a fuzzy measure of
coherence. The constraint satisfaction approach has also been applied in the
context of \emph{cognitive dissonance }\cite{CogDis1,CogDis2}. It would be
interesting to apply the theory of cognitive dissonance to reasoning about
self-referential propositions, but as far as we know this has not been done
until now.

Last but not least, the first chapter of Hofstadter's book \cite{Hofstadter}
has a large collection of self-referential sentences.\ Some of these are of a
quite different flavor from the Liar and are not easily formalized; however,
we recommend the book to the reader for a more general view on the problem of self-reference.

\section{Conclusion}

\label{sec07}

In this paper we have presented a fuzzy-logic formulation which can describe a
large family of collections of self-referential sentences; in particular it
can accommodate the Liar, the inconsistent dualist, the consistent dualist and
the \emph{strengthened }Liar (see the Appendix, Section \ref{secA07}). In our
formulation, subject to some mild continuity conditions, every member of this
family admits at least one consistent assignment of fuzzy truth values. Hence
the main contribution of this paper is to expand Zadeh's analysis and to thus
show that the Liar and related self referential collections cease to be
paradoxical in the context of fuzzy logic by the simple expedient of expanding
the set of possible solutions from $\left\{  0,1\right\}  ^{M}$ to $\left[
0,1\right]  ^{M}$. If we accept the price of admitting fuzzy solutions, i.e.
truth values which fall short of certainty, then we can at one stroke resolve
a large number of potential paradoxes.

Furthermore, we have presented several computational approaches to the problem
of finding consistent truth values. In addition to the rather standard
approaches of nonlinear equation solving by Newton-Raphson and by minimization
we have presented an algorithm inspired by control theory which, as
demonstrated by a number of numerical experiments, appears to strike a good
balance between speed of execution, convergence and psychological plausibility
(see also the Appendix, Section \ref{secA08}).

Many issues remain open; in the following paragraphs we list several groups of
questions which we would like to investigate in the future.

First, there are mathematical questions regarding the \emph{solutions of the
truth value equations}. Examples of such questions include existence and
uniqueness of boundary and vertex solutions and invariance of the solutions
under different implementations of the logical operators and the truth
function Tr$\left(  \cdot\right)  $.\ Also, there are questions regarding the
\emph{algorithmic }behavior, for instance the stability of a \ solution under
a particular algorithm. The complete resolution of these questions appears to
be a difficult problem. Perhaps a fruitful first step will be the study of a
restricted family of self-referential collections (for example the study of
self-referential \emph{Boolean }truth value assessments).

A second group of questions concerns \emph{optimization issues}. There is
scope for experimentation with alternative implementations of the Tr$\left(
\cdot\right)  $ function, alternative inconsistency functions (for example
using absolute values rather than squares) and so on. We have also mentioned
that truth value assignment could be formulated as \emph{constrained}
optimization, i.e. as the minimization of some fuzziness criterion subject to
zero inconsistency.

However, we believe that the most interesting variations of the optimization
problem are the ones which attempt to address the issue of \emph{bounded
rationality}. Whatever the attractions of infinite valued logic may \ be, we
find rather implausible that humans reason with a continuum of truth values. A
(perhaps small) step to increase the relevance of our approach to actual human
reasoning would be to minimize inconsistency using a finite (and
small)\ number of truth values. This is an intermdeiate step between the
classical approach of searching for solutions in $\left\{  0,1\right\}  ^{M}$
and the fuzzy logic approach of searching in $\left[  0,1\right]  ^{M}$. While
the formulation of the problem is identical to the one we have presented here,
a major difference is that now there is no guarantee for the existence of a
zero inconsistency solution. Hence a number of mathematical questions arise,
regarding the properties of the minima of the inconsistency function; from the
computational point of view, the use of a finite set of truth values may
require the use of entirely different optimization algorithms.

Let us also note that there are other possibilities for addressing the issue
of bounded rationality, for example the use of interval-valued fuzzy sets, the
description of truth values in terms of tolerance classes etc. These issues
require further research.

Finally, we are very interested in the \emph{coherence }approach to
self-referential sentences. Since self-referential sentences make claims about
each other's truth values it is rather straightforward to setup a network with
one node per sentence and connections which are either reinforcing or
inhbiting (depending on what each sentence says about each other). We have
performed preliminary experiments based on a Boltzmann machine formulation of
the problem; we use a variation of Schoch's fuzzy coherence measure
\cite{Schoch}. These results will be reported in a future publication.

\bigskip\newpage

\appendix    

\section{Appendix: Examples and Experiments}

\label{sec05}

In this appendix we present several examples of our formulation of
self-referential collections. Whenever possible, we solve the truth value
equations analytically; in addition we apply the three algorithms of Section
\ref{sec04} to numerically compute truth values.

\subsection{Example 1: The Liar}

\label{sec0500}

We present our first example very briefly since it has already been addressed
by Zadeh and several other authors. The Liar sentence is
\[
A=\text{``}A\text{ is false''.}%
\]
According to our previous remarks, we define
\[
C=\text{``Tr}\left(  A\right)  \text{=0''}%
\]
and then set
\[
A=C
\]
which implies
\begin{align}
\text{Tr}\left(  A\right)   &  =\text{Tr}\left(  C\right)  \Rightarrow
\nonumber\\
\text{Tr}\left(  A\right)   &  =1-\left|  \text{Tr}\left(  A\right)
-0\right|  \Rightarrow\nonumber\\
x  &  =1-\left|  x-0\right|  \Rightarrow\nonumber\\
x  &  =1-x \label{eq0502}%
\end{align}
where we have set $x=$Tr$\left(  A\right)  $. (\ref{eq0502}) is the truth
value equation and it has the unique solution $x=1/2$.

\subsection{Example 2: The Inconsistent Dualist}

\label{sec0501}

Our next example is the inconsistent dualist:
\begin{align}
A_{1}  &  :\text{``}A_{2}\text{ is true''}\label{eq0511}\\
A_{2}  &  :\text{``}A_{1}\text{ is false''.} \label{eq0512}%
\end{align}

In this case the 1st level elementary sentences are $\mathbf{V}_{1}=\left\{
A_{1},A_{2}\right\}  $. We will use two 2nd level elementary sentences
\begin{align*}
C_{1}  &  :\text{``Tr}\left(  A_{2}\right)  =1\text{''}\\
C_{2}  &  :\text{``Tr}\left(  A_{1}\right)  =0\text{''.}%
\end{align*}
Note that $C_{1}$ and $C_{2}$ are Boolean elementary truth value assessments,
i.e. $C_{1},C_{2}\in\widetilde{\mathbf{V}}_{2}\subseteq\widetilde{\mathbf{S}%
}_{2}$. The translation of (\ref{eq0511}), (\ref{eq0512}) consists in mapping
$A_{1}$ to $C_{1}$ and $A_{2}$ to $C_{2}$; then we have the logical equations
\begin{align*}
A_{1}  &  =C_{1}\\
A_{2}  &  =C_{2}%
\end{align*}
from which follows
\begin{align*}
\text{Tr}\left(  A_{1}\right)   &  =\text{Tr}\left(  C_{1}\right)  =1-\left|
\text{Tr}\left(  A_{2}\right)  -1\right| \\
\text{Tr}\left(  A_{2}\right)   &  =\text{Tr}\left(  C_{2}\right)  =1-\left|
\text{Tr}\left(  A_{1}\right)  -0\right|  .
\end{align*}
Substituting Tr$\left(  A_{m}\right)  $ $\ $with $x_{m}$ ($m=1,2$) we get
\begin{align*}
x_{1}  &  =1-\left|  x_{2}-1\right| \\
x_{2}  &  =1-\left|  x_{1}-0\right|
\end{align*}
and, since $x_{1},x_{2}\in\left[  0,1\right]  $, we finally get the truth
value equations
\begin{align}
x_{1}  &  =x_{2}\label{eq051a}\\
x_{2}  &  =1-x_{1}. \label{eq051b}%
\end{align}
Obviously, eqs. (\ref{eq051a})--(\ref{eq051b})\ have the unique solution
$\overline{x}=\left(  \overline{x}_{1},\overline{x}_{2}\right)  =\left(
1/2,1/2\right)  $.

The inconsistency function is
\[
J\left(  x_{1},x_{2}\right)  =\left(  x_{1}-x_{2}\right)  ^{2}+\left(
x_{1}+x_{2}-1\right)  ^{2}%
\]
and has partial derivatives
\begin{align*}
\frac{\partial J}{\partial x_{1}}  &  =\frac{\partial}{\partial x_{1}}\left(
\left(  x_{1}-x_{2}\right)  ^{2}+\left(  x_{1}+x_{2}-1\right)  ^{2}\right)
=\allowbreak4x_{1}-2\\
\frac{\partial J}{\partial x_{2}}  &  =\frac{\partial}{\partial x_{2}}\left(
\left(  x_{1}-x_{2}\right)  ^{2}+\left(  x_{1}+x_{2}-1\right)  ^{2}\right)
=\allowbreak4x_{2}-2.
\end{align*}
Setting the partial derivatives equal to zero we obtain
\begin{align}
\frac{\partial J}{\partial x_{1}}  &  =\allowbreak4x_{1}-2=0\label{eq051i}\\
\frac{\partial J}{\partial x_{2}}  &  =4x_{2}-2=0 \label{eq051j}%
\end{align}
and, obviously, eqs. (\ref{eq051i})--(\ref{eq051j}) have the unique solution
$\overline{x}=\left(  \overline{x}_{1},\overline{x}_{2}\right)  =\left(
1/2,1/2\right)  $ which gives the unique truth value assignment of zero inconsistency.

Furthermore, using the steepest descent algorithm we obtain the the dynamical
system
\begin{align}
x_{1}\left(  t+1\right)   &  =x_{1}\left(  t\right)  -k\cdot\left(
4x_{1}\left(  t\right)  -2\right) \label{eq051g}\\
x_{2}\left(  t+1\right)   &  =x_{2}\left(  t\right)  -k\cdot\left(
4x_{2}\left(  t\right)  -2\right)  . \label{eq051f}%
\end{align}
This is a linear system which can be written in matrix notation as
\[
x\left(  t+1\right)  =Ax\left(  t\right)  +b
\]
with
\[
A=\left(
\begin{array}
[c]{cc}%
1-4k & 0\\
0 & 1-4k
\end{array}
\right)  ,\qquad b=\left(
\begin{array}
[c]{c}%
2k\\
2k
\end{array}
\right)
\]
$A$ has double eigenvalue: $\lambda_{1}=\lambda_{2}=1-4k$. We have (for
$i=1,2$)
\[
\left|  \lambda_{i}\right|  =1-4k<1
\]
for sufficiently small $k$. Hence (\ref{eq051g})--(\ref{eq051f}) is stable.
The unique fixed point $\overline{x}$ is obtained by solving
\[
x=Ax+b.
\]
The solution is $\overline{x}=\left(  1/2,1/2\right)  $. Because of stability
we have for every $x\left(  0\right)  $ that
\[
\lim_{t\rightarrow\infty}x\left(  t\right)  =\overline{x}.
\]

This result is verified by numerical simulation. Using $k=0.1$ and starting
with random initial conditions we have performed several simulations of
(\ref{eq051g})--(\ref{eq051f}). A typical run is illustrated in Figure 1.
Specifically, Fig.1.a illustrates the evolution of the truth values
$x_{1}\left(  t\right)  $ (solid line) and $x_{2}\left(  t\right)  $
(dash-dotted line) as computed by the steepest descent minimization algorithm
(\ref{eq051g})--(\ref{eq051f}) with $k=0.1$; Fig.1.b illustrates the
corresponding evolution of inconsistency.

\begin{center}%
\begin{tabular}
[c]{cc}%
\textbf{Fig. 1.a} & \textbf{Fig. 1.b}\\
\scalebox{0.5}{\includegraphics{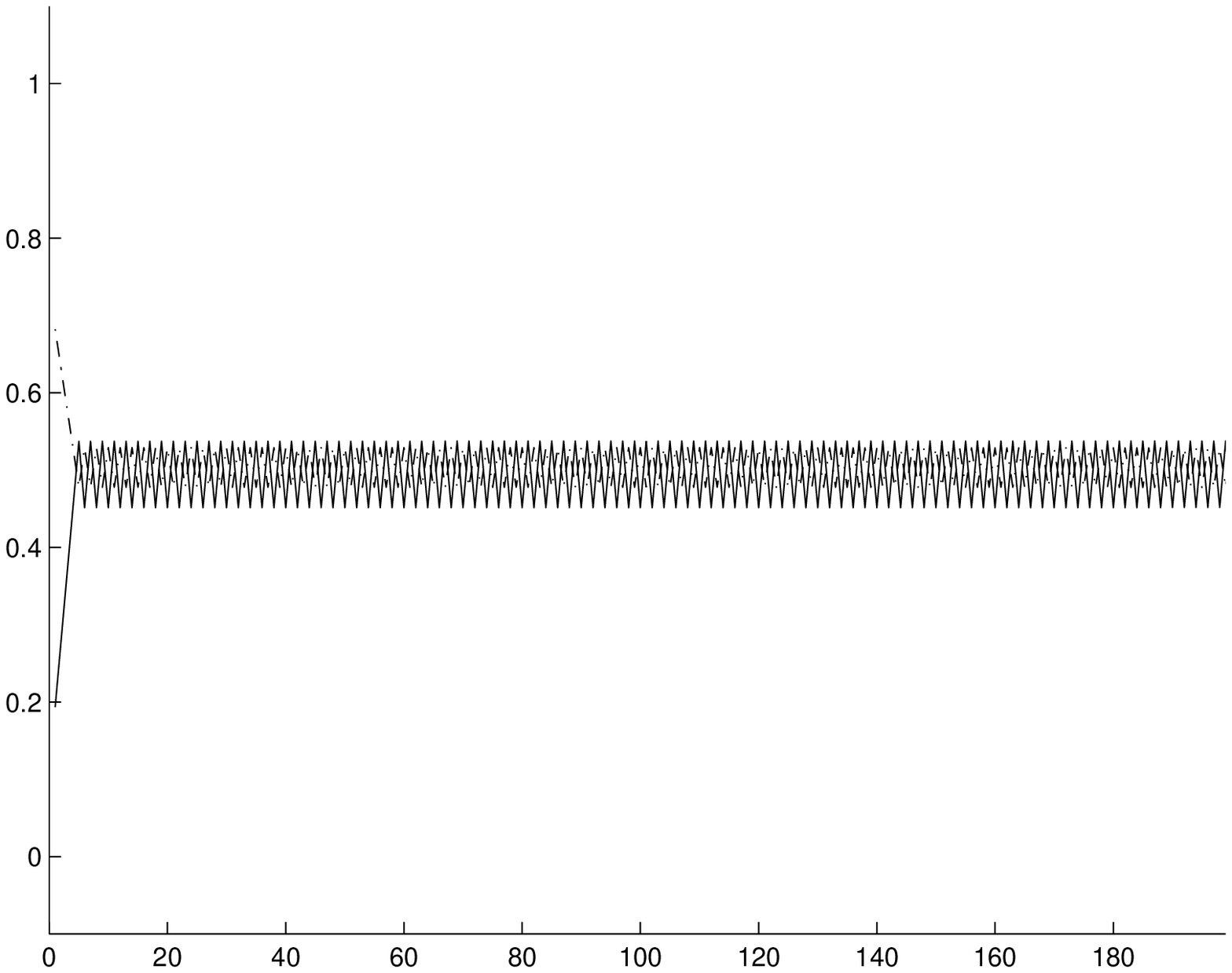}}  & \scalebox{0.5}%
{\includegraphics{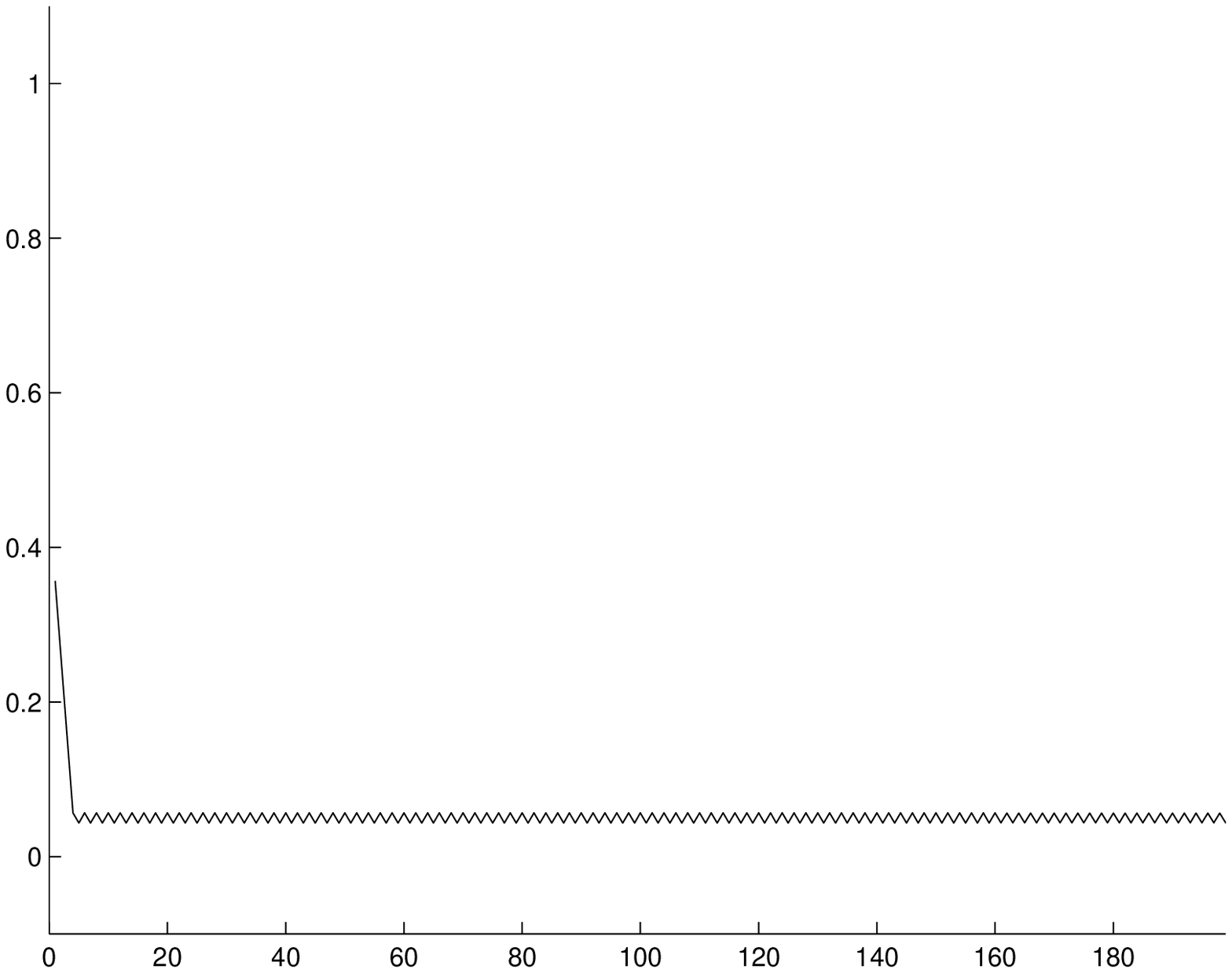}}
\end{tabular}
\end{center}

\medskip

We observe that the algorithm gets to the neighborhood of the optimal truth
values $\overline{x}$ in a few steps and thereafter oscillates around
$\overline{x}$; note also that at equilbrium the algorithm yields an
inconsistency value larger than zero. This behavior is related to the step
size $k$. Using smaller step size we can decrease the amplitude of the
oscillation and bring the inconsistency closer to zero, at the expense of
slower convergence; a typical simulation with $k=0.01$ appears in Figure 2.
Note that inconsistency is now a lot closer to zero.

\begin{center}%
\begin{tabular}
[c]{cc}%
\textbf{Fig. 2.a} & \textbf{Fig. 2.b}\\
\scalebox{0.5}{\includegraphics{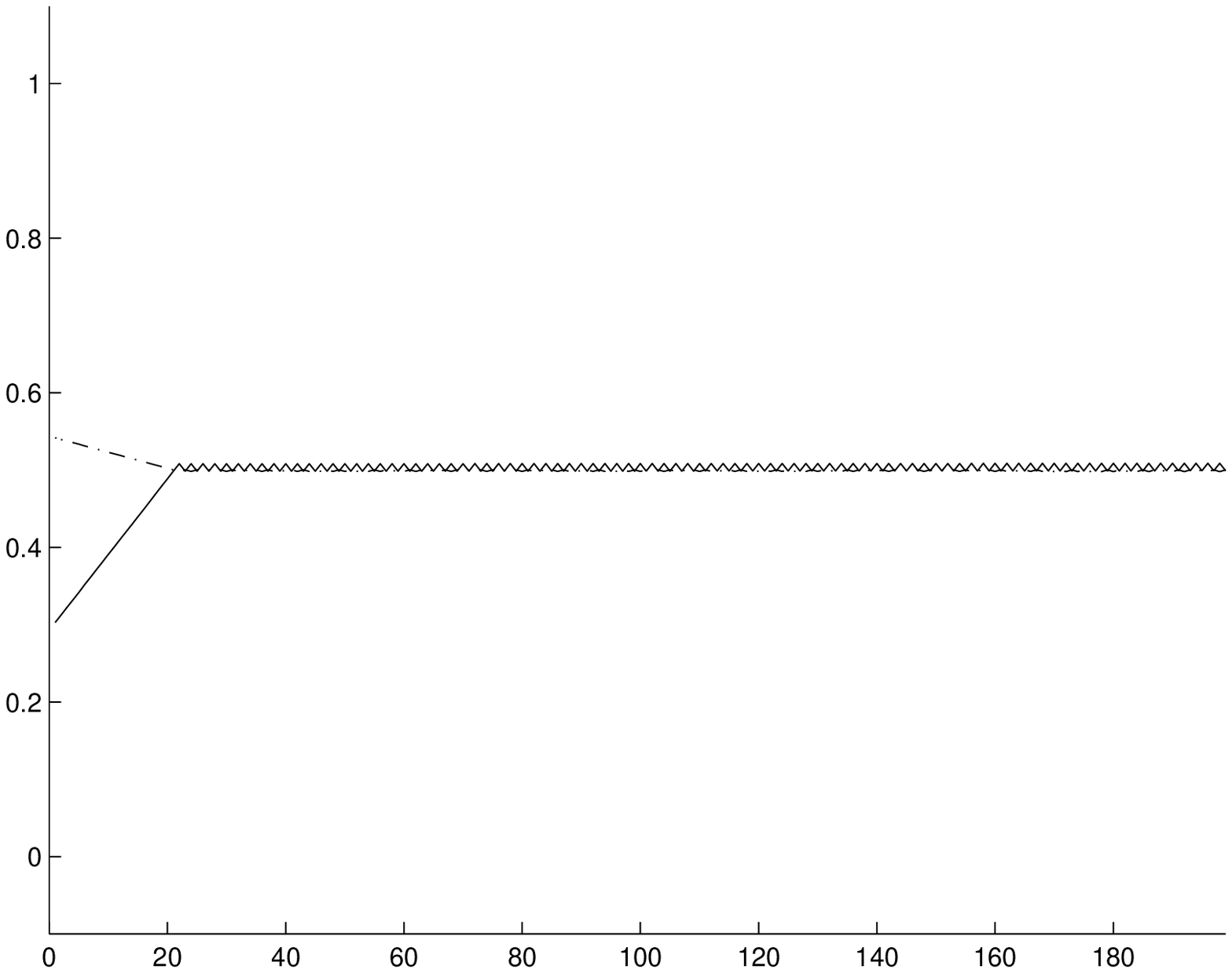}}  & \scalebox{0.5}%
{\includegraphics{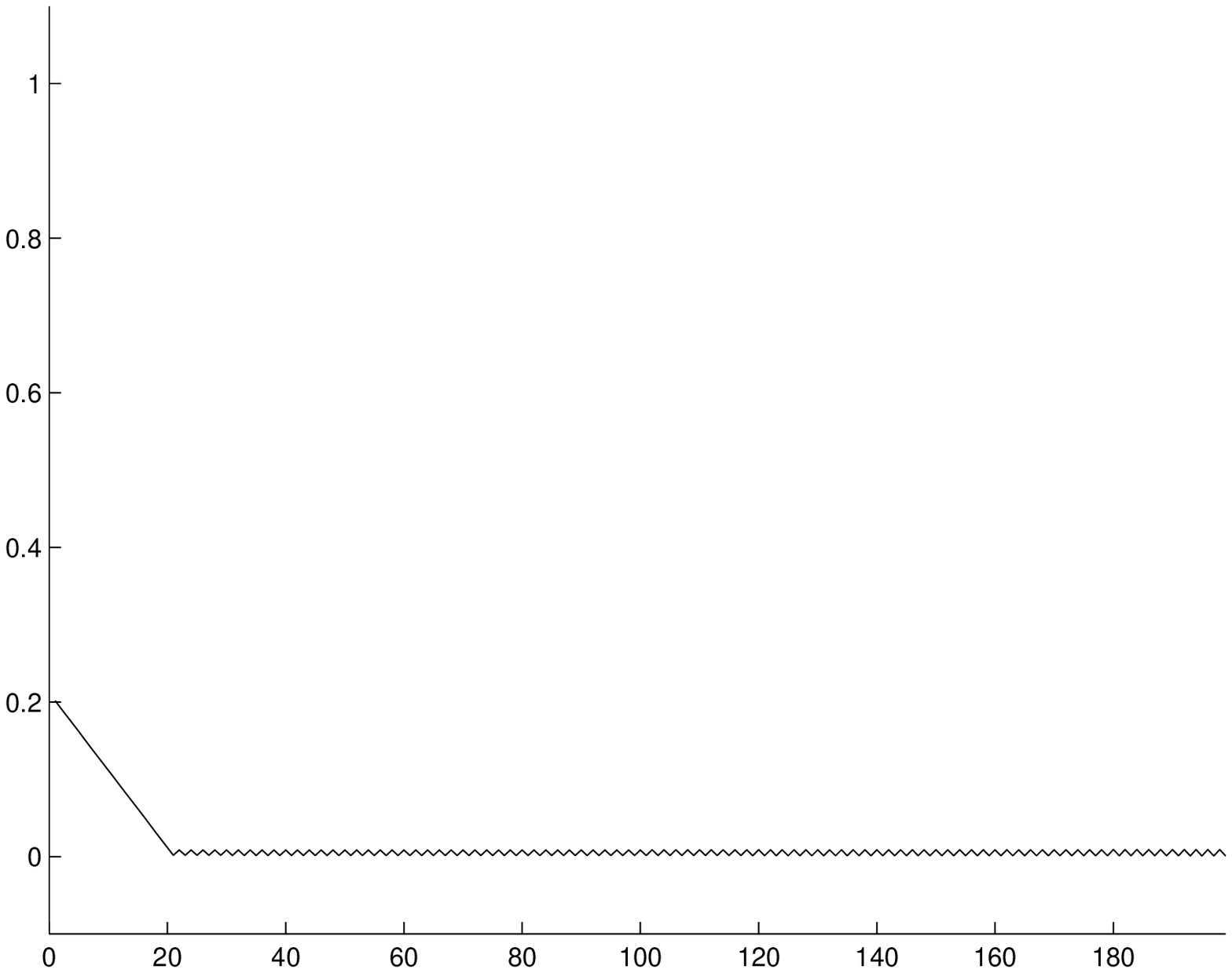}}
\end{tabular}
\end{center}

\medskip

Finally, using the control algorithm, we obtain the dynamical system
\begin{align}
x_{1}\left(  t+1\right)   &  =x_{1}\left(  t\right)  -k\cdot\left(
x_{1}\left(  t\right)  -x_{2}\left(  t\right)  \right) \label{eq051c}\\
x_{2}\left(  t+1\right)   &  =x_{2}\left(  t\right)  -k\cdot\left(
x_{2}\left(  t\right)  -1+x_{1}\left(  t\right)  \right)  . \label{eq051d}%
\end{align}
This is a linear system which can be written in matrix notation as
\[
x\left(  t+1\right)  =Ax\left(  t\right)  +b
\]
with
\[
A=\left(
\begin{array}
[c]{cc}%
1-k & k\\
-k & 1-k
\end{array}
\right)  ,\qquad b=\left(
\begin{array}
[c]{c}%
0\\
k
\end{array}
\right)
\]
$A$ has eigenvalues: $\lambda_{1}=-k+1+ik$,\allowbreak\ $\lambda_{2}=-k+1-ik$.
We have (for $i=1,2$)
\[
\left|  \lambda_{i}\right|  =\sqrt{1-2k+2k^{2}}<1
\]
for sufficiently small $k$. Hence (\ref{eq051c})--(\ref{eq051d})\ is stable.
The unique fixed point $\overline{x}$ is obtained by solving
\[
x=Ax+b.
\]
The solution is $\overline{x}=\left(  1/2,1/2\right)  $. Because of stability
we have for every $x\left(  0\right)  $ that
\[
\lim_{t\rightarrow\infty}x\left(  t\right)  =\overline{x}.
\]

This result is verified by numerical simulation. Using $k=0.1$ and starting
with random initial conditions we have performed several simulations of
(\ref{eq051c})--(\ref{eq051d}). A typical run is illustrated in Figure 3.
Again, Fig.3.a illustrates the evolution of the truth values $x_{1}\left(
t\right)  $ (solid line) and $x_{2}\left(  t\right)  $ (dash-dotted line) and
Fig.3.b illustrates the corresponding inconsistency (as given by (\ref{eq0421})).

\begin{center}%
\begin{tabular}
[c]{cc}%
\textbf{Fig. 3.a} & \textbf{Fig. 3.b}\\
\scalebox{0.5}{\includegraphics{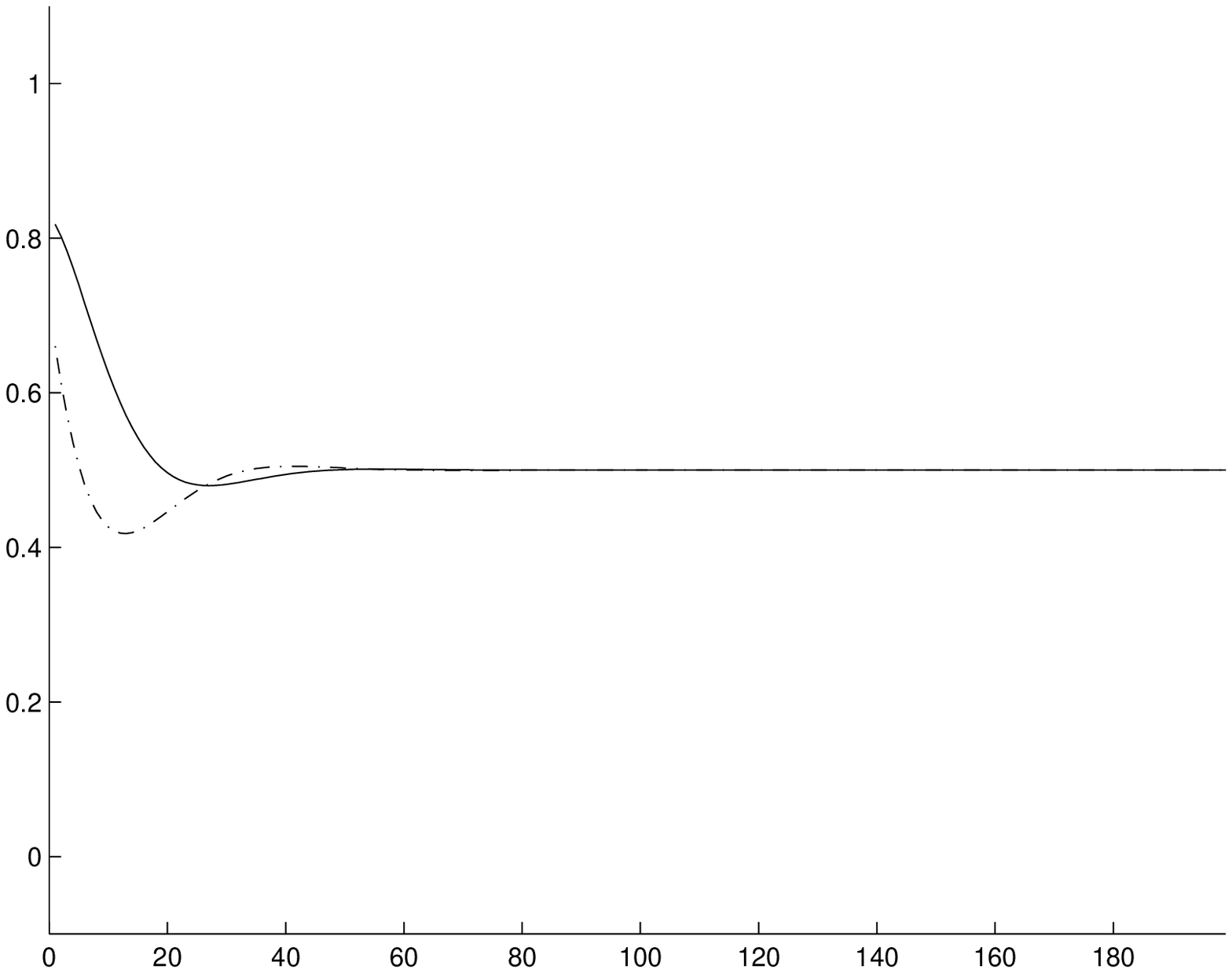}}  & \scalebox{0.5}%
{\includegraphics{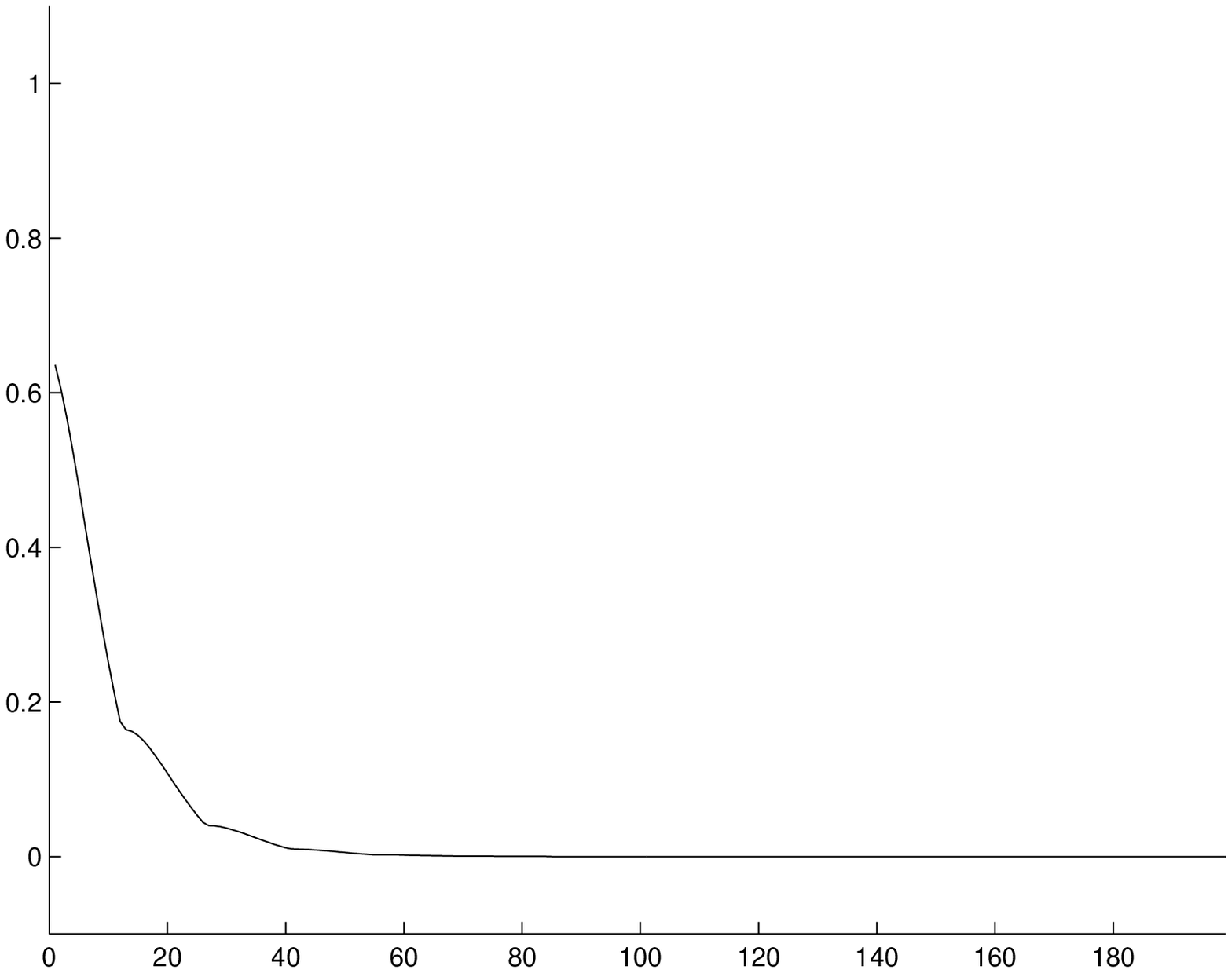}}
\end{tabular}
\end{center}

\ We observe that the control algorithm locates the solution of the truth
value equations with higher accuracy than the steepest descent method and
almost as fast.

\subsection{Example 3:\ The Consistent Dualist}

\label{sec0502}

Our next example is the consistent dualist:
\begin{align*}
A_{1}  &  :\text{``}A_{2}\text{ is true''}\\
A_{2}  &  :\text{``}A_{1}\text{ is true''.}%
\end{align*}
Again the 1st level elementary sentences are $\mathbf{V}_{1}=\left\{
A_{1},A_{2}\right\}  $; the 2nd level elementary sentences are
\begin{align*}
C_{1}  &  :\text{``Tr}\left(  A_{2}\right)  =1\text{''}\\
C_{2}  &  :\text{``Tr}\left(  A_{1}\right)  =1\text{''}%
\end{align*}
which are Boolean elementary truth value assessments, i.e. $C_{1},C_{2}%
\in\widetilde{\mathbf{V}}_{2}\subseteq\widetilde{\mathbf{S}}_{2}$. Setting
\begin{align*}
A_{1}  &  =C_{1}\\
A_{2}  &  =C_{2}%
\end{align*}
we obtain
\begin{align*}
\text{Tr}\left(  A_{1}\right)   &  =\text{Tr}\left(  C_{1}\right)  =1-\left|
\text{Tr}\left(  A_{2}\right)  -1\right| \\
\text{Tr}\left(  A_{2}\right)   &  =\text{Tr}\left(  C_{2}\right)  =1-\left|
\text{Tr}\left(  A_{1}\right)  -1\right|
\end{align*}
and then the truth value equations
\begin{align*}
x_{1}  &  =1-\left|  x_{2}-1\right| \\
x_{2}  &  =1-\left|  x_{1}-1\right|
\end{align*}
which can be written in simple form as
\begin{align*}
x_{1}  &  =x_{2}\\
x_{2}  &  =x_{1}.
\end{align*}
Any vector of the form $\overline{x}=\left(  \beta,\beta\right)  $ ($\beta
\in\left[  0,1\right]  $) is a solution; i.e. there is an infinite number of
consistent truth value assignments including complete truth (Tr$\left(
A_{1}\right)  =$Tr$\left(  A_{2}\right)  =1$) and complete falsity (Tr$\left(
A_{1}\right)  =$Tr$\left(  A_{2}\right)  =0$); in accordance to Proposition
\ref{cnt0304}, $\left(  1/2,1/2\right)  $ is also a solution.

The inconsistency function is
\[
J=\left(  x_{1}-x_{2}\right)  ^{2}+\left(  x_{1}-x_{2}\right)  ^{2}%
\]
and has partial derivatives
\begin{align*}
\frac{\partial J}{\partial x_{1}}  &  =\allowbreak4x_{1}-4x_{2}\\
\frac{\partial J}{\partial x_{2}}  &  =4x_{2}-4x_{1}.
\end{align*}
Setting the partial derivatives equal to zero we obtain
\begin{align}
\frac{\partial J}{\partial x_{1}}  &  =\allowbreak4x_{1}-4x_{2}%
=0\label{eq0521}\\
\frac{\partial J}{\partial x_{2}}  &  =4x_{2}-4x_{1}=0 \label{eq0522}%
\end{align}
and it is immediate that (\ref{eq0521})--(\ref{eq0522}) has the family of
solutions $x=\left(  \beta,\beta\right)  $, each of which gives zero
inconsistency. Furthermore, using the steepest descent algorithm we obtain the
linear dynamical system
\begin{align}
x_{1}\left(  t+1\right)   &  =x_{1}\left(  t\right)  -k\cdot\left(
4x_{1}\left(  t\right)  -4x_{2}\left(  t\right)  \right) \label{eq0523}\\
x_{2}\left(  t+1\right)   &  =x_{2}\left(  t\right)  -k\cdot\left(
4x_{2}\left(  t\right)  -4x_{1}\left(  t\right)  \right)  \label{eq0524}%
\end{align}
which can be written in matrix notation as
\[
x\left(  t+1\right)  =Ax\left(  t\right)
\]
with
\[
A=\left(
\begin{array}
[c]{cc}%
1-4k & 4k\\
4k & 1-4k
\end{array}
\right)
\]
$A$ has eigenvalues: $\lambda_{1}=1,\allowbreak\lambda_{2}=1-8k$. We have
\[
\left|  \lambda_{1}\right|  =1\text{ and }\left|  \lambda_{2}\right|
\sqrt{1-8k}<1
\]
hence (\ref{eq0523})--(\ref{eq0524})\ will generally be convergent, but will
have oscillatory behavior for certain initial conditions. The fixed points are
obtained by solving
\[
x=Ax.
\]
The family of solutions is $x=\left(  \beta,\beta\right)  $. These results are
verified by numerical simulation. Using $k=0.1$ and starting with random
initial conditions we have performed several simulations of (\ref{eq0523}%
)--(\ref{eq0524}). Three typical runs are illustrated in Figures 4, 5 and 6;
the left panels indicate the evolution of the truth values and the right
panels illustrate the corresponding inconsistency. We observe that the actual
equilibrium reached depends on the initial conditions.

\begin{center}%
\begin{tabular}
[c]{cc}%
\textbf{Fig. 4.a} & \textbf{Fig. 4.b}\\
\scalebox{0.5}{\includegraphics{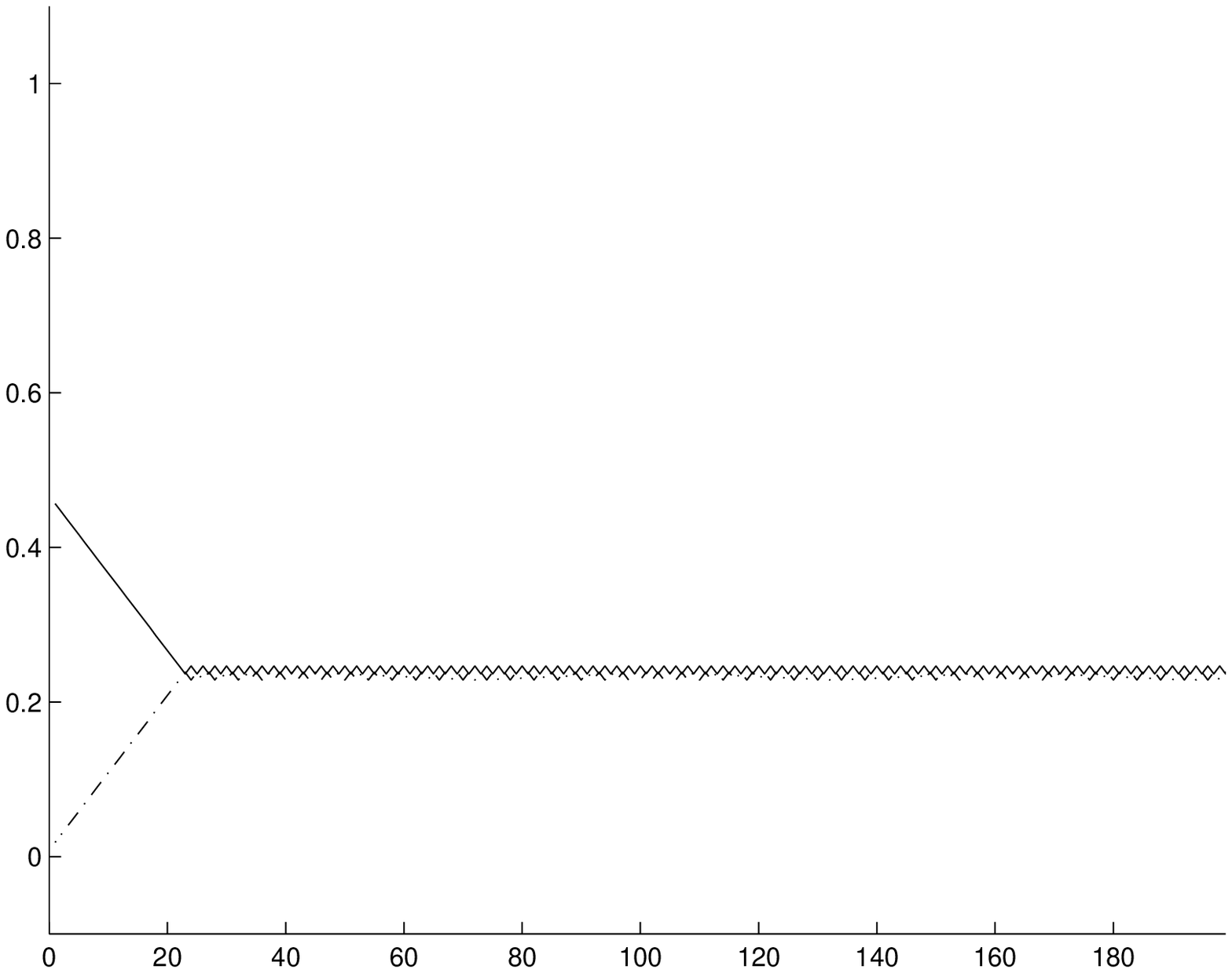}}  & \scalebox{0.5}%
{\includegraphics{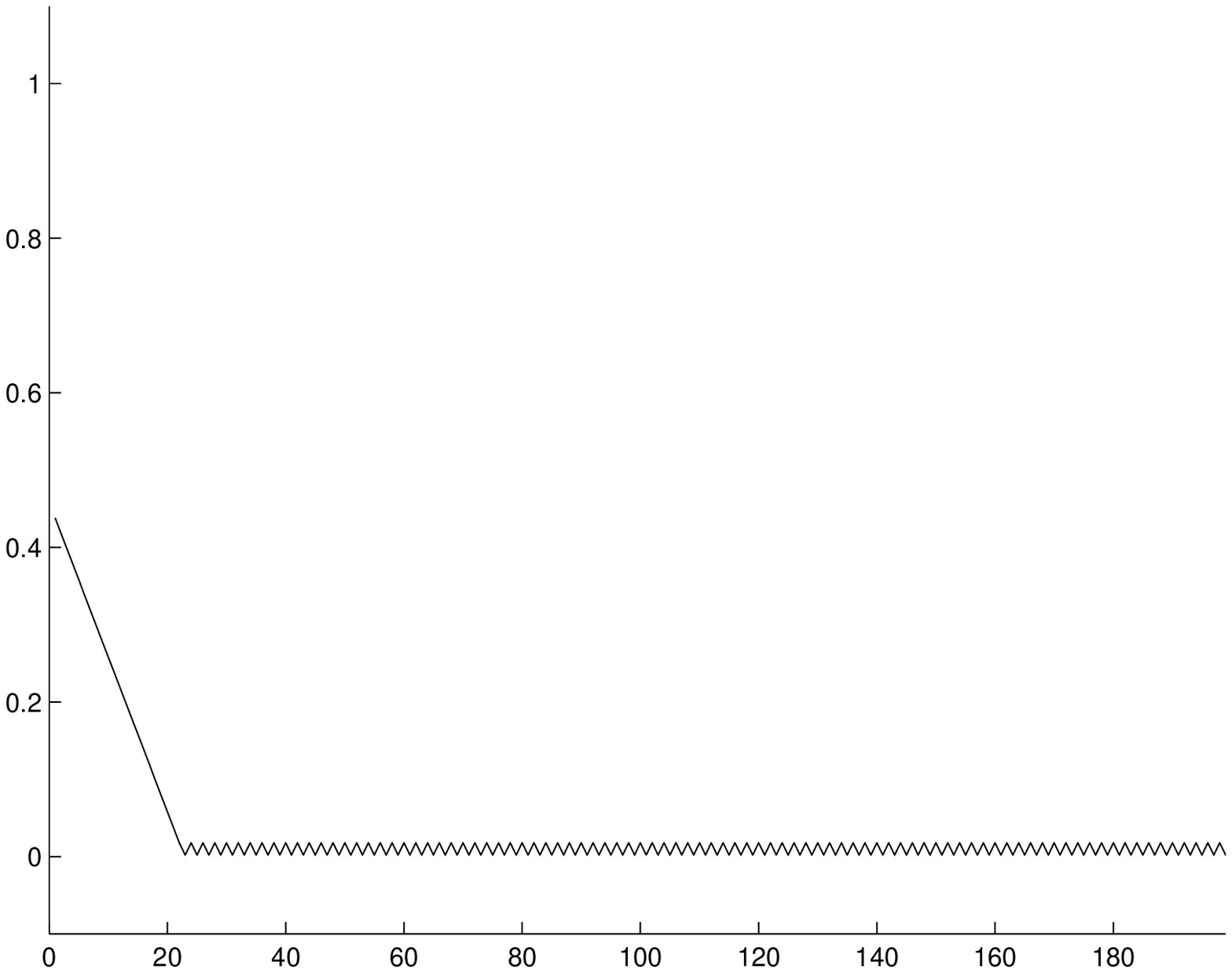}}
\end{tabular}%

\begin{tabular}
[c]{cc}%
\textbf{Fig. 5.a} & \textbf{Fig. 5.b}\\
\scalebox{0.5}{\includegraphics{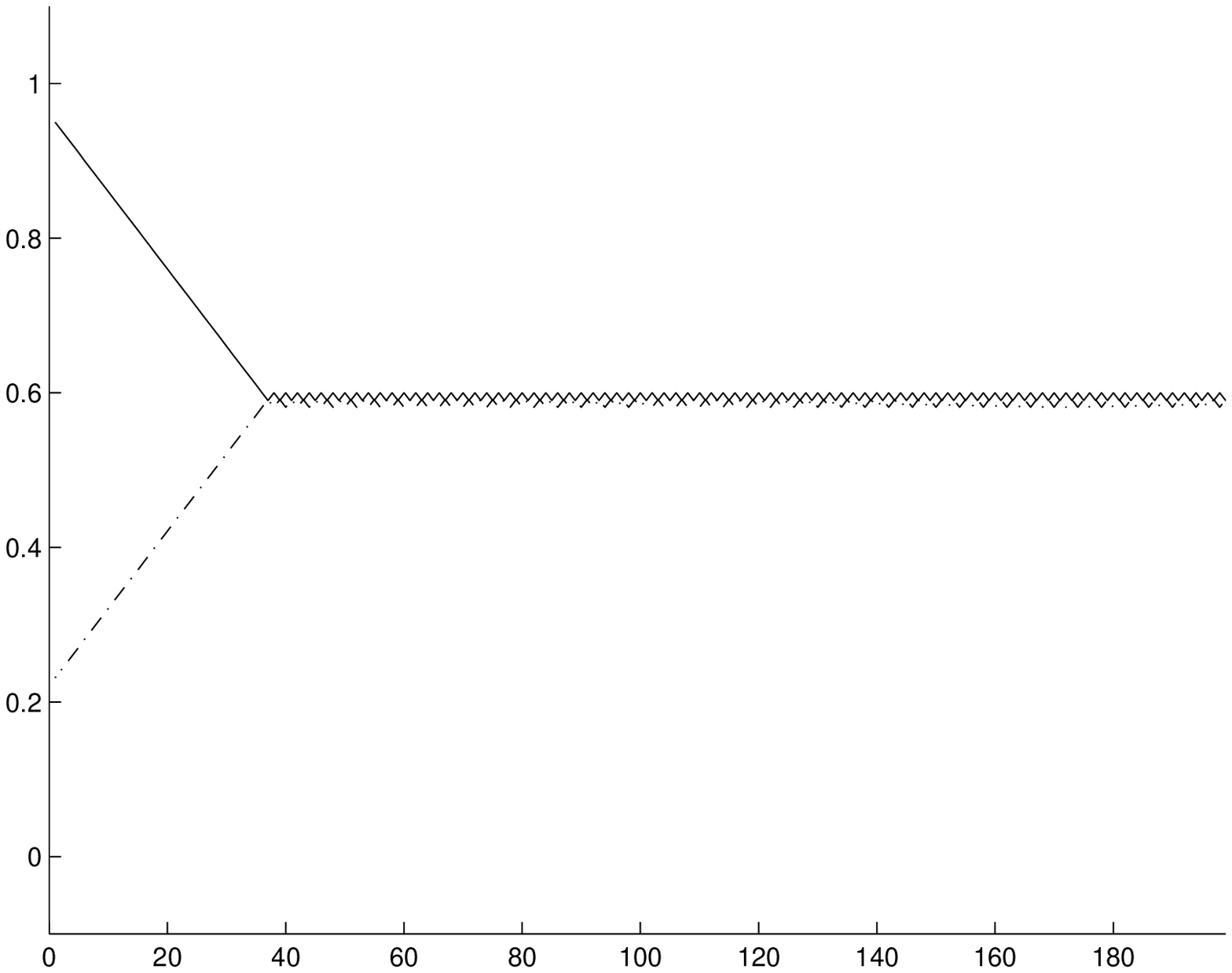}}  & \scalebox{0.5}%
{\includegraphics{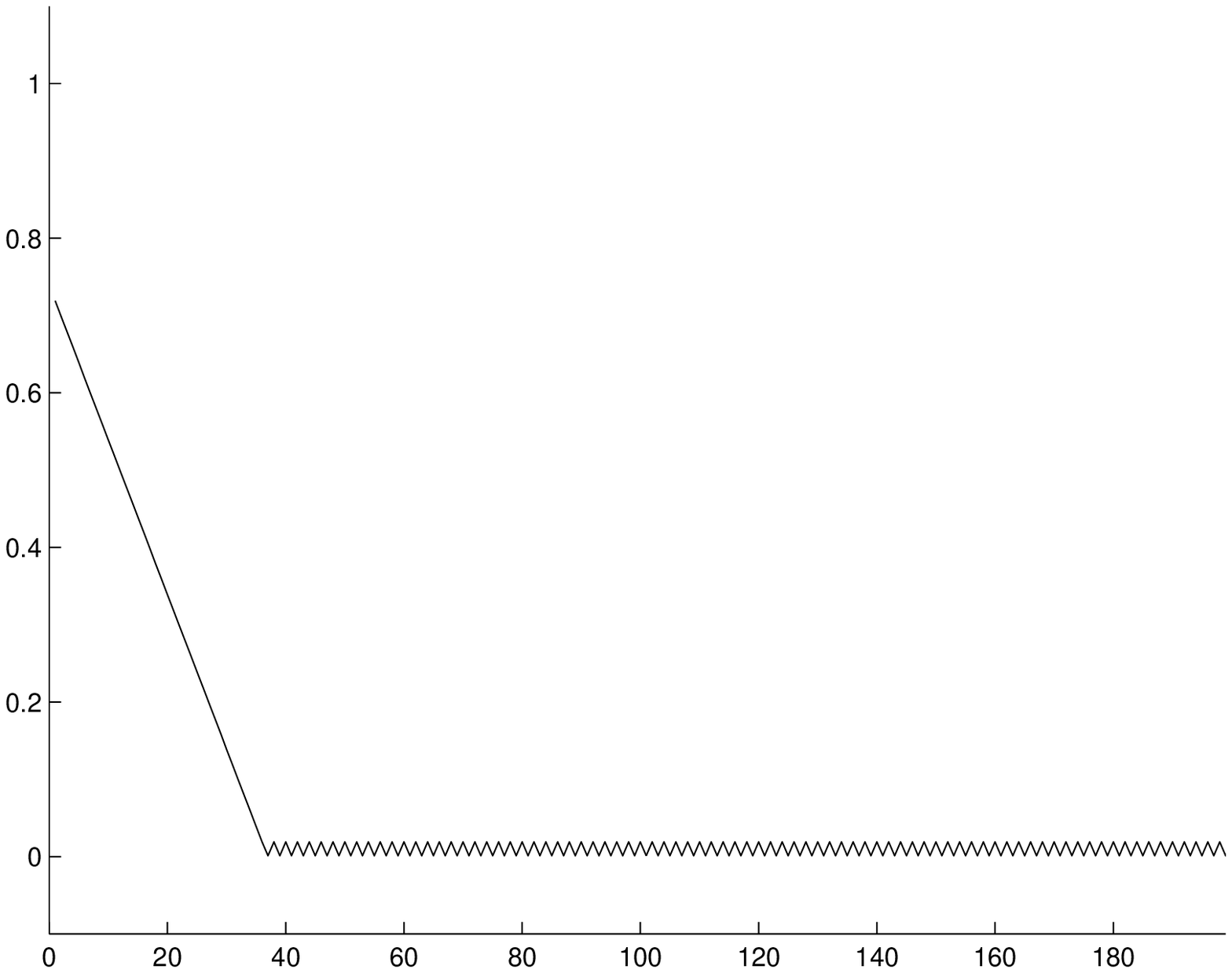}}
\end{tabular}%

\begin{tabular}
[c]{cc}%
\textbf{Fig. 6.a} & \textbf{Fig. 6.b}\\
\scalebox{0.5}{\includegraphics{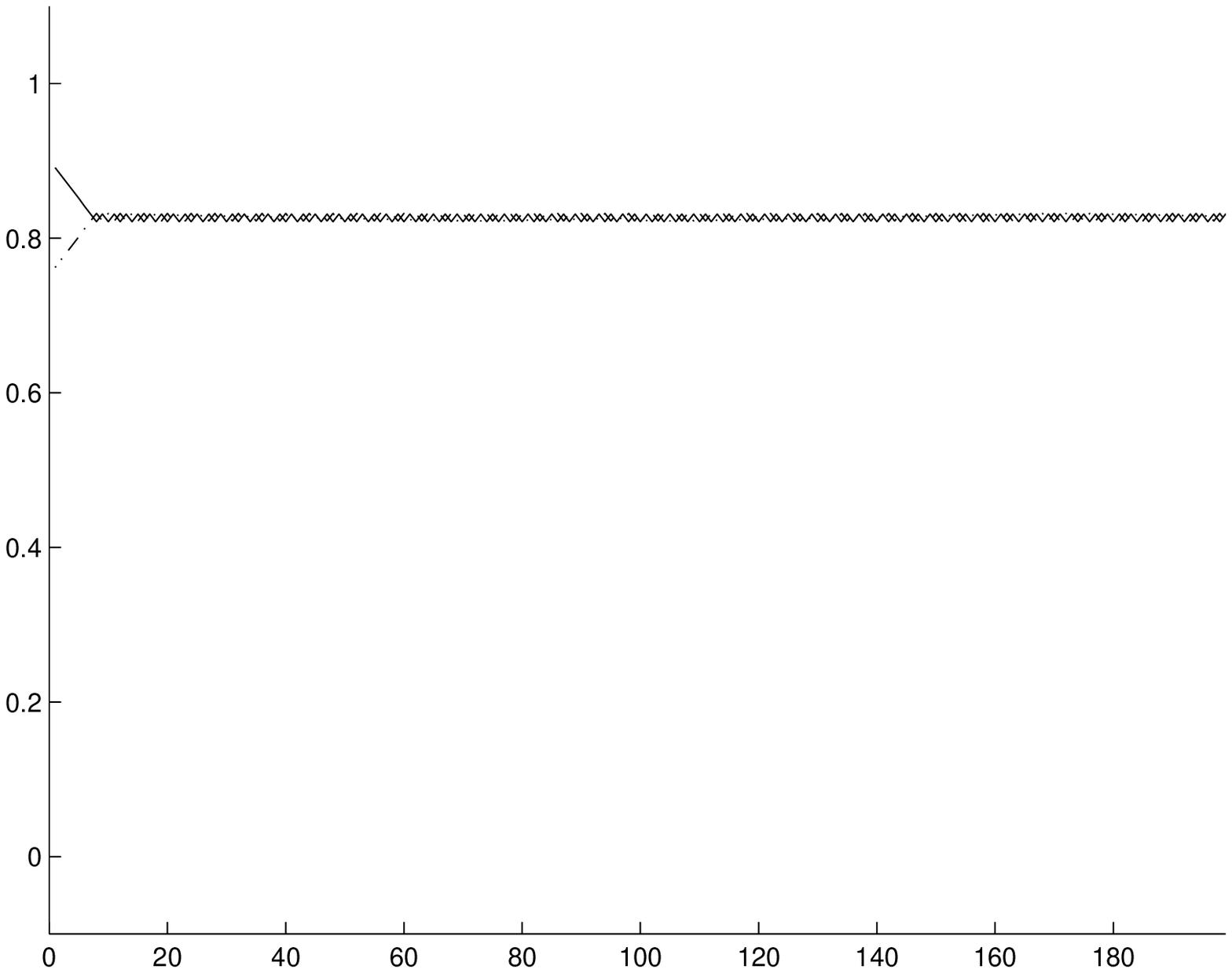}}  & \scalebox{0.5}%
{\includegraphics{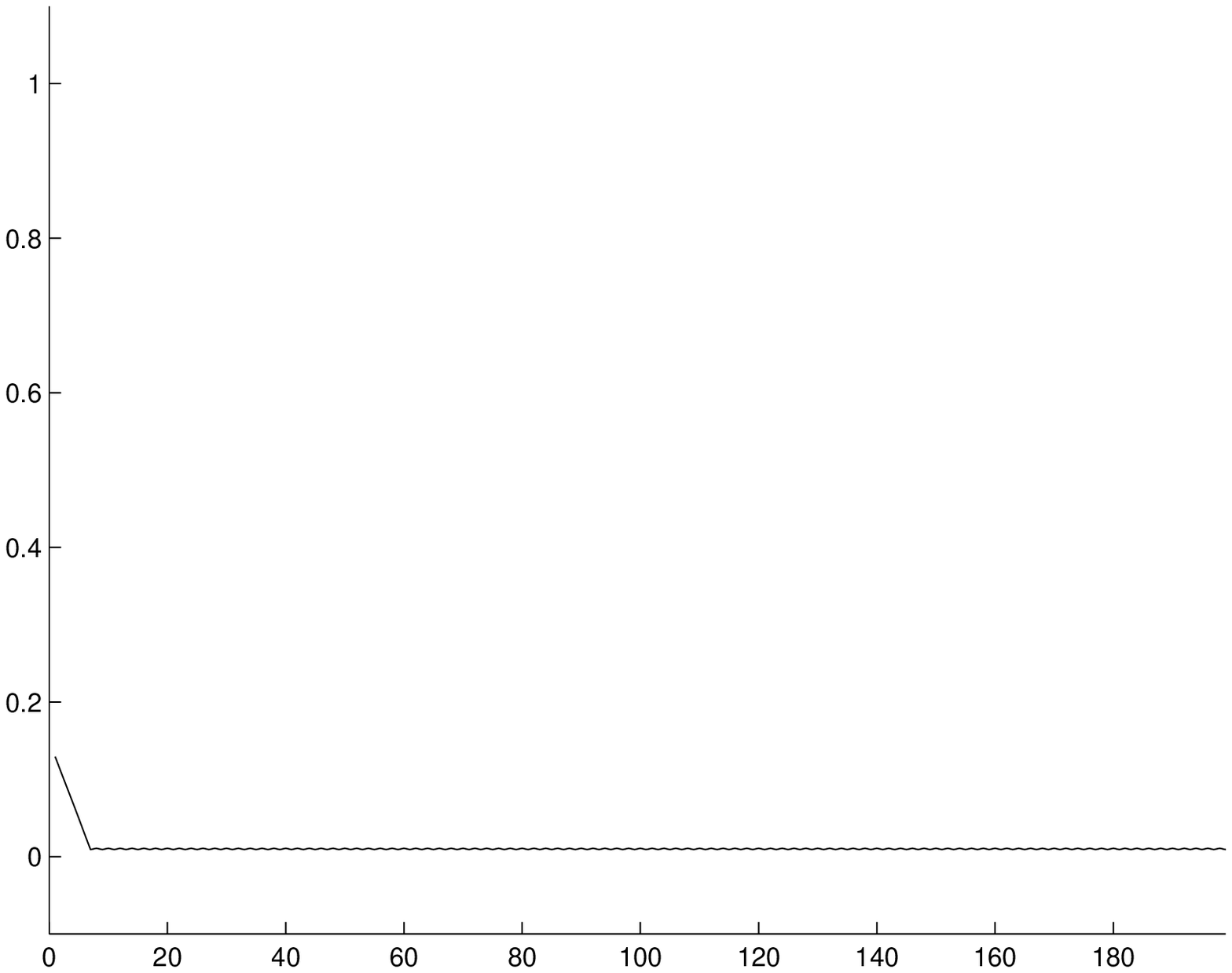}}
\end{tabular}
\end{center}

Finally, using the control algorithm, we obtain the dynamical system
\begin{align}
x_{1}\left(  t+1\right)   &  =x_{1}\left(  t\right)  -k\cdot\left(
x_{1}\left(  t\right)  -x_{2}\left(  t\right)  \right) \label{eq0525}\\
x_{2}\left(  t+1\right)   &  =x_{2}\left(  t\right)  -k\cdot\left(
x_{2}\left(  t\right)  -x_{1}\left(  t\right)  \right)  . \label{eq0526}%
\end{align}
This is a linear system which can be written in matrix notation as
\[
x\left(  t+1\right)  =Ax\left(  t\right)
\]
with
\[
A=\left(
\begin{array}
[c]{cc}%
1-k & k\\
k & 1-k
\end{array}
\right)
\]
$A$ has eigenvalues: $\lambda_{1}=1$,\allowbreak\ $\lambda_{2}=1-2k$. We have
for sufficiently small $k$
\[
\left|  \lambda_{1}\right|  =1\text{ and }\left|  \lambda_{2}\right|
\sqrt{1-2k}<1
\]
hence (\ref{eq0525})--(\ref{eq0526})\ can have oscillatory behavior (for
certain initial conditions). Also, the fixed points of are obtained by
solving
\[
x=Ax.
\]
The family of solutions is, as expected, $x=\left(  \beta,\beta\right)  .$ The
results are verified by numerical simulation. Using $k=0.1$ and starting with
random initial conditions we have performed several simulations of
(\ref{eq0525})--(\ref{eq0526}). Typical runs are illustrated in Figures 7, 8,
9 (again, the left panels indicate the evolution of the truth values and the
right panels illustrate the corresponding inconsistency).

\begin{center}%
\begin{tabular}
[c]{cc}%
\textbf{Fig. 7.a} & \textbf{Fig. 7.b}\\
\scalebox{0.5}{\includegraphics{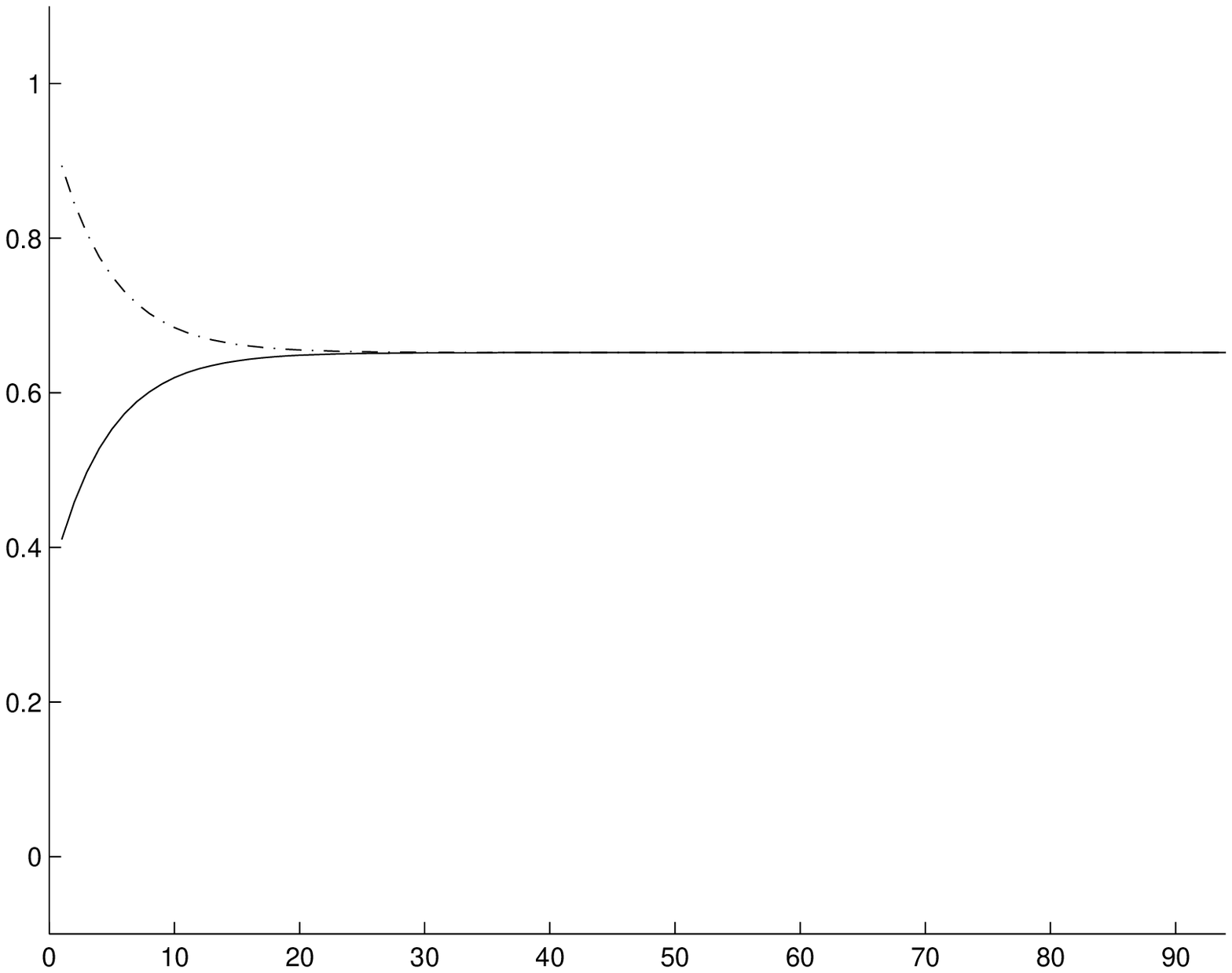}}  & \scalebox{0.5}%
{\includegraphics{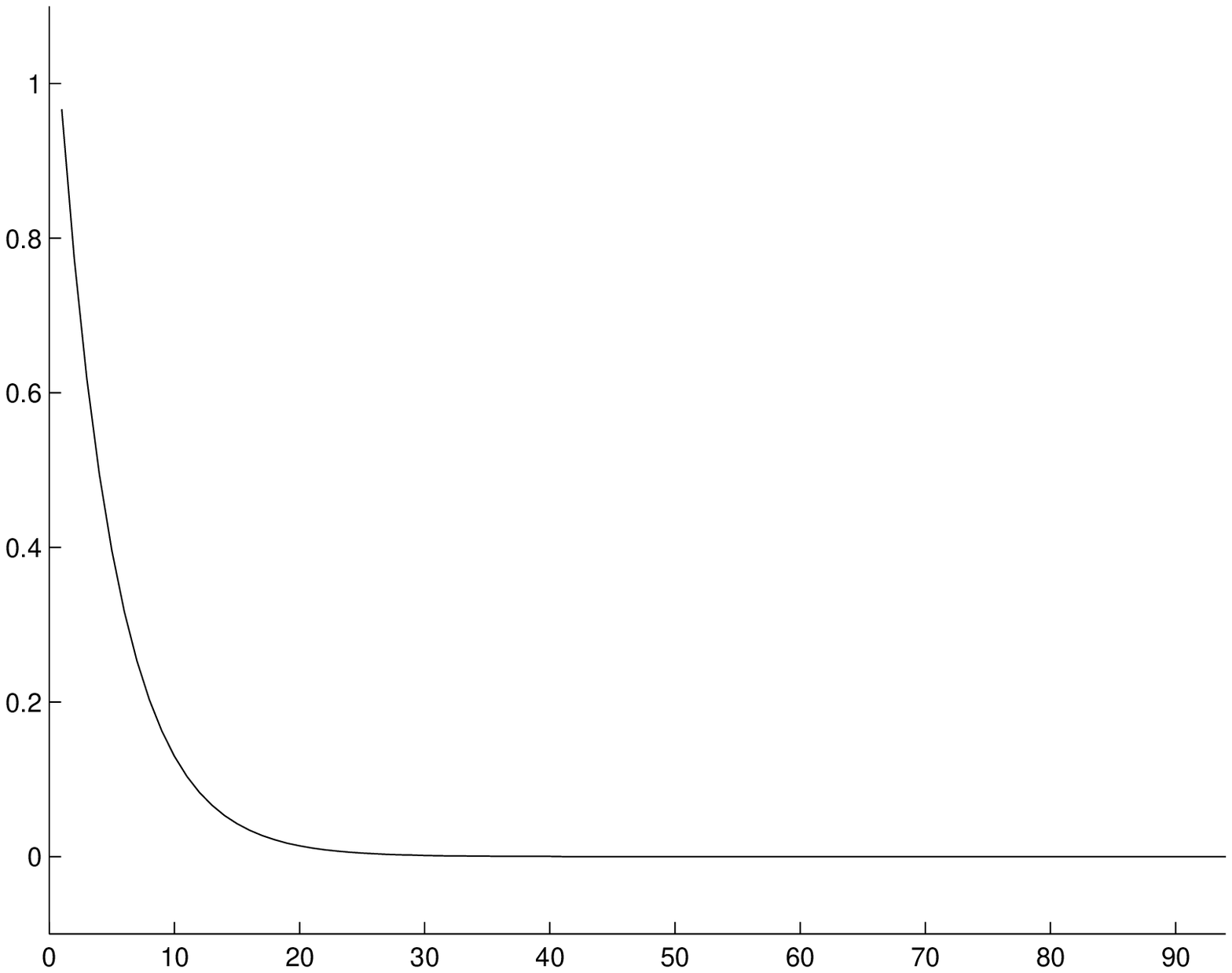}}
\end{tabular}%

\begin{tabular}
[c]{cc}%
\textbf{Fig. 8.a} & \textbf{Fig. 8.b}\\
\scalebox{0.5}{\includegraphics{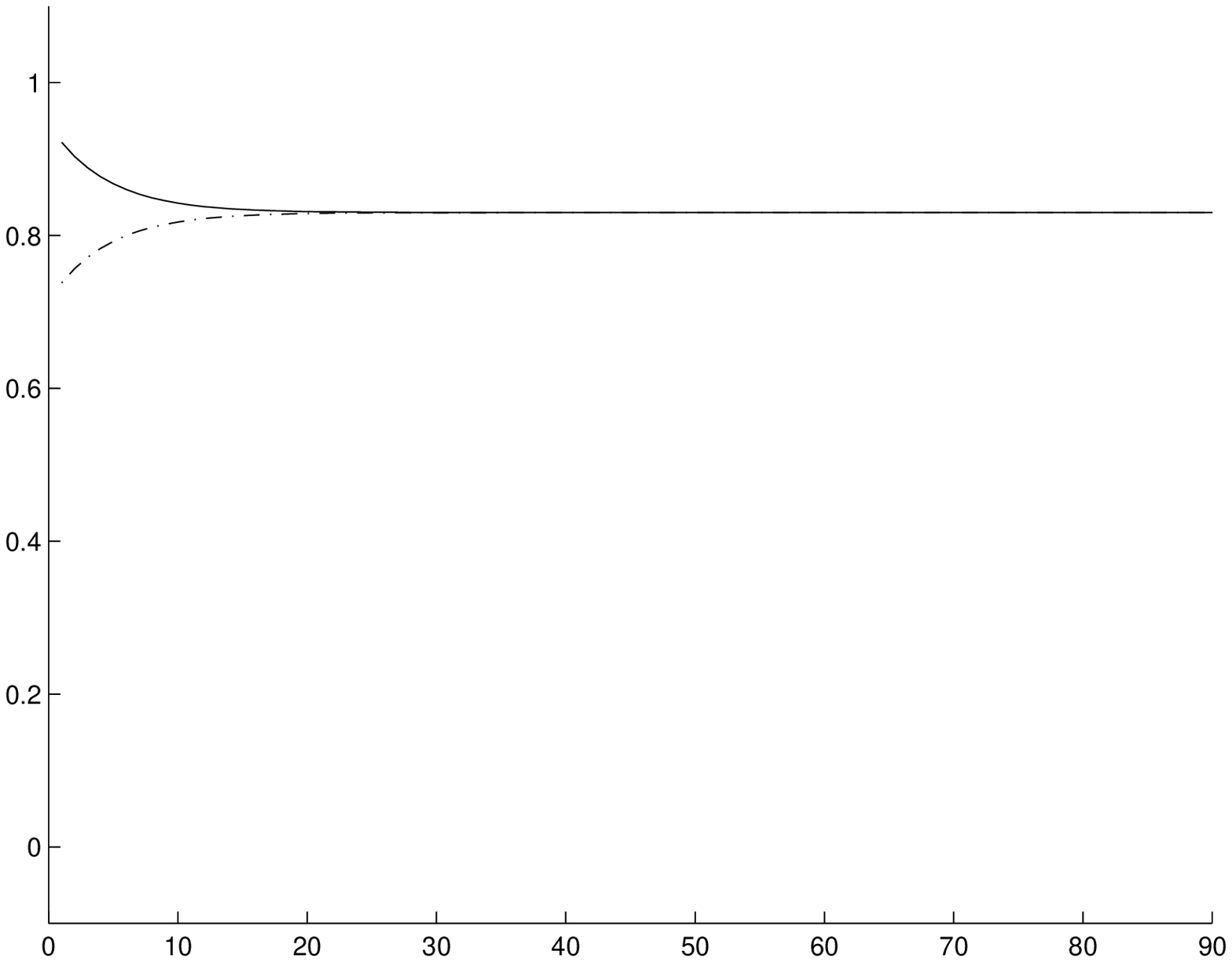}}  & \scalebox{0.5}%
{\includegraphics{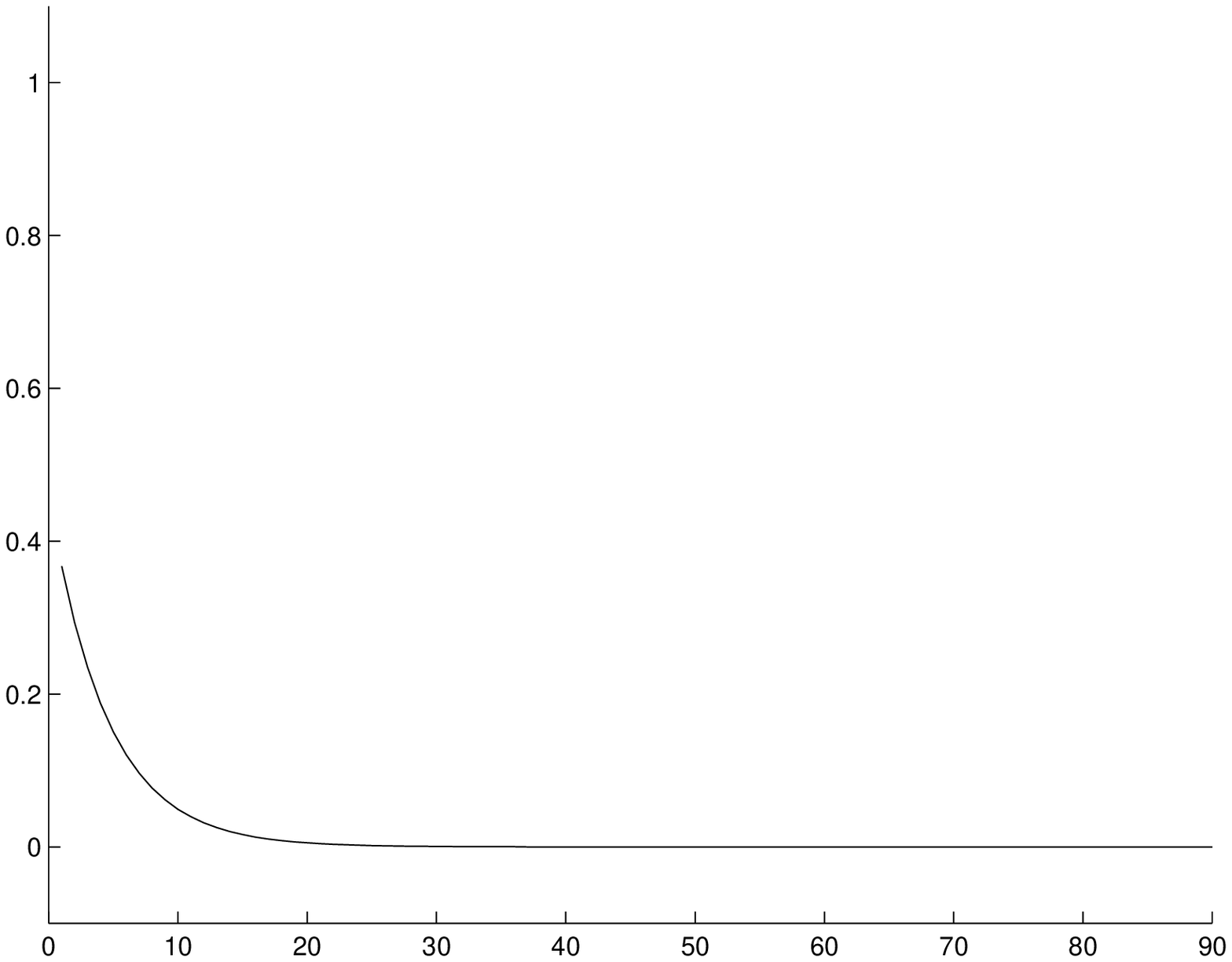}}
\end{tabular}%

\begin{tabular}
[c]{cc}%
\textbf{Fig. 9.a} & \textbf{Fig. 9.b}\\
\scalebox{0.5}{\includegraphics{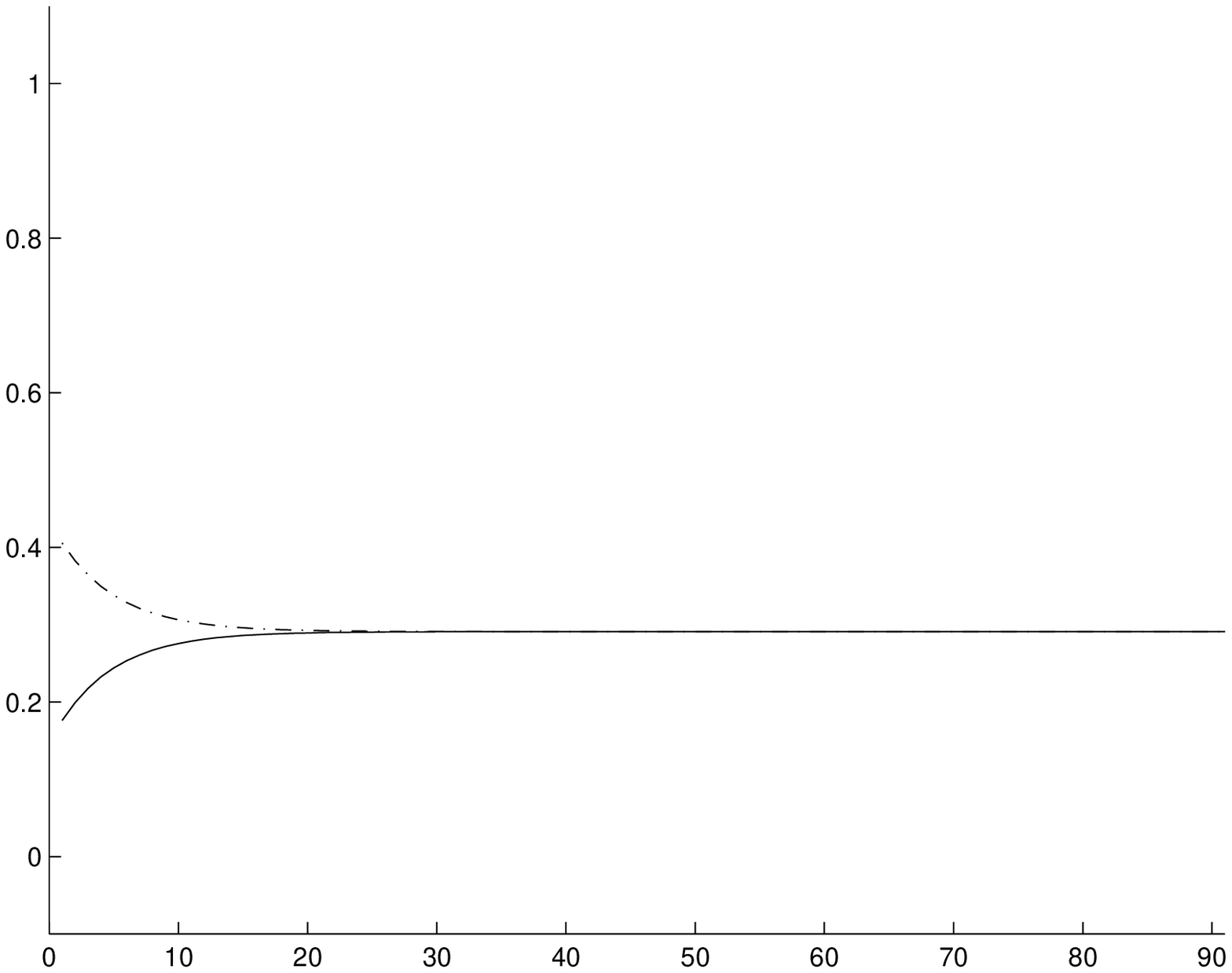}}  & \scalebox{0.5}%
{\includegraphics{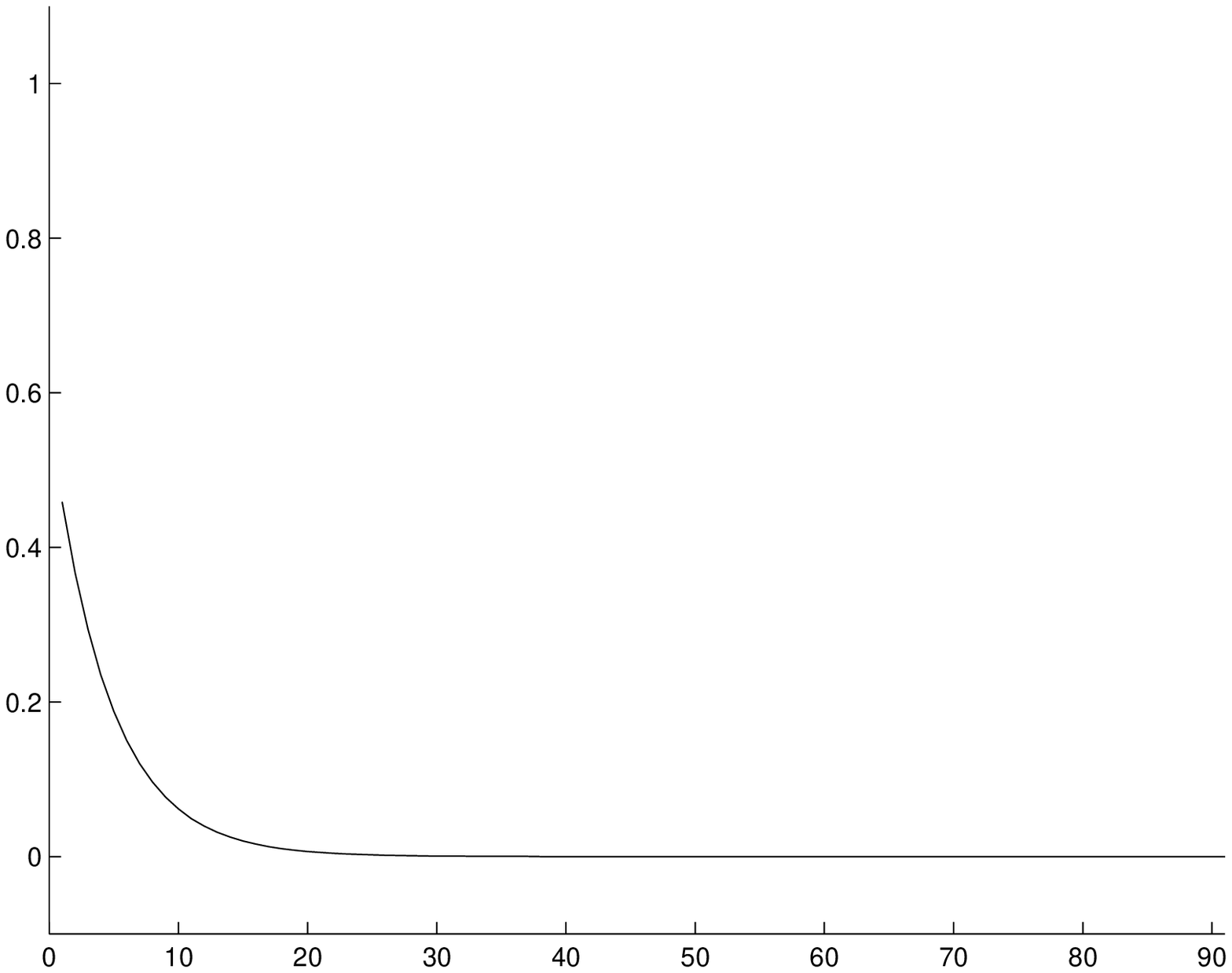}}
\end{tabular}
\end{center}

\subsection{Example 4}

\label{sec0503}

Our third example is somewhat more complicated but still involves only
$\widetilde{\mathbf{V}}_{2}$ sentences:
\begin{align}
A_{1}  &  :\text{``}A_{2}\text{ is true and }A_{3}\text{ is false
''}\label{eq0531}\\
A_{2}  &  :\text{``}A_{1}\text{ is true and }A_{3}\text{ is false
''}\label{eq0532}\\
A_{3}  &  :\text{``}A_{1}\text{ is false''.} \label{eq0533}%
\end{align}
(\ref{eq0531})--(\ref{eq0533}) \ translates to
\begin{align*}
A_{1}  &  =D_{1}=C_{1}\wedge C_{2}\text{ }\\
A_{2}  &  =D_{2}=C_{3}\wedge C_{2}\\
A_{3}  &  =D_{3}=C_{4}%
\end{align*}
where
\begin{align*}
C_{1}  &  :\text{``Tr}\left(  A_{2}\right)  =1\text{''}\\
C_{2}  &  :\text{``Tr}\left(  A_{3}\right)  =0\text{''}\\
C_{3}  &  :\text{``Tr}\left(  A_{1}\right)  =1\text{''}\\
C_{4}  &  :\text{``Tr}\left(  A_{1}\right)  =0\text{''.}%
\end{align*}

We will consider two different implemementations of $\wedge$.

\subsubsection{$\wedge$ Implemented by Minimum}

If we implement $\wedge$ by the min t-norm, the truth value equations become
\begin{align}
x_{1}  &  =\min\left[  x_{2},\left(  1-x_{3}\right)  \right] \label{eq0537}\\
x_{2}  &  =\min\left[  x_{1},\left(  1-x_{3}\right)  \right] \label{eq0538}\\
x_{3}  &  =1-x_{1} \label{eq0539}%
\end{align}
and the inconsistency function is
\begin{equation}
J=\left(  x_{1}-\min\left[  x_{2},\left(  1-x_{3}\right)  \right]  \right)
^{2}+\left(  x_{2}-\min\left[  x_{1},\left(  1-x_{3}\right)  \right]  \right)
^{2}+\left(  x_{3}-\left(  1-x_{1}\right)  \right)  ^{2}. \label{eq053a}%
\end{equation}

The truth value equations can be solved analytically. From (\ref{eq0539})\ we
obtain
\[
x_{1}=1-x_{3}%
\]
and then (\ref{eq0537}) -- (\ref{eq0538})\ become
\[
x_{1}=\min\left[  x_{2},x_{1}\right]  ,\qquad x_{2}=\min\left[  x_{1}%
,x_{1}\right]
\]
from which follows that
\[
x_{1}=x_{2},\text{\qquad\ }x_{3}=1-x_{1}.
\]
In other words, the general solution of (\ref{eq0537}) -- (\ref{eq0539})\ is
\[
x=\left(  \beta,\beta,1-\beta\right)
\]
with $\beta\in\left[  0,1\right]  $. Note that this includes the extremal
solutions $\left(  1,1,0\right)  $ and $\left(  0,0,1\right)  $ as well as the
mid-point solution $\left(  1/2,1/2,1/2\right)  $. Also, using (\ref{eq0539}%
)\ \ the inconsistency function becomes
\[
J=\left(  x_{1}-\min\left[  x_{2},x_{1}\right]  \right)  ^{2}+\left(
x_{2}-\min\left[  x_{1},x_{1}\right]  \right)  ^{2}+\left(  x_{3}-\left(
1-x_{1}\right)  \right)  ^{2}%
\]
which attains the minimum value of 0 for every $x=\left(  \beta,\beta
,1-\beta\right)  $ with $\beta\in\left[  0,1\right]  $. For the steepest
descent algorithm the dynamical system is
\begin{equation}
x\left(  t+1\right)  =x\left(  t\right)  -k\cdot\frac{\partial J}{\partial x}.
\label{eq053c}%
\end{equation}
We cannot write equations for the gradient explicitly, because the expressions
$\min\left[  x_{2},1-x_{3}\right]  $ and $\min\left[  x_{1},1-x_{3}\right]  $
are not everywhere differentiable. However, using an approximate numerical
differentiation we can simulate (\ref{eq053c}). In Figures 10 and 11 we
present results for two typical simulations with $k=0.01$.

\begin{center}%
\begin{tabular}
[c]{cc}%
\textbf{Fig. 10.a} & \textbf{Fig. 11.b}\\
\scalebox{0.5}{\includegraphics{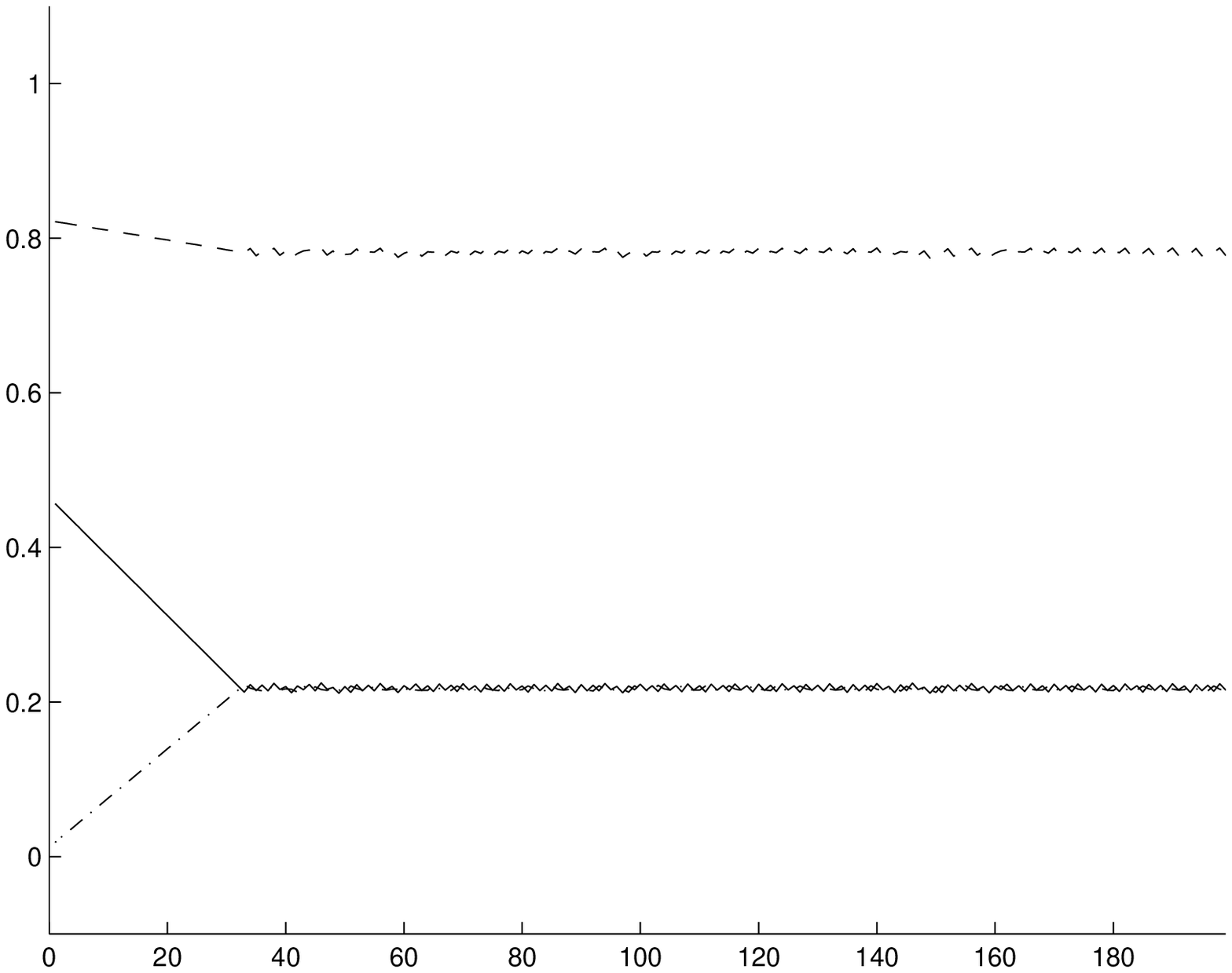}}  & \scalebox{0.5}%
{\includegraphics{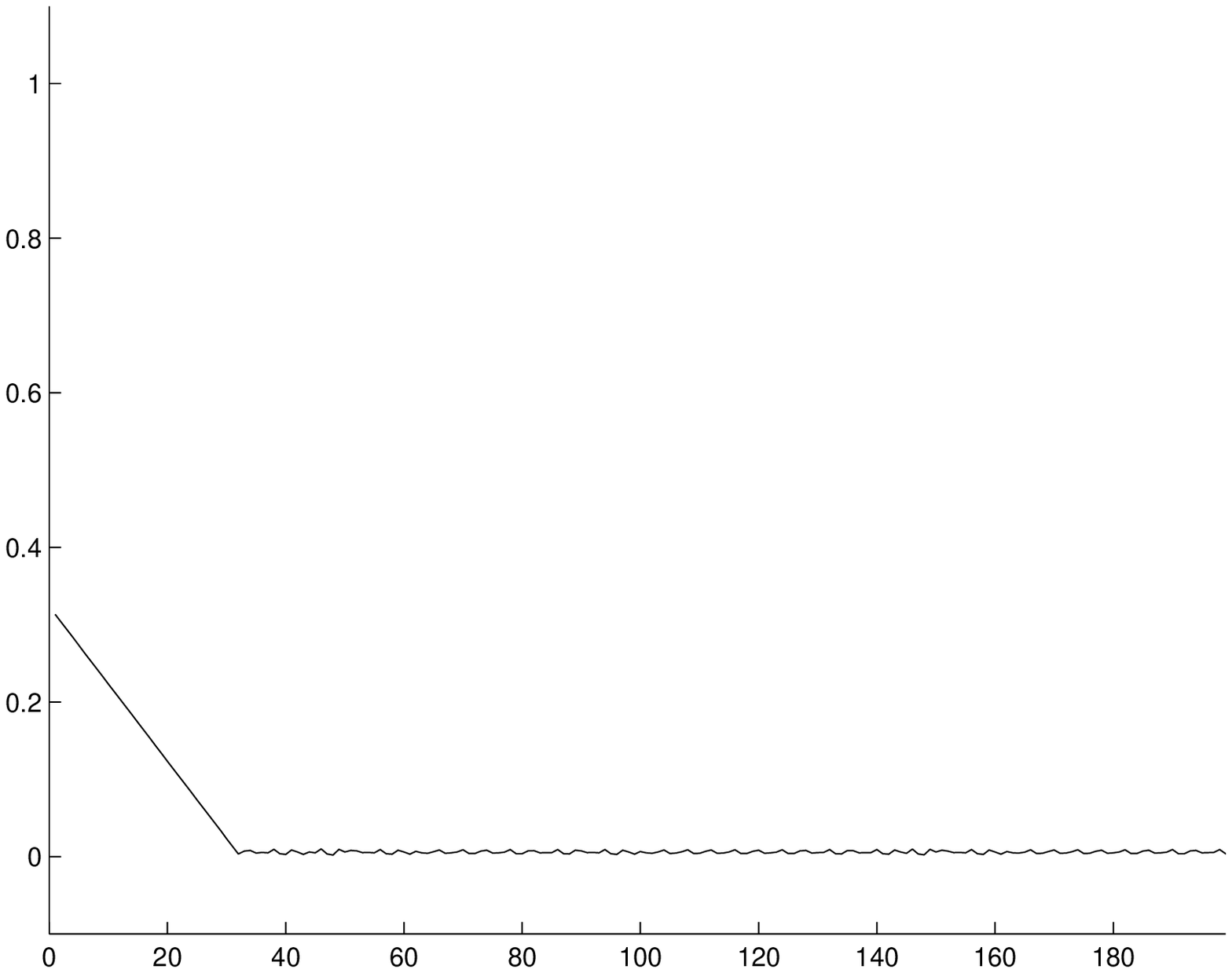}}
\end{tabular}%

\begin{tabular}
[c]{cc}%
\textbf{Fig. 11.a} & \textbf{Fig. 11.b}\\
\scalebox{0.5}{\includegraphics{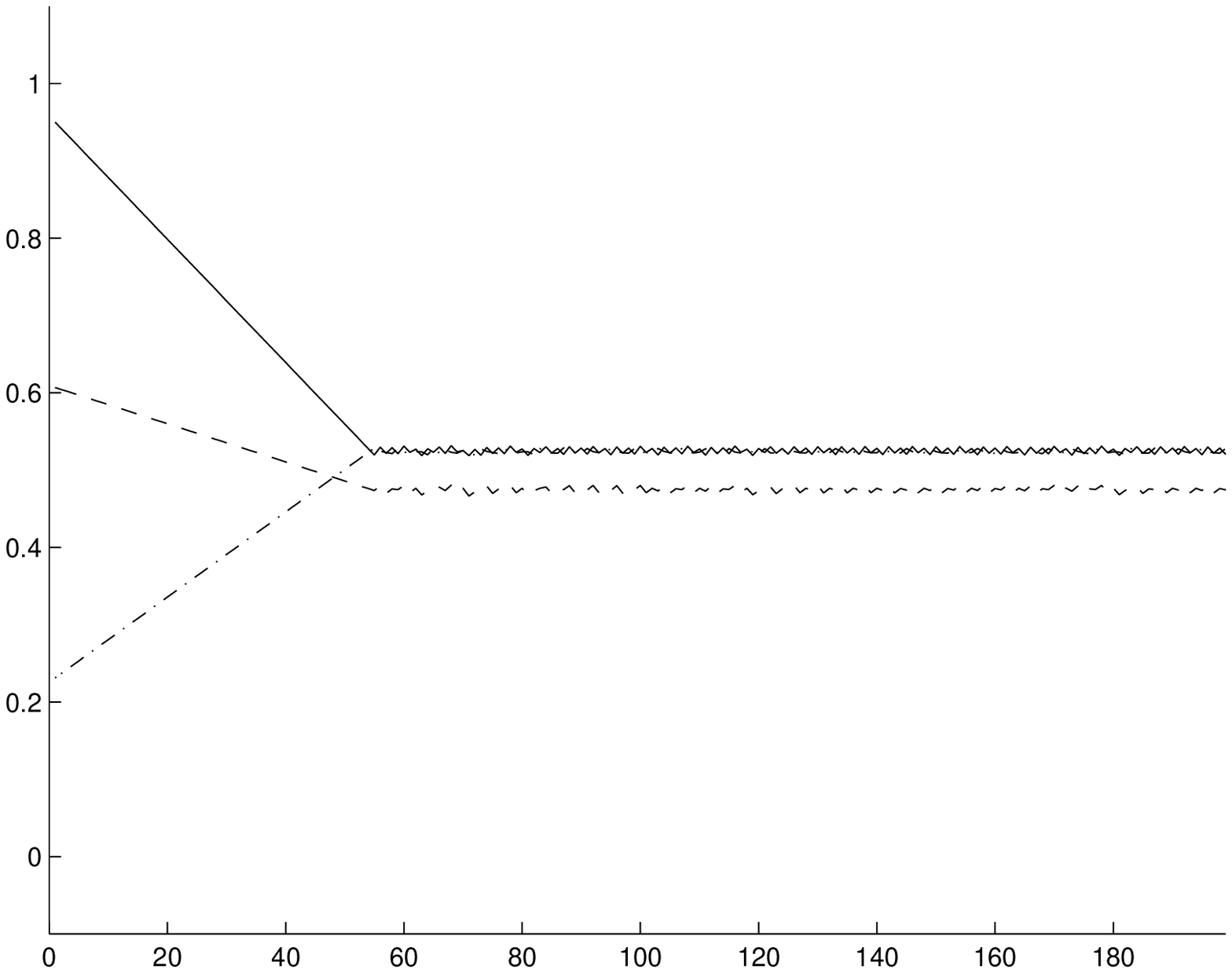}}  & \scalebox{0.5}%
{\includegraphics{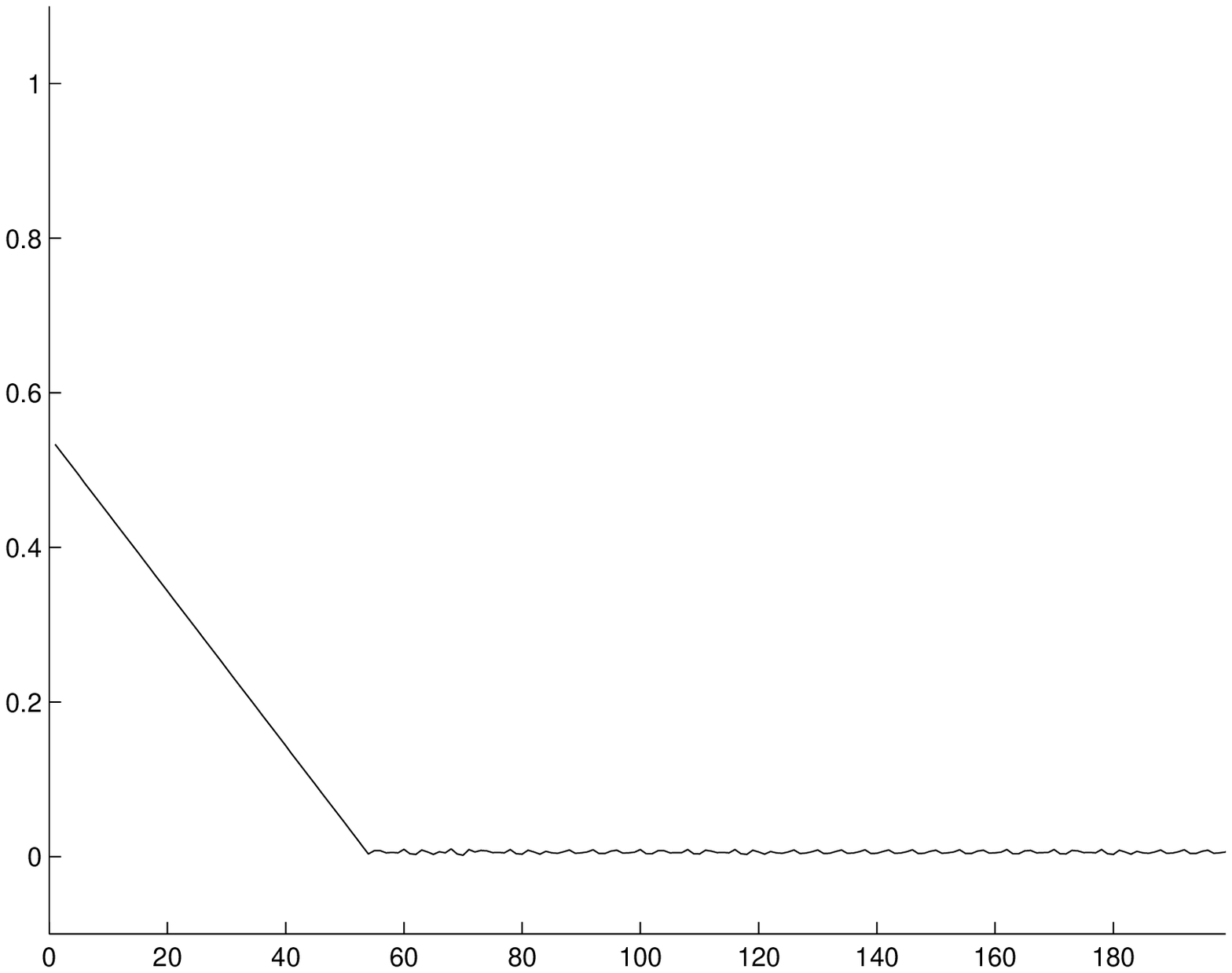}}
\end{tabular}
\end{center}

Let us also present the results of the control theoretic algorithm; the
dynamical system is
\begin{align*}
x_{1}\left(  t+1\right)   &  =x_{1}\left(  t\right)  -k\cdot\left(  x_{2}%
-\min\left[  x_{2}\left(  t\right)  ,1-x_{3}\left(  t\right)  \right]  \right)
\\
x_{2}\left(  t+1\right)   &  =x_{2}\left(  t\right)  -k\cdot\left(  x_{2}%
-\min\left[  x_{1}\left(  t\right)  ,1-x_{3}\left(  t\right)  \right]  \right)
\\
x_{3}\left(  t+1\right)   &  =x_{3}\left(  t\right)  -k\cdot\left(
x_{3}-\left(  1-x_{1}\left(  t\right)  \right)  \right)  .
\end{align*}
This system is non-linear. In Figures 12 and 13 we present results for two
typical simulations with $k=0.1$.

\begin{center}%
\begin{tabular}
[c]{cc}%
\textbf{Fig.12.a} & \textbf{Fig.12.b}\\
\scalebox{0.5}{\includegraphics{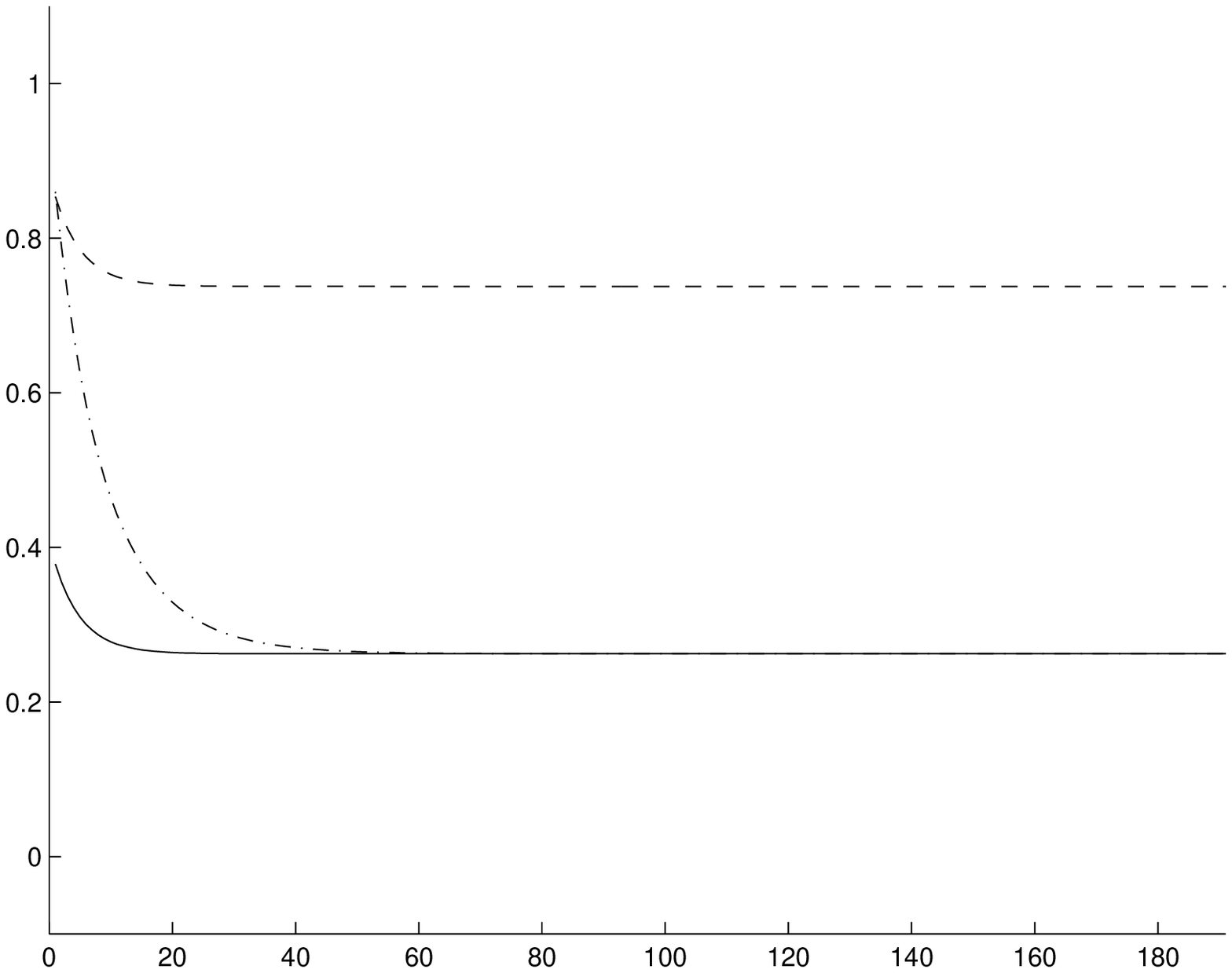}}  & \scalebox{0.5}%
{\includegraphics{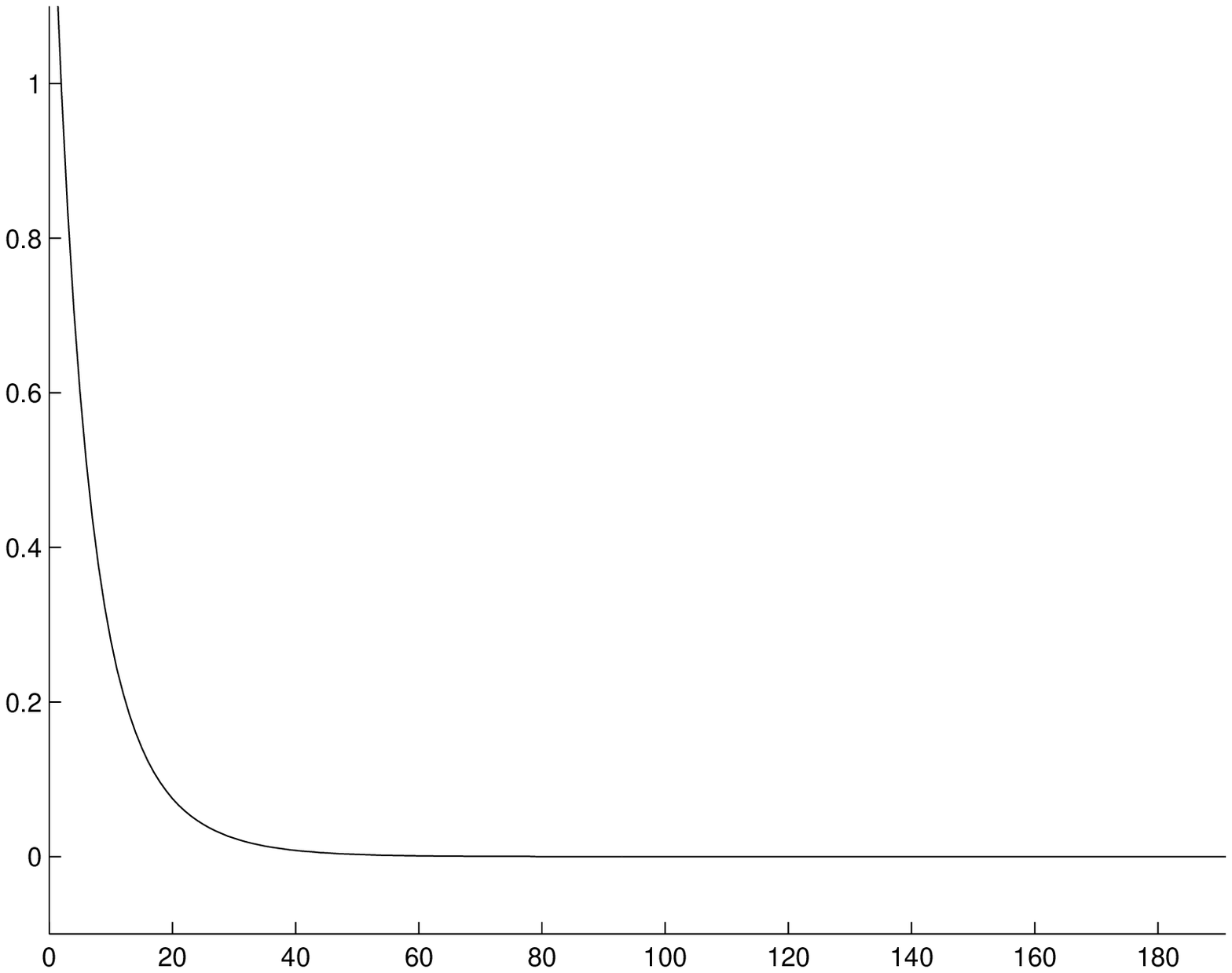}}
\end{tabular}%

\begin{tabular}
[c]{cc}%
\textbf{Fig.13.a} & \textbf{Fig.13.b}\\
\scalebox{0.5}{\includegraphics{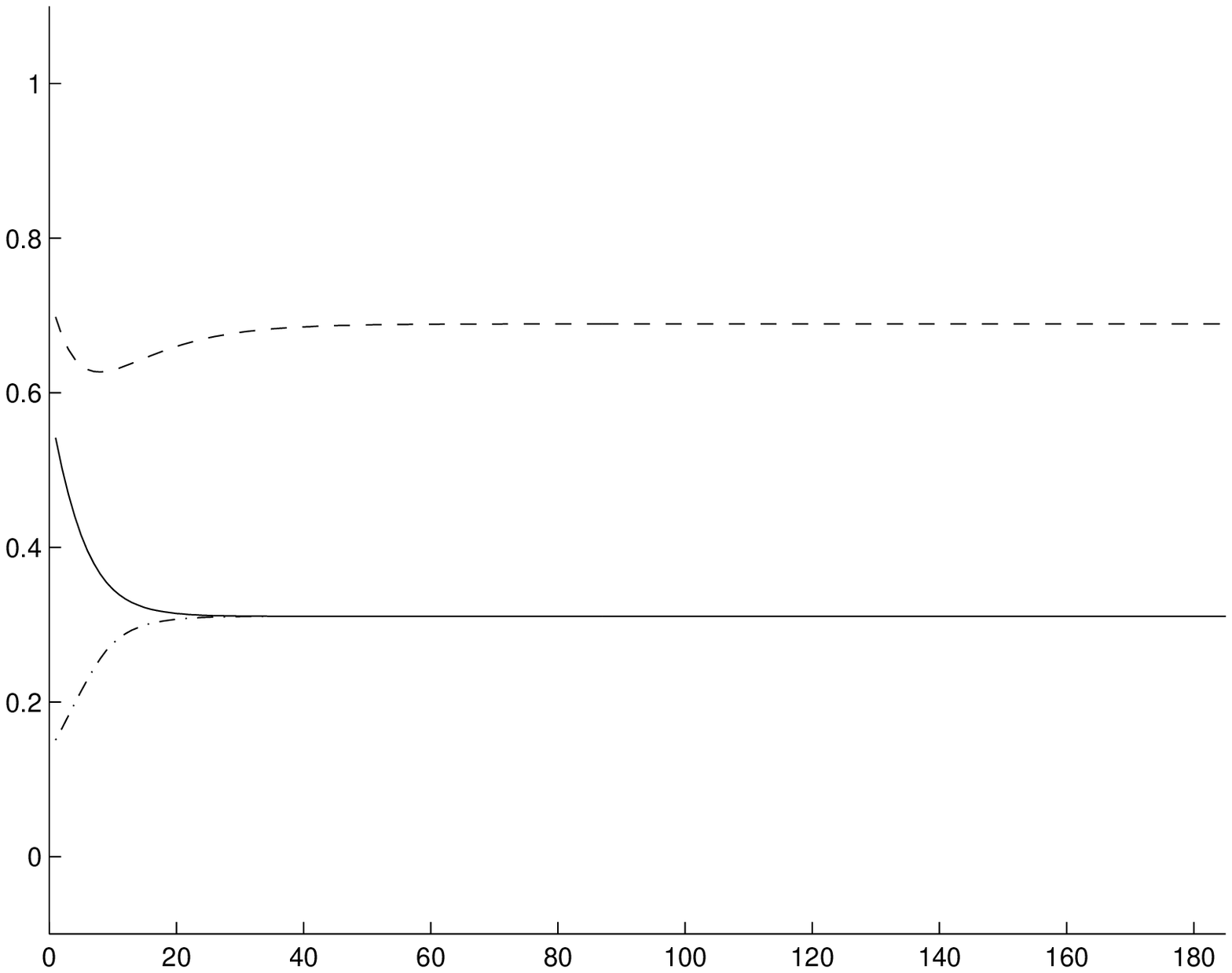}}  & \scalebox{0.5}%
{\includegraphics{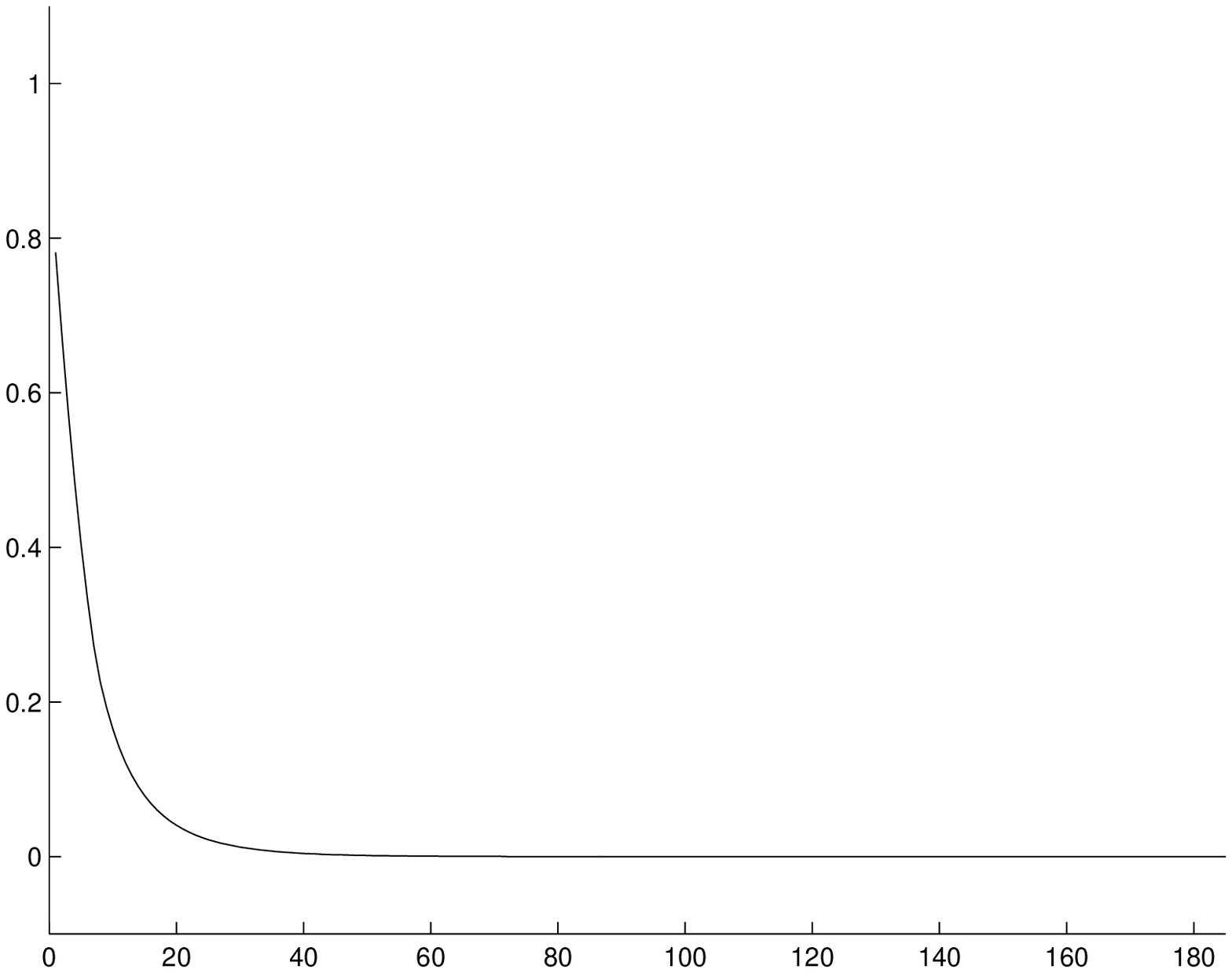}}
\end{tabular}
\end{center}

\subsubsection{$\wedge$ Implemented by Product}

Implementing $\wedge$ by the product t-norm we obtain the truth value
equations
\begin{align}
x_{1}  &  =x_{2}\cdot\left(  1-x_{3}\right) \nonumber\\
x_{2}  &  =x_{1}\cdot\left(  1-x_{3}\right) \nonumber\\
x_{3}  &  =1-x_{1}\nonumber
\end{align}
We can still use $x_{1}=1-x_{3}$ to simplify the truth value equations to
\[
x_{1}=x_{2}\cdot x_{1},\qquad x_{2}=x_{1}^{2},\qquad x_{3}=1-x_{1}%
\]
from which we obtain
\[
x_{1}=x_{1}^{3},\qquad x_{2}=x_{1}^{2},\qquad x_{3}=1-x_{1}%
\]
and finally
\[
x_{1}\cdot\left(  1-x_{1}^{2}\right)  =0,\qquad x_{2}=x_{1}^{2},\qquad
x_{3}=1-x_{1}.
\]
This has the solutions
\[
\left(  0,0,1\right)  ,\qquad\left(  1,1,0\right)  ,\qquad\left(
-1,1,0\right)  ;
\]
the last solution, however, is inadmissible as a truth value assignment. Hence
we see that for the same self-referential collection, the product
implementation of $\wedge$ yields a subset of the solutions obtained through
the min implementation.

The inconsistency function is
\[
J=\left(  x_{1}-x_{2}\cdot\left(  1-x_{3}\right)  \right)  ^{2}+\left(
x_{2}-x_{1}\cdot\left(  1-x_{3}\right)  \right)  ^{2}+\left(  x_{3}-\left(
1-x_{1}\right)  \right)  ^{2}%
\]
Because the product operator is everywhere differentiable, we can write
explicitly the gradient equations. We have%

\begin{align}
\frac{\partial J}{\partial x_{1}}  &  =-4x_{1}+4x_{2}+4x_{1}x_{3}-4x_{2}%
x_{3}+2x_{2}x_{3}^{2}\nonumber\\
\frac{\partial J}{\partial x_{2}}  &  =6x_{1}-4x_{2}+2x_{3}+4x_{2}x_{3}%
-4x_{1}x_{3}+2x_{1}x_{3}^{2}-2\nonumber\\
\frac{\partial J}{\partial x_{3}}  &  =2x_{1}+2x_{3}+4x_{1}x_{2}-2x_{2}%
^{2}+2x_{2}^{2}x_{3}-2x_{1}^{2}+2x_{1}^{2}x_{3}-2.\nonumber
\end{align}
Using the above expressions and the steepest descent algorithm we obtain
results of the type presented in Figure 14. Note that in the product
implementation the steepest descent algorithm requires a considerably larger
number of steps to converge than for the min implementation.

\begin{center}%
\begin{tabular}
[c]{cc}%
\textbf{Fig. 14.a} & \textbf{Fig. 14.b}\\
\scalebox{0.5}{\includegraphics{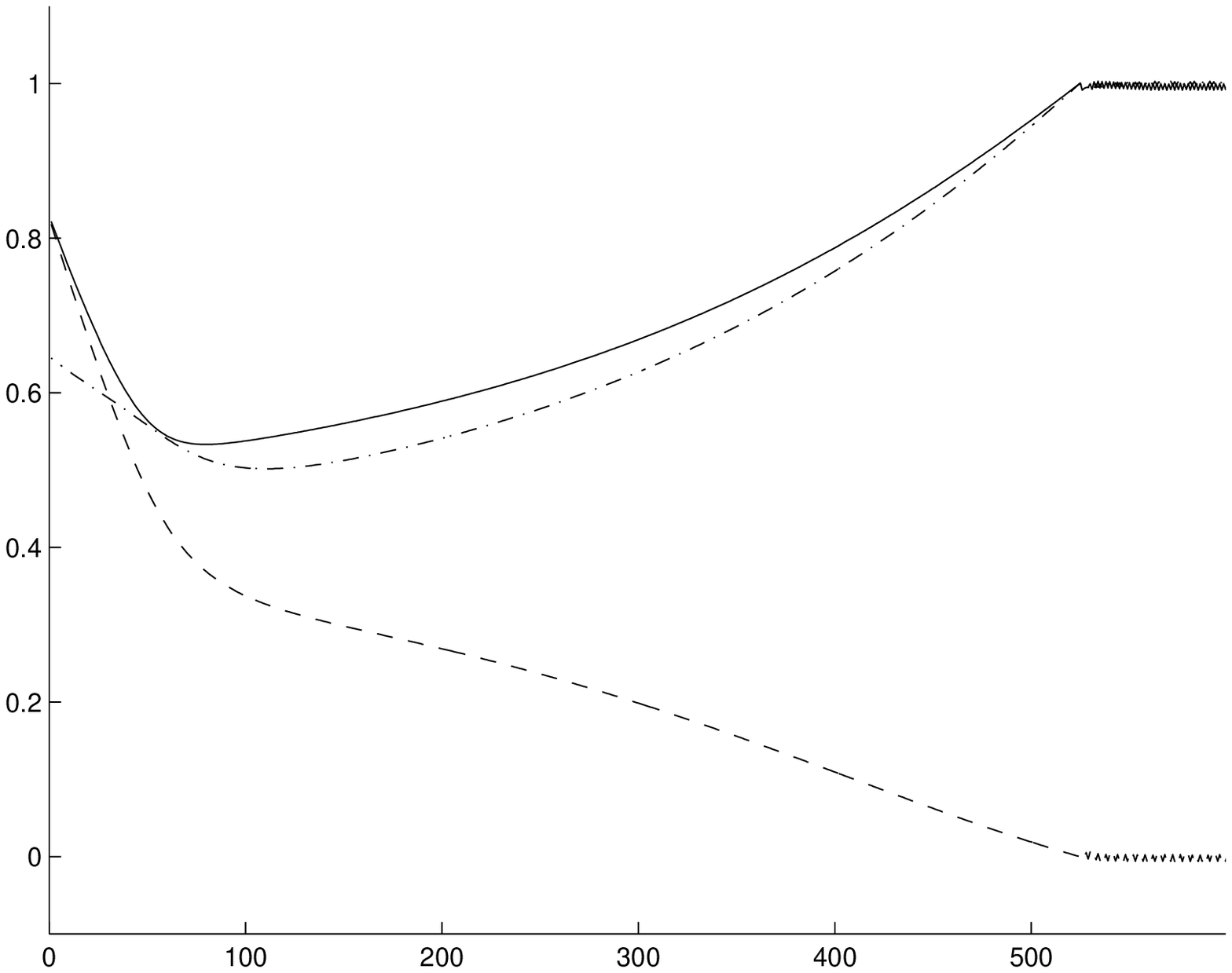}}  & \scalebox{0.5}%
{\includegraphics{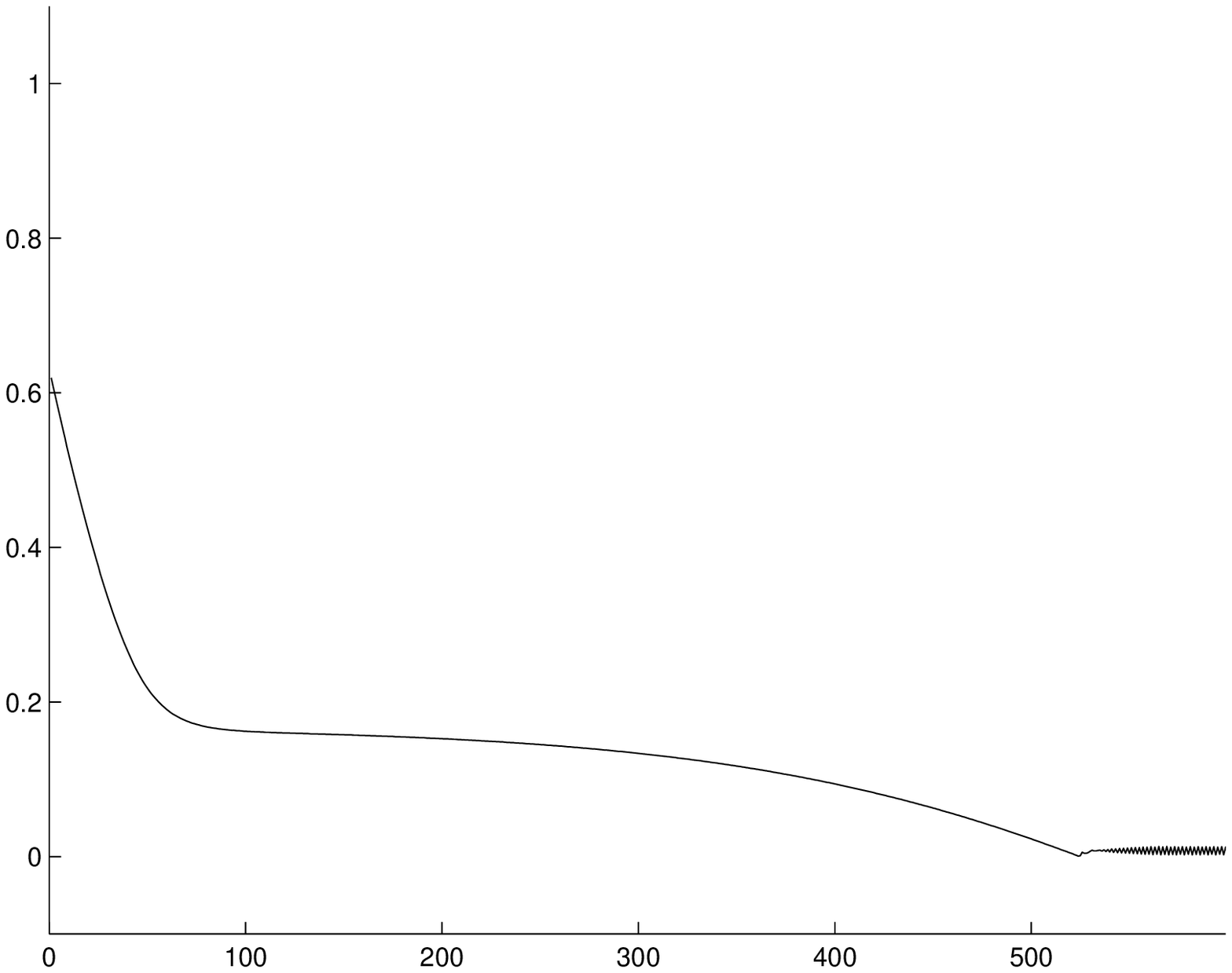}}
\end{tabular}
\end{center}

For the the control theoretic algorithm the dynamical system is
\begin{align}
x_{1}\left(  t+1\right)   &  =x_{1}\left(  t\right)  -k\cdot\left(
x_{1}\left(  t\right)  -x_{2}\left(  t\right)  \cdot\left(  1-x_{3}\left(
t\right)  \right)  \right) \nonumber\\
x_{2}\left(  t+1\right)   &  =x_{2}\left(  t\right)  -k\cdot\left(
x_{2}\left(  t\right)  -x_{1}\left(  t\right)  \cdot\left(  1-x_{3}\left(
t\right)  \right)  \right) \nonumber\\
x_{3}\left(  t+1\right)   &  =x_{3}\left(  t\right)  -k\cdot\left(
1-x_{3}\left(  t\right)  \right)  .\nonumber
\end{align}
In Figure 15 we present a typical simulation result. Note the fast convergence
to equilibirum.

\begin{center}%
\begin{tabular}
[c]{cc}%
\textbf{Fig. 15.a} & \textbf{Fig. 15.b}\\
\scalebox{0.5}{\includegraphics{fig101.eps}}  & \scalebox{0.5}%
{\includegraphics{fig102.eps}}
\end{tabular}
\end{center}

\subsection{Example 5}

\label{sec0504}

Let us now look at an example similar to the previous one, but involving
sentences outside of $\widetilde{\mathbf{S}}_{2}$:
\begin{align}
A_{1}  &  :\text{``The truth value of }A_{2}\text{ is 0.90 and the truth value
of }A_{3}\text{ is 0.20''}\label{eq0541}\\
A_{2}  &  :\text{``The truth value of }A_{1}\text{ is 0.80 and the truth value
of }A_{3}\text{ is 0.30''}\label{eq0542}\\
A_{3}  &  :\text{``The truth value of }A_{1}\text{ is 0.10''.} \label{eq0543}%
\end{align}
(\ref{eq0541})-(\ref{eq0543}) \ translates to
\begin{align*}
A_{1}  &  =D_{1}=C_{1}\wedge C_{2}\text{ }\\
A_{2}  &  =D_{2}=C_{3}\wedge C_{4}\\
A_{3}  &  =D_{3}=C_{5}%
\end{align*}
where
\begin{align*}
C_{1}  &  :\text{``Tr}\left(  A_{2}\right)  =0.90\text{''}\\
C_{2}  &  :\text{``Tr}\left(  A_{3}\right)  =0.20\text{''}\\
C_{3}  &  :\text{``Tr}\left(  A_{1}\right)  =0.80\text{''}\\
C_{4}  &  :\text{``Tr}\left(  A_{3}\right)  =0.30\text{''}\\
C_{5}  &  :\text{``Tr}\left(  A_{1}\right)  =0.10\text{''.}%
\end{align*}

We will consider two different implemementations of $\wedge$.

\subsubsection{$\wedge$ Implemented by Minimum}

If we implement $\wedge$ with the min operator the truth value equations
become
\begin{align*}
x_{1}  &  =\min\left[  1-\left|  x_{2}-0.90\right|  ,1-\left|  x_{3}%
-0.20\right|  \right] \\
x_{2}  &  =\min\left[  1-\left|  x_{2}-0.80\right|  ,1-\left|  x_{3}%
-0.30\right|  \right] \\
x_{3}  &  =1-\left|  x_{1}-0.10\right|  .
\end{align*}
These equations cannot be further reduced and while in principle they can be
solved analytically by distinguishing cases, this requires an inordinate
amount of work. Instead, we will solve them numerically, using both
Newton-Raphson and the control algorithm. Similarly, we will use steepest
descent to minimize the inconsistency function
\[
J\left(  x_{1},x_{2},x_{3}\right)  =\sum_{m=1}^{3}J_{m}\left(  x_{1}%
,x_{2},x_{3}\right)
\]
where
\begin{align*}
J_{1}\left(  x_{1},x_{2},x_{3}\right)   &  =\left(  x_{1}-\min\left[
1-\left|  x_{2}-0.90\right|  ,1-\left|  x_{3}-0.20\right|  \right]  \right)
^{2}\\
J_{2}\left(  x_{1},x_{2},x_{3}\right)   &  =\left(  x_{2}-\min\left[
1-\left|  x_{2}-0.80\right|  ,1-\left|  x_{3}-0.30\right|  \right]  \right)
^{2}\\
J_{3}\left(  x_{1},x_{2},x_{3}\right)   &  =\left(  x_{3}-1-\left|
x_{1}-0.10\right|  \right)  ^{2};
\end{align*}
the dynamical system corresponding to the steepest descent algorithm is%
\begin{align*}
x_{1}\left(  t+1\right)   &  =x_{1}\left(  t\right)  -k\cdot\frac{\partial
J}{\partial x_{1}}\\
x_{2}\left(  t+1\right)   &  =x_{2}\left(  t\right)  -k\cdot\frac{\partial
J}{\partial x_{2}}\\
x_{3}\left(  t+1\right)   &  =x_{3}\left(  t\right)  -k\cdot\frac{\partial
J}{\partial x_{3}}.
\end{align*}
For a numerical implementation the partial derivatives can be approximated
numerically. Finally, for the control theoretic algorithm the dynamical system
is
\begin{align*}
x_{1}\left(  t+1\right)   &  =x_{1}\left(  t\right)  -k\cdot\left(
x_{1}\left(  t\right)  -\min\left[  1-\left|  x_{2}\left(  t\right)
-0.90\right|  ,1-\left|  x_{3}\left(  t\right)  -0.20\right|  \right]  \right)
\\
x_{2}\left(  t+1\right)   &  =x_{2}\left(  t\right)  -k\cdot\left(
x_{2}\left(  t\right)  -\min\left[  1-\left|  x_{1}\left(  t\right)
-0.90\right|  ,1-\left|  x_{3}\left(  t\right)  -0.20\right|  \right]  \right)
\\
x_{3}\left(  t+1\right)   &  =x_{3}\left(  t\right)  -k\cdot\left(
x_{3}\left(  t\right)  -1+\left|  x_{3}\left(  t\right)  -0.20\right|
\right)  .
\end{align*}

In Figure 16 we present some simulation results for the Newton-Raphson
algorithm, in Figure 17 we present results for the steepest descent algorithm
and in Figure 18 for the control theoretic algorithm. Both Newton-Raphson and
the control algorithm discover the same solution, namely $\overline{x}=\left(
0.95,0.85,0.15\right)  ^{T}$. Repeated simulations (not presented here) give
always the same solution, so it is possible that this is the \emph{unique}
consistent truth value assignment for this problem. The steepest descent
algorithm gets trapped at a local minimum, namely $\overline{x}=\left(
0.56,0.71,0.59\right)  ^{T}$, which does \emph{not} yield zero inconsistency.
This is not accidental; we have noticed after repeated simulations (not
presented here) that in general the steepest descent algorithm in this case
does not yield zero inconsistency, except when started quite close to the true
solution. Note the very fast convergence of Newton-Raphson:\ it reaches
equilibrium in 7 steps, as compared to the approximately 130 steps required by
the control algorithm.

\begin{center}%
\begin{tabular}
[c]{cc}%
\textbf{Fig. 16.a} & \textbf{Fig. 16.b}\\
\scalebox{0.5}{\includegraphics{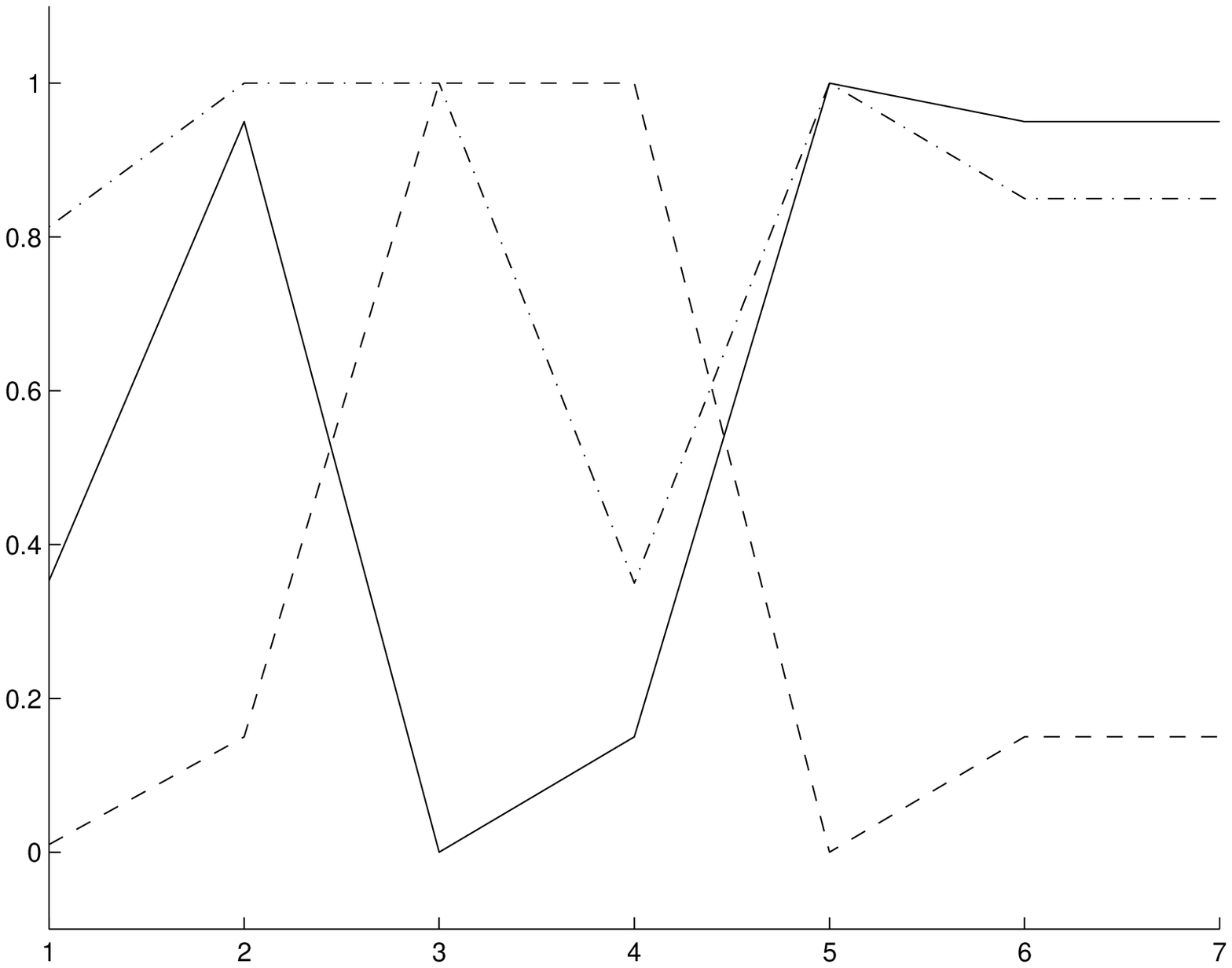}}  & \scalebox{0.5}%
{\includegraphics{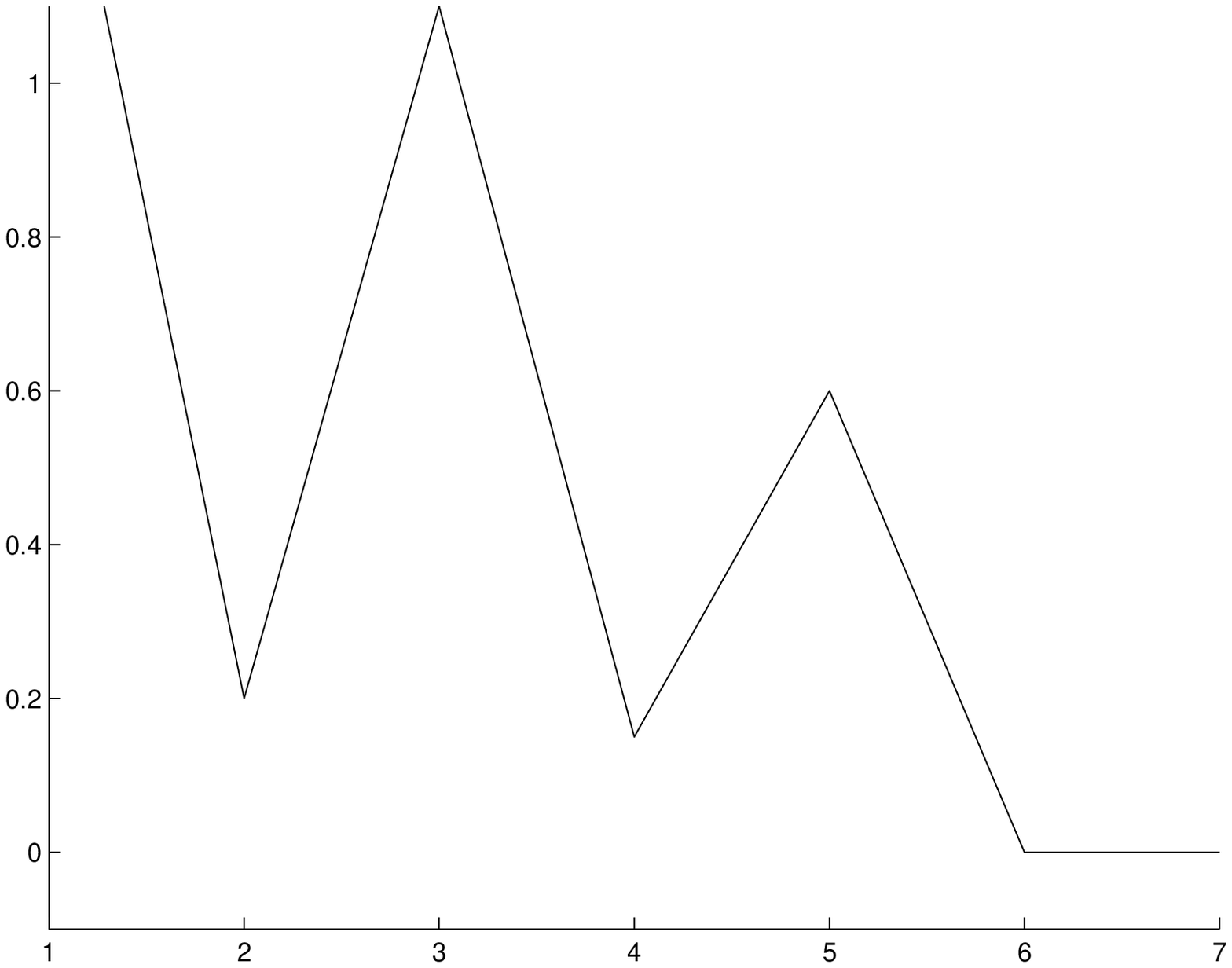}}
\end{tabular}%

\begin{tabular}
[c]{cc}%
\textbf{Fig. 17.a} & \textbf{Fig. 17.b}\\
\scalebox{0.5}{\includegraphics{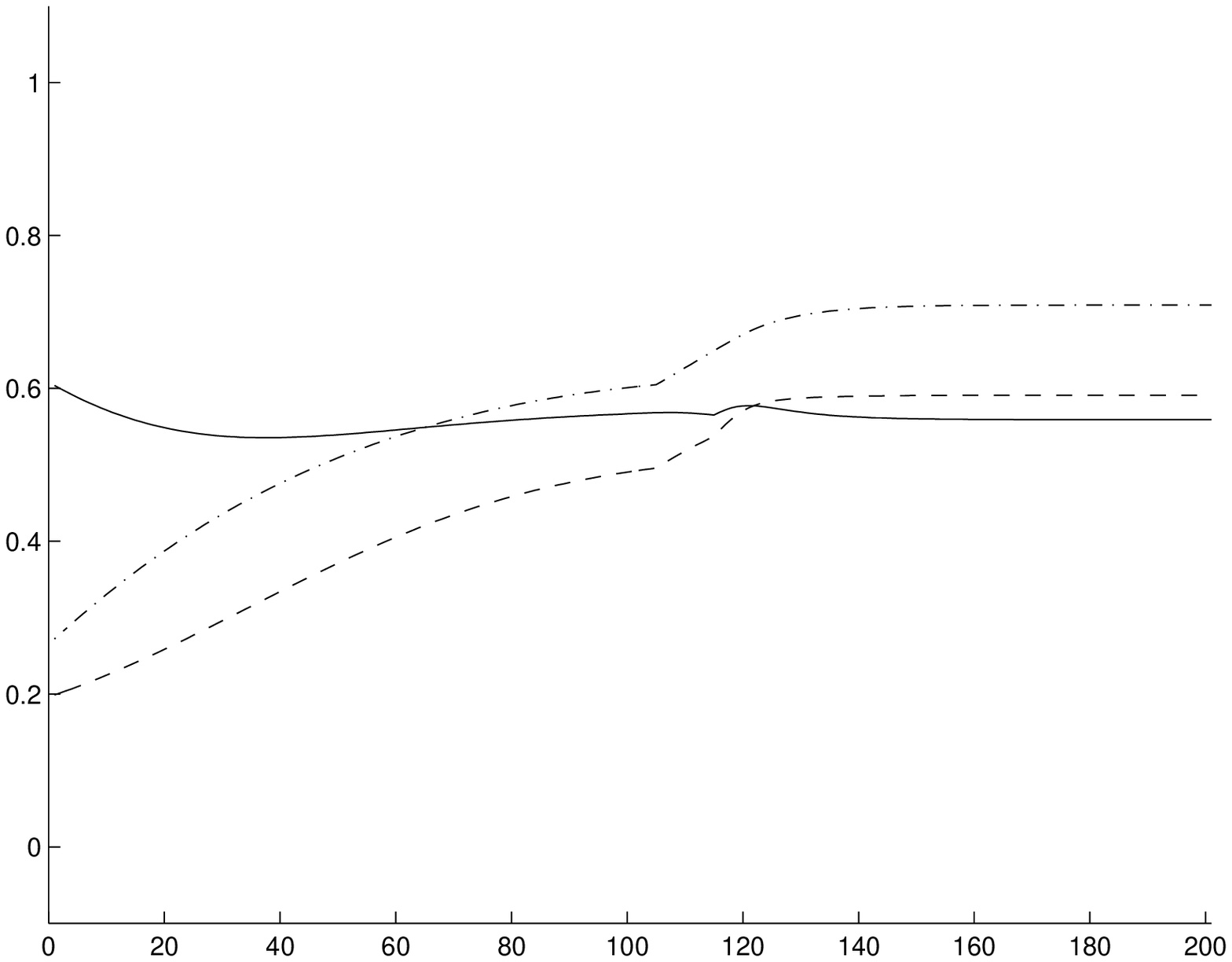}}  & \scalebox{0.5}%
{\includegraphics{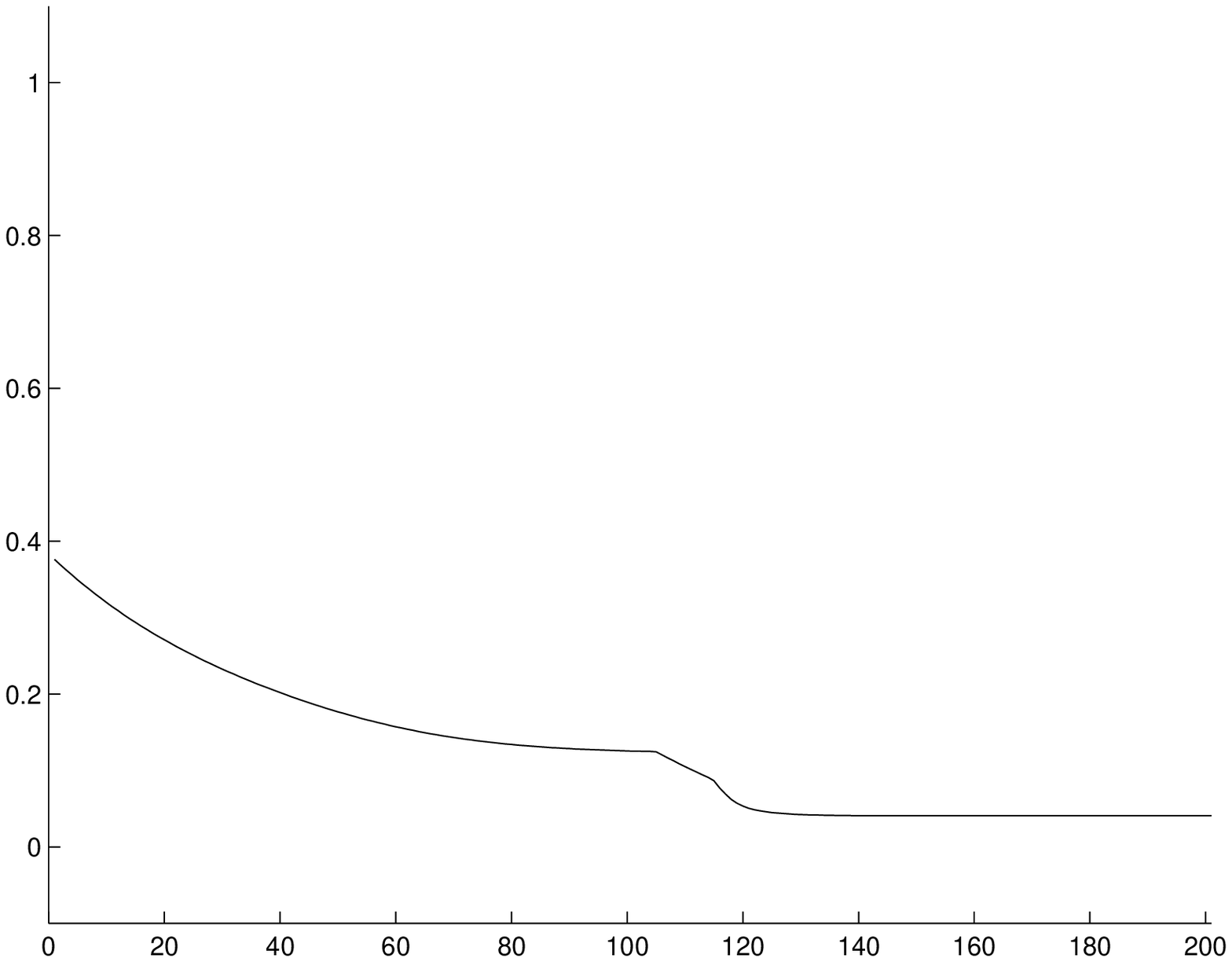}}
\end{tabular}%

\begin{tabular}
[c]{cc}%
\textbf{Fig. 18.a} & \textbf{Fig. 18.b}\\
\scalebox{0.5}{\includegraphics{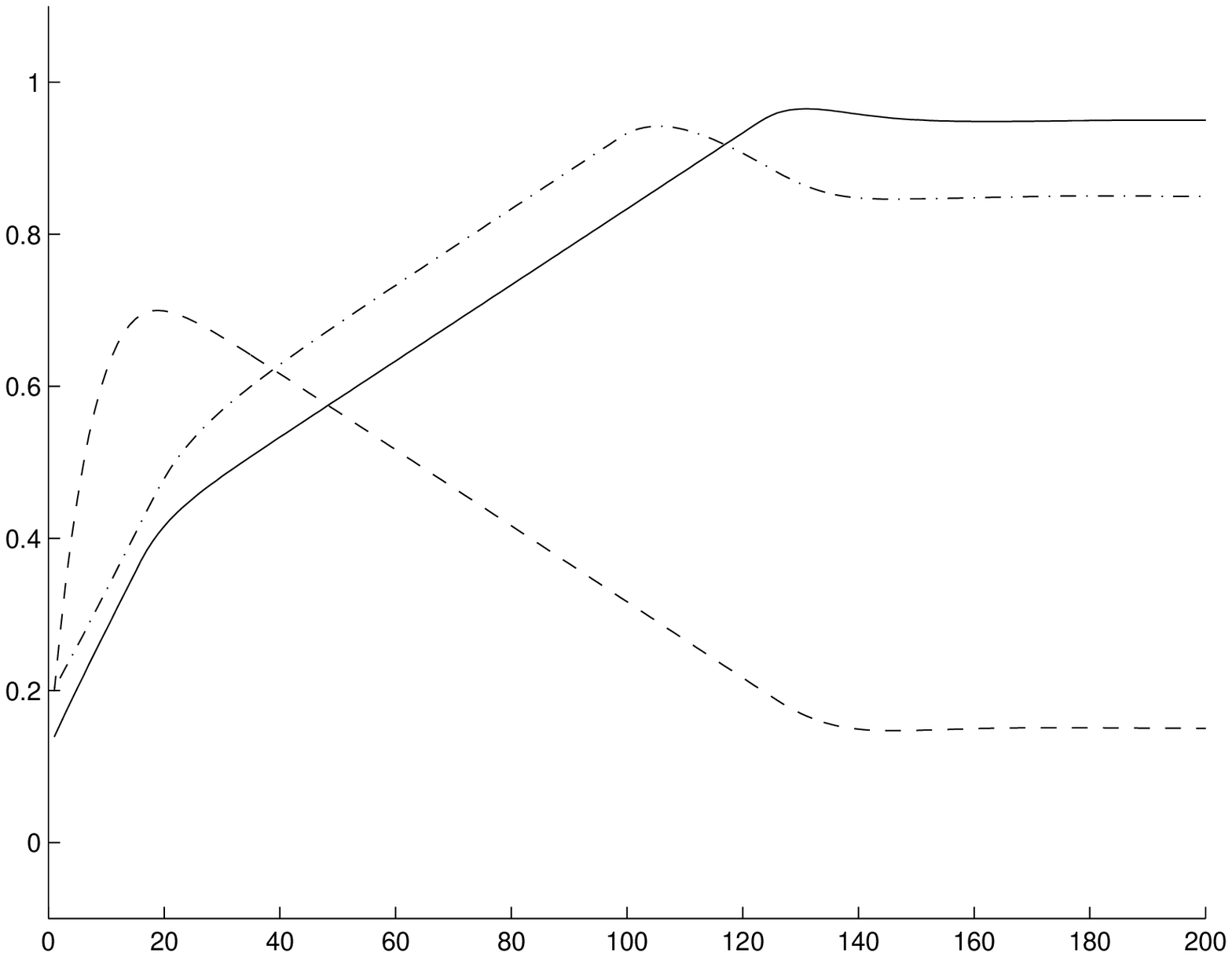}}  & \scalebox{0.5}%
{\includegraphics{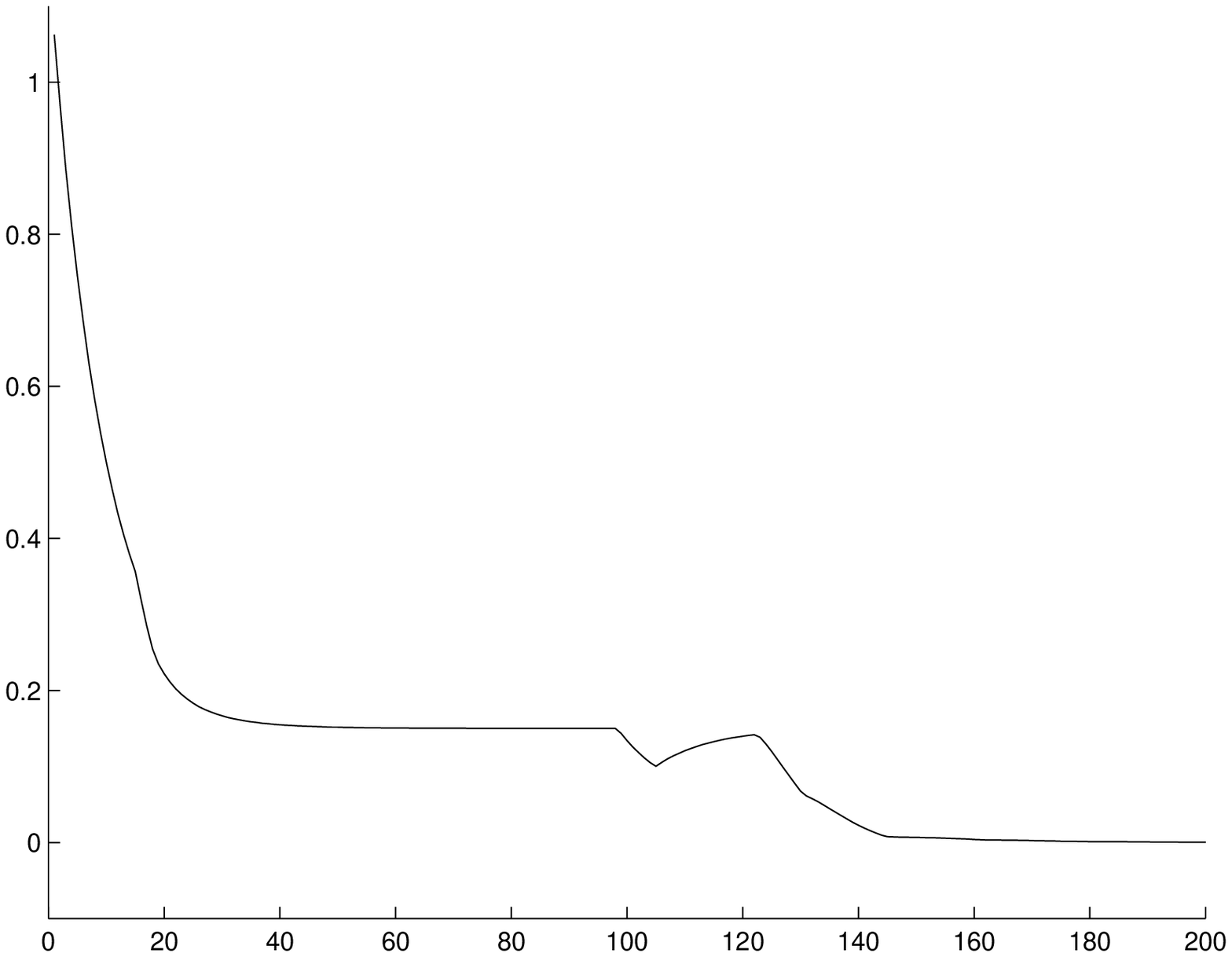}}
\end{tabular}
\end{center}

\subsubsection{$\wedge$ Implemented by Product}

The situation is similar when we implement $\wedge$ by product.. The truth
value equations become
\begin{align*}
x_{1}  &  =\left(  1-\left|  x_{2}-0.90\right|  \right)  \cdot\left(
1-\left|  x_{3}-0.20\right|  \right) \\
x_{2}  &  =\left(  1-\left|  x_{2}-0.80\right|  \right)  \cdot\left(
1-\left|  x_{3}-0.30\right|  \right) \\
x_{3}  &  =1-\left|  x_{1}-0.10\right|
\end{align*}
We omit the details which are very similar to previous cases and present the
results of numerical simulation.

In Figure 19 we present some simulation results for the Newton-Raphson
algorithm, in Figure 20 we present results for the steepest descent algorithm
and in Figure 21 for the control theoretic algorithm. The Newton-Raphson and
steepest descent algorithms reach the solution $\overline{x}=\left(
0.6784,0.7715,0.4216\right)  ^{T}$, while the control algorithm reaches the
solution $\widetilde{x}=\left(  0.0473,0.0872,0.9473\right)  ^{T}$. Both of
these are consistent truth value assignments; other simulations (not presented
here)\ have yielded additional solutions. Note that in this case the steepest
descent algorithm reaches zero inconsistency. Otherwise, the remarks
previously made for the min implementation also hold for the product implementation.

\begin{center}%
\begin{tabular}
[c]{cc}%
\textbf{Fig. 19.a} & \textbf{Fig. 19.b}\\
\scalebox{0.5}{\includegraphics{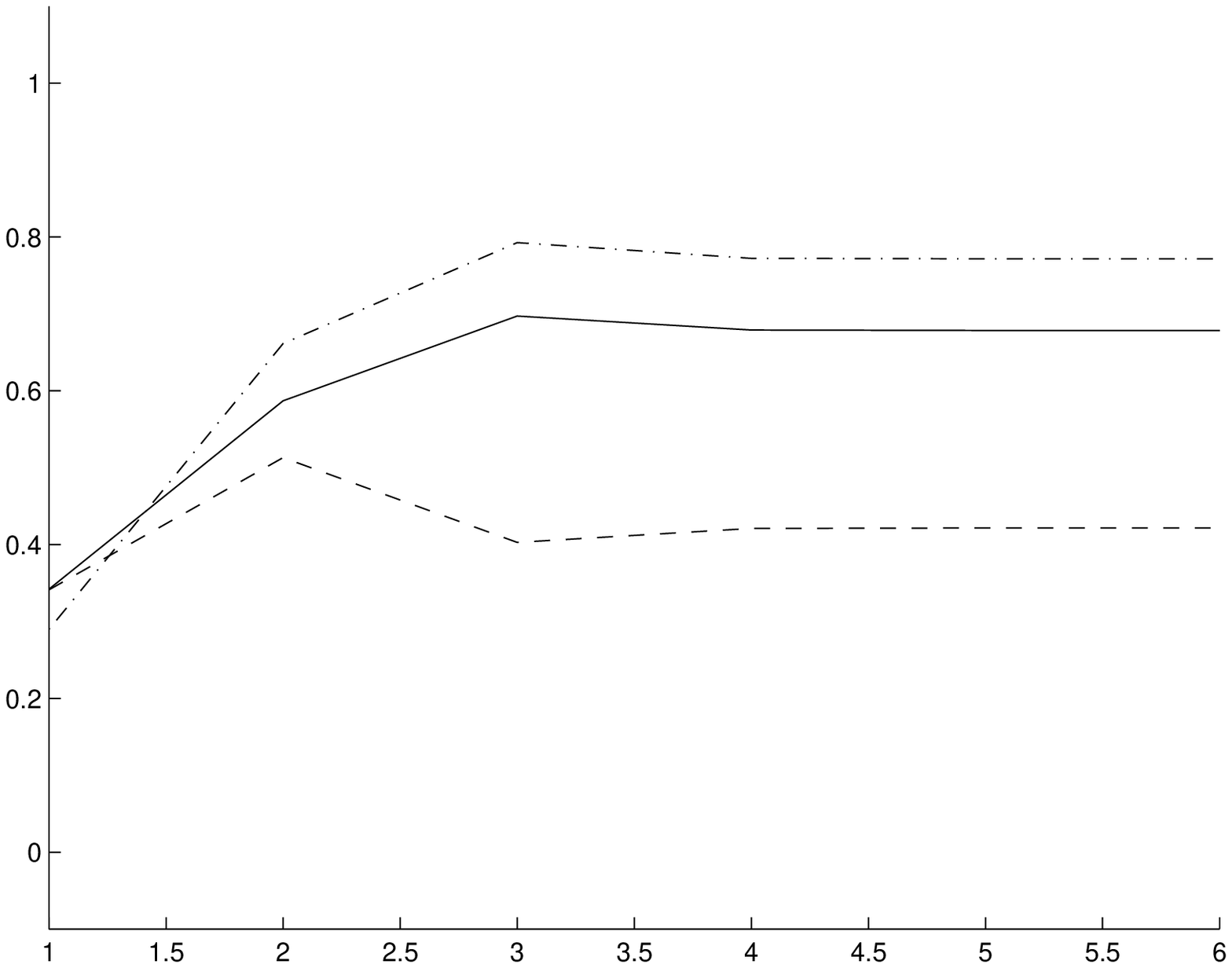}}  & \scalebox{0.5}%
{\includegraphics{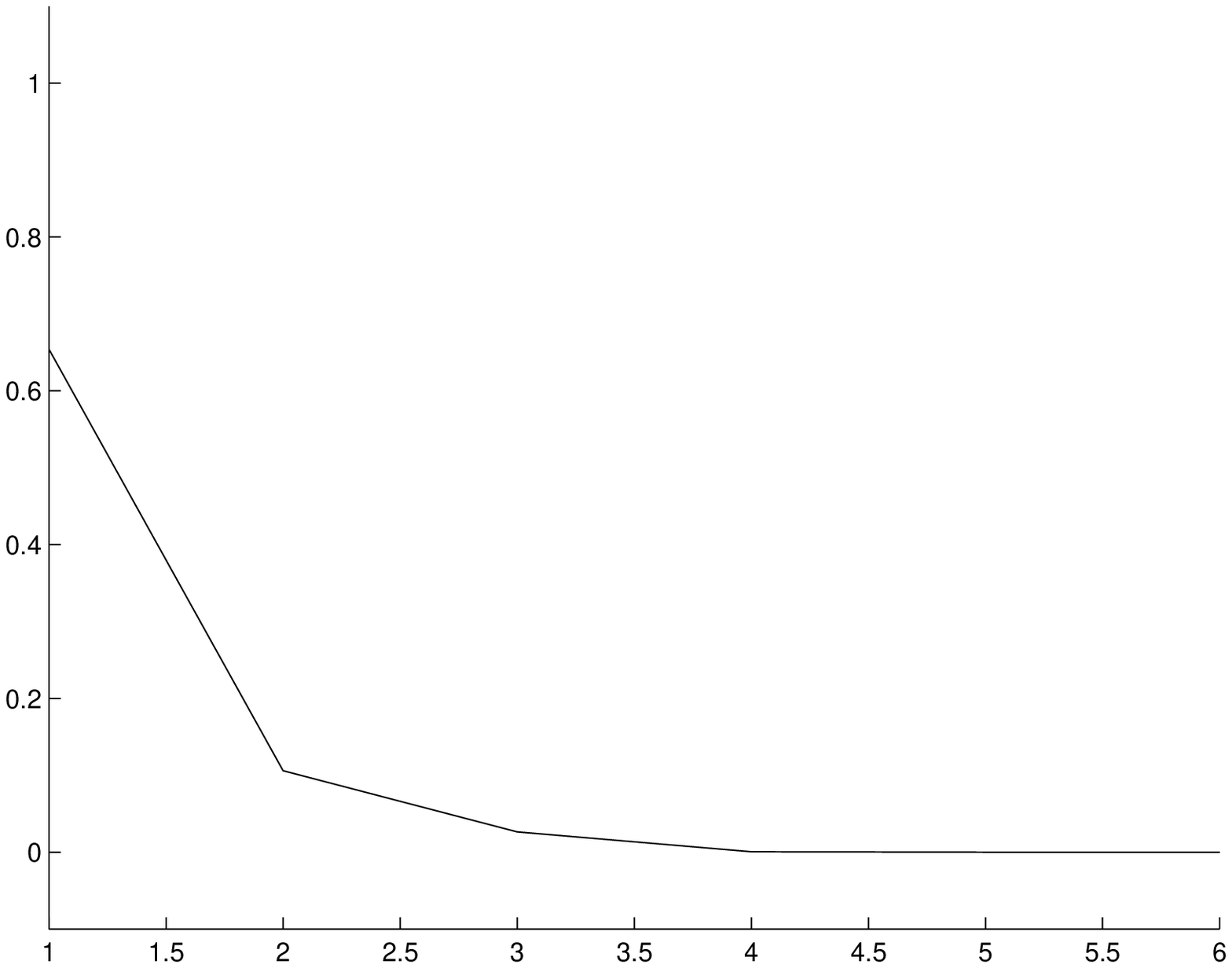}}
\end{tabular}%

\begin{tabular}
[c]{cc}%
\textbf{Fig. 20.a} & \textbf{Fig. 20.b}\\
\scalebox{0.5}{\includegraphics{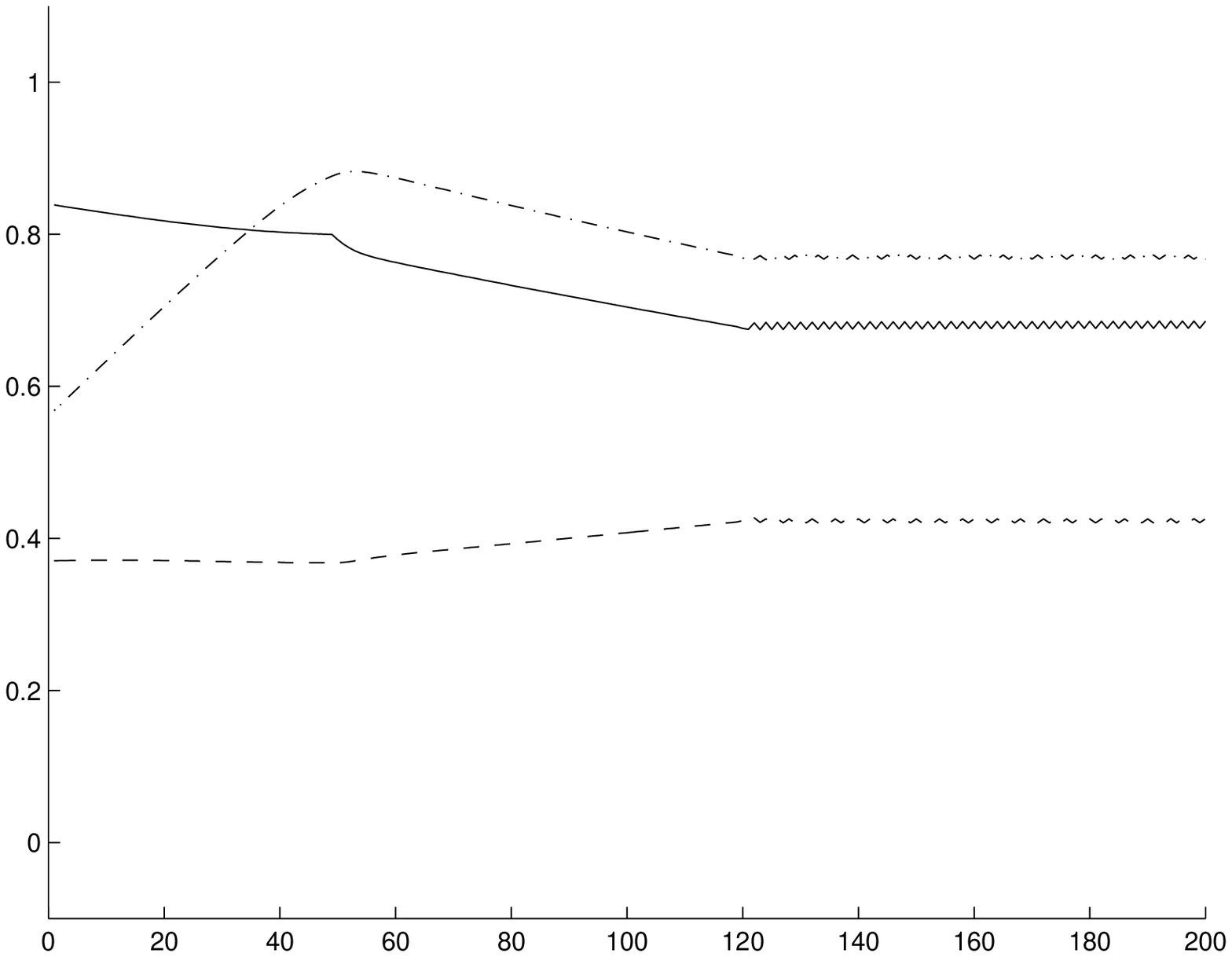}}  & \scalebox{0.5}%
{\includegraphics{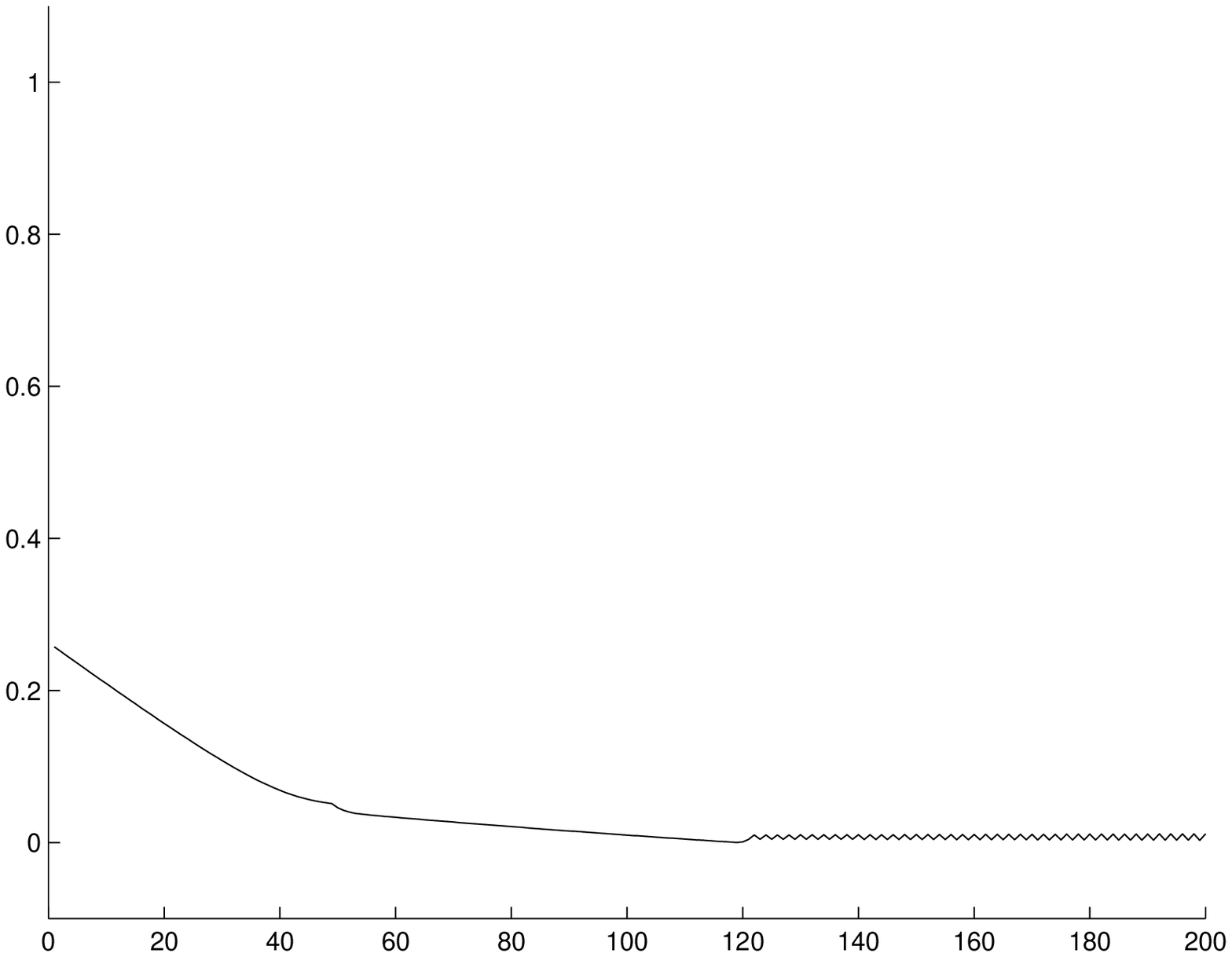}}
\end{tabular}%

\begin{tabular}
[c]{cc}%
\textbf{Fig. 21.a} & \textbf{Fig. 21.b}\\
\scalebox{0.5}{\includegraphics{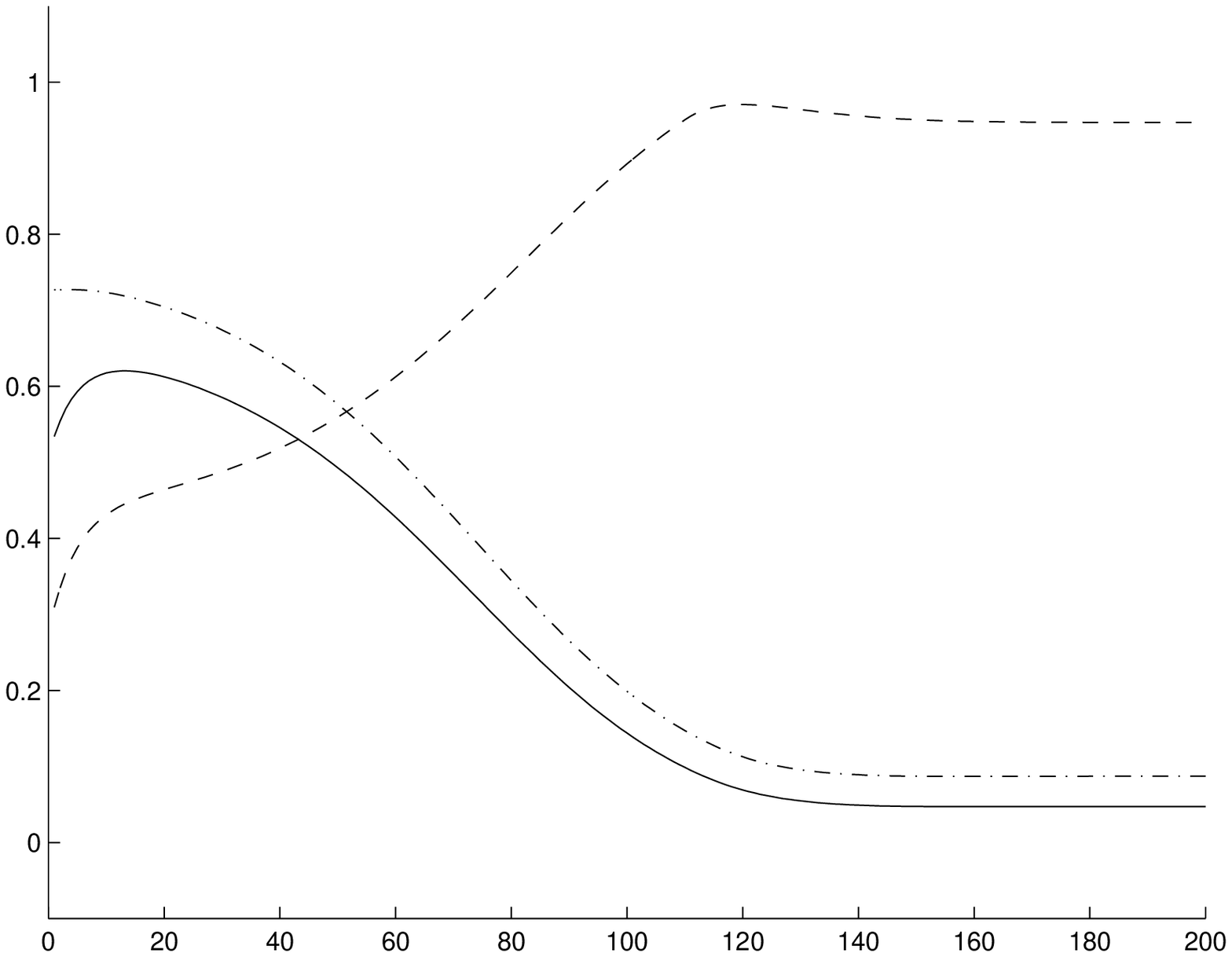}}  & \scalebox{0.5}%
{\includegraphics{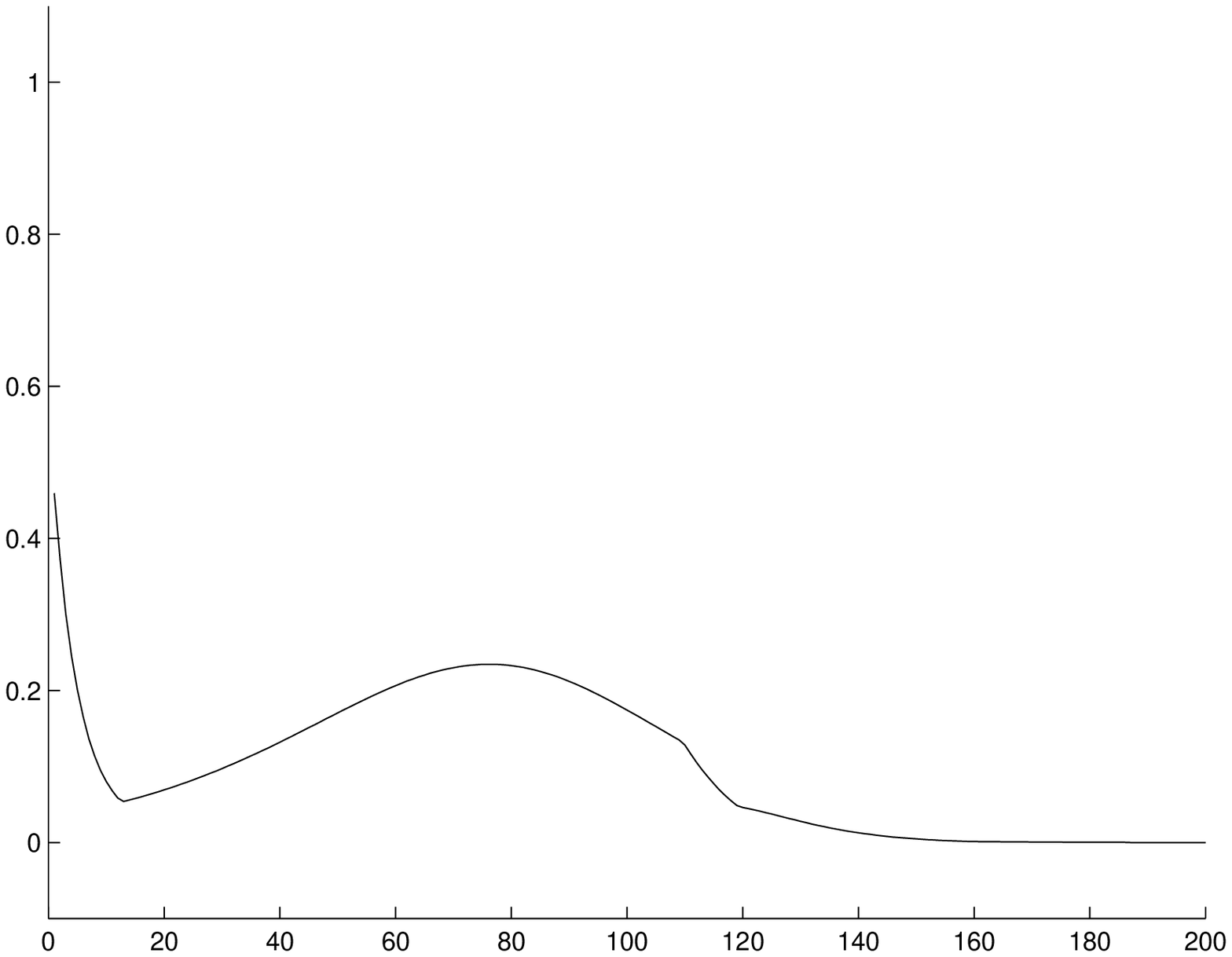}}
\end{tabular}
\end{center}

\subsection{Example 6}

\label{sec0505}

Here is a more complicated example:
\begin{align*}
A_{1}  &  :\text{(``}A_{1}\text{ has truth value 0.75'' and ``}A_{2}\text{ has
truth value 0.35'') or ``}A_{4}\text{ has truth value 1.00''}\\
A_{2}  &  :\text{(``}A_{1}\text{ or }A_{3}\text{ has truth value 1.00'') and
``}A_{4}\text{ has truth value 0.10''}\\
A_{3}  &  :\text{``}A_{2}\text{ has truth value 0.00'' and ``}A_{3}\text{ has
truth value 0.35''}\\
A_{4}  &  :\text{``The opposite of }A_{1}\text{ has truth value 0.25''}%
\end{align*}
This translates to
\begin{align*}
A_{1}  &  =\left(  C_{1}\wedge C_{2}\right)  \vee C_{3}\\
A_{2}  &  =C_{4}\wedge C_{5}\\
A_{3}  &  =C_{6}\wedge C_{7}\\
A_{4}  &  =C_{8}%
\end{align*}
where%

\begin{align*}
C_{1}  &  :\text{``The truth value of }A_{1}\text{ is 0.75''\quad}\\
C_{2}  &  :\text{``The truth value of }A_{2}\text{ is 0.35''}\\
C_{3}  &  :\text{``The truth value of }A_{4}\text{ is 1.00''}\\
C_{4}  &  :\text{``The truth value of }A_{1}\vee A_{3}\text{ is 1.00''}\\
C_{5}  &  :\text{``The truth value of }A_{4}\text{ is 0.10''}\\
C_{6}  &  :\text{``The truth value of }A_{2}\text{ is 0.00''}\\
C_{7}  &  :\text{``The truth value of }A_{3}\text{ is 0.35''}\\
C_{8}  &  :\text{``The truth value of }A_{1}^{\prime}\text{ is 0.25''.}%
\end{align*}
Note that $C_{4}$ belongs to $\mathbf{S}_{2}$ proper, i.e. it is \emph{not} an
elementary truth value assessment.

We consider two alternative systems of truth value equations; in the first
case we implement $\vee$, $\wedge$ by the standard implementation; in the
second case we implement $\vee$, $\wedge$ by the algebraic implementation.

\subsubsection{$\wedge$ Implemented by Minimum, $\vee$ Implemented by Maximum}

We present the truth value equations:
\begin{align}
x_{1}  &  =\max\left[  \min\left(  1-\left|  x_{1}-0.75\right|  ,1-\left|
x_{2}-0.35\right|  \right)  ,1-\left|  x_{4}-1.00\right|  \right]
\label{eq101}\\
x_{2}  &  =\min\left[  1-\left|  \max\left(  x_{1},x_{3}\right)  -1.00\right|
,1-\left|  x_{4}-0.10\right|  \right] \label{eq102}\\
x_{3}  &  =\min\left[  1-\left|  x_{2}-0.00\right|  ,1-\left|  x_{3}%
-0.35\right|  \right] \label{eq103}\\
x_{4}  &  =1-\left|  1-x_{1}-0.25\right|  . \label{eq104}%
\end{align}
We omit further details and directly present the results of numerical simulation.

In Figures 22, 23 we present some simulation results for the Newton-Raphson
algorithm, with various initial conditions. Note that in both figures the
algorithm fails to locate a consistent truth value assignment; this is true
for most of the simulations we have run, i.e. usually Newton-Raphson fails to
solve the truth value equations (\ref{eq101})--(\ref{eq104}).

\begin{center}%
\begin{tabular}
[c]{cc}%
\textbf{Fig. 22.a} & \textbf{Fig. 22.b}\\
\scalebox{0.5}{\includegraphics{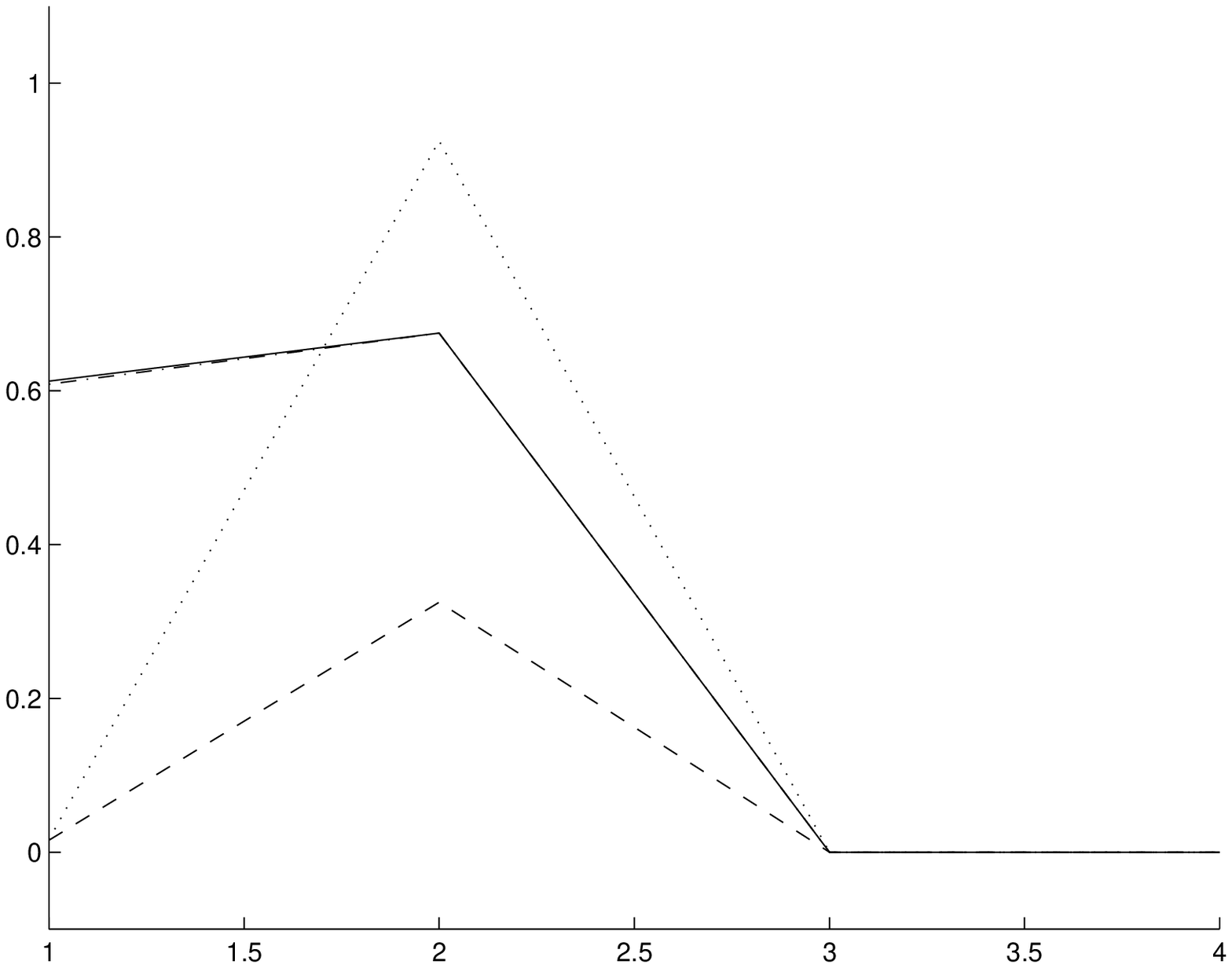}}  & \scalebox{0.5}%
{\includegraphics{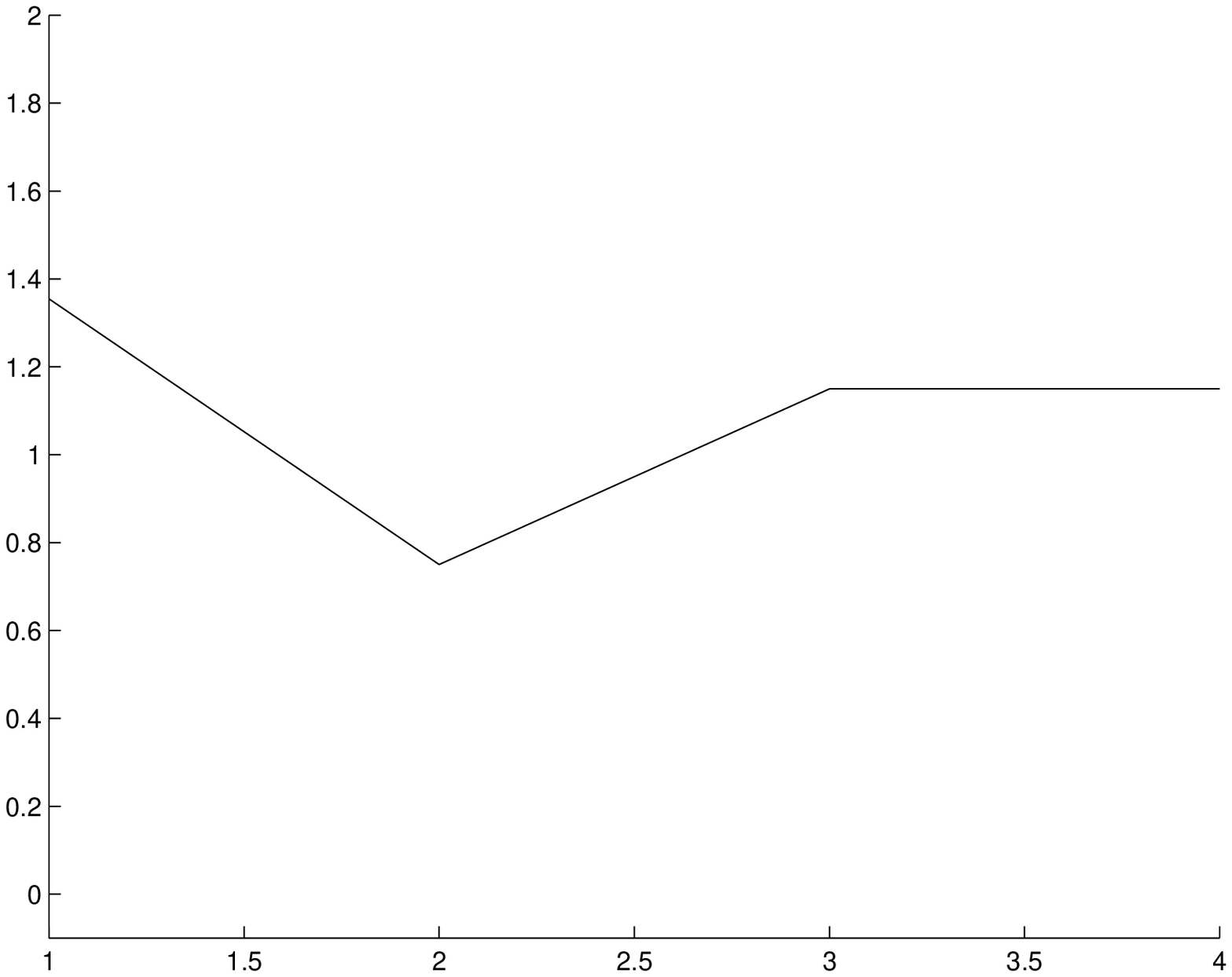}}
\end{tabular}%

\begin{tabular}
[c]{cc}%
\textbf{Fig. 23.a} & \textbf{Fig. 23.b}\\
\scalebox{0.5}{\includegraphics{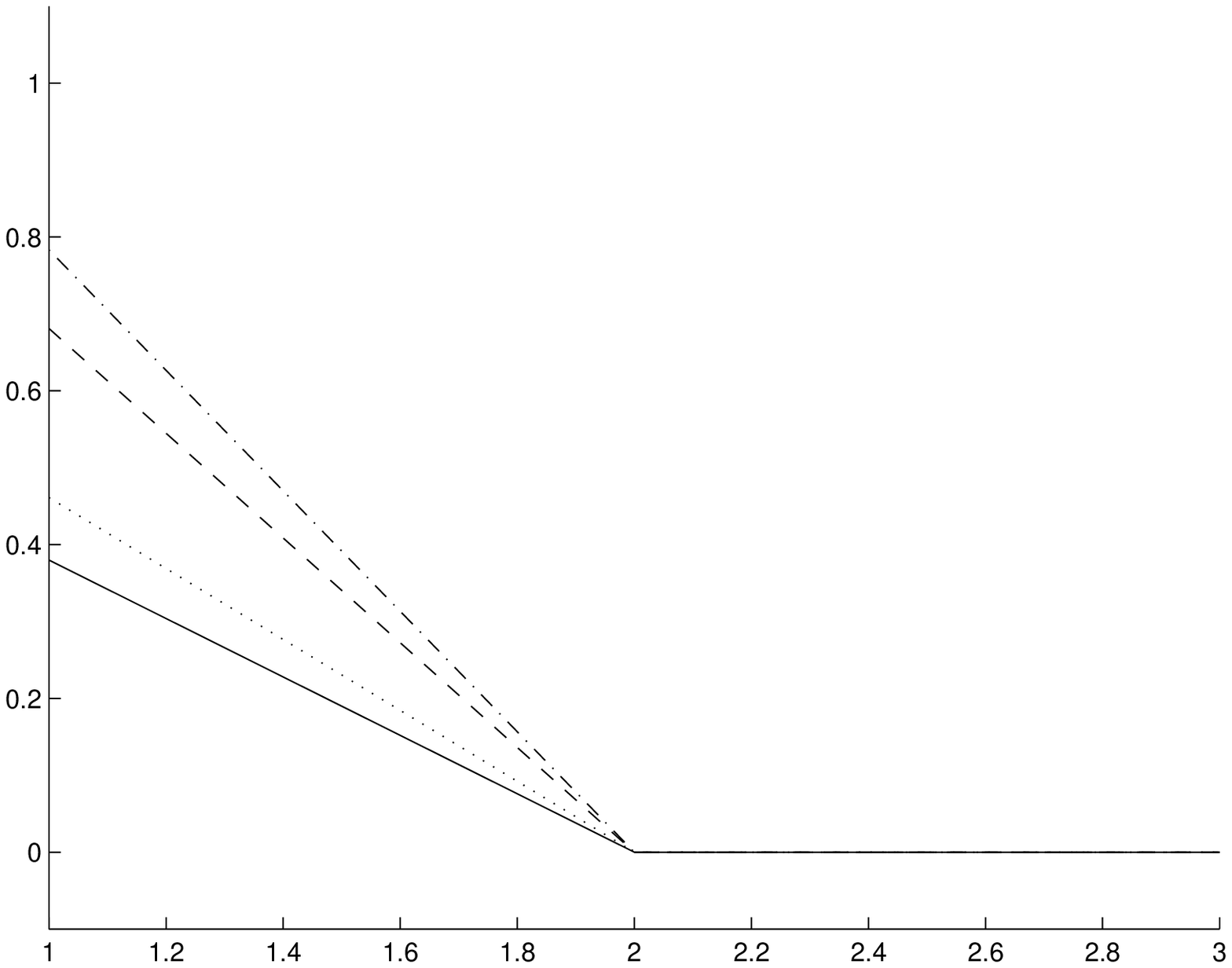}}  & \scalebox{0.5}%
{\includegraphics{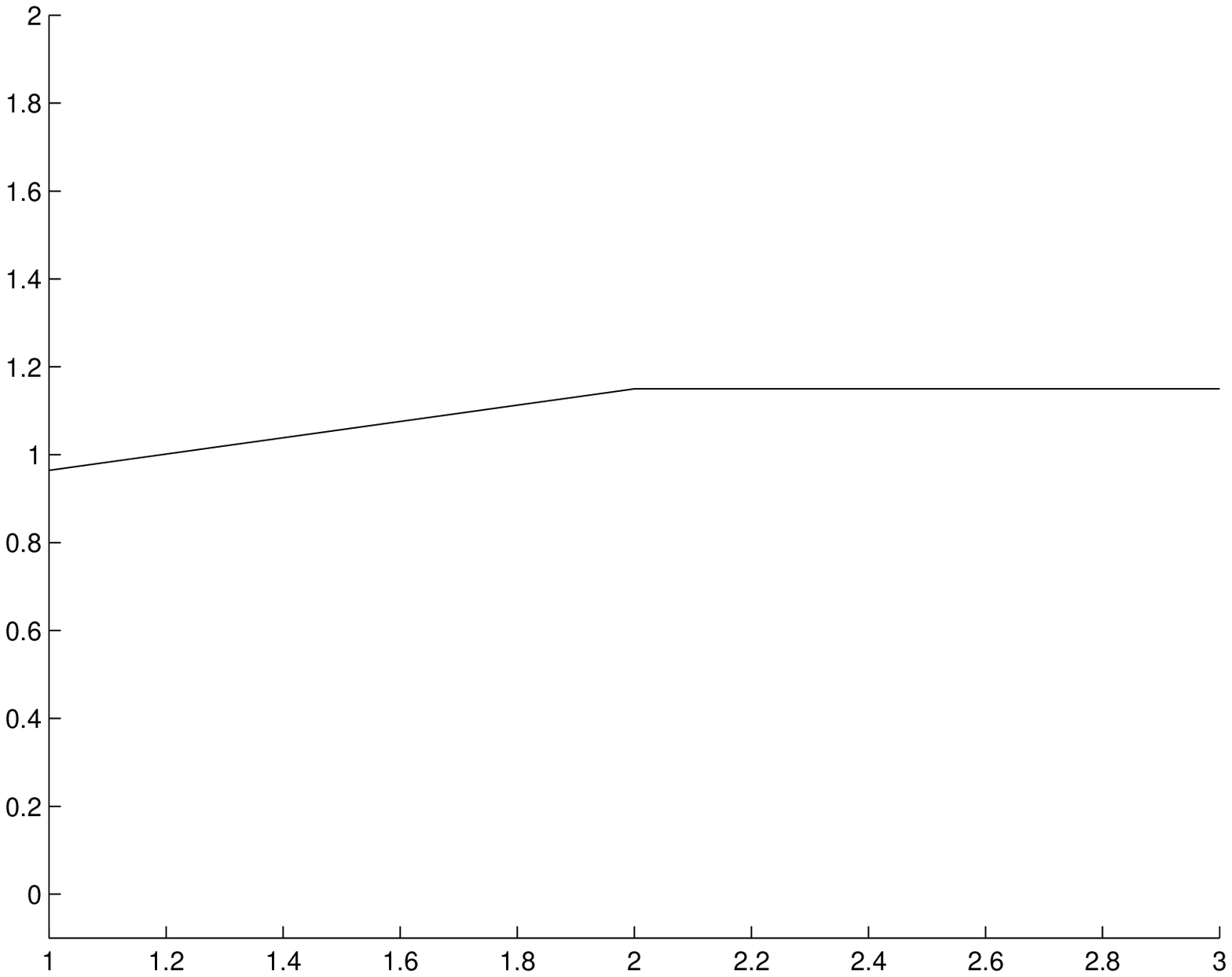}}
\end{tabular}
\end{center}

\medskip In Figures 24, 25 we present some simulation results for the steepest
descent algorithm, with various initial conditions. In Figure 24 the algorithm
converges to $\overline{x}$=$\left(  0.875,0.225,0.675,0.875\right)  ^{T}$
which yields zero total inconsistency; this is not the case in Figure 25,
where the algorithm converges to a local minimum and the truth value equations
are not satisfied. This has occurred in several simualtions, i.e. the steepest
descent algorithm does not reliably solve the truth value equations
(\ref{eq101})--(\ref{eq104}).

\begin{center}%
\begin{tabular}
[c]{cc}%
\textbf{Fig. 24.a} & \textbf{Fig. 24.b}\\
\scalebox{0.5}{\includegraphics{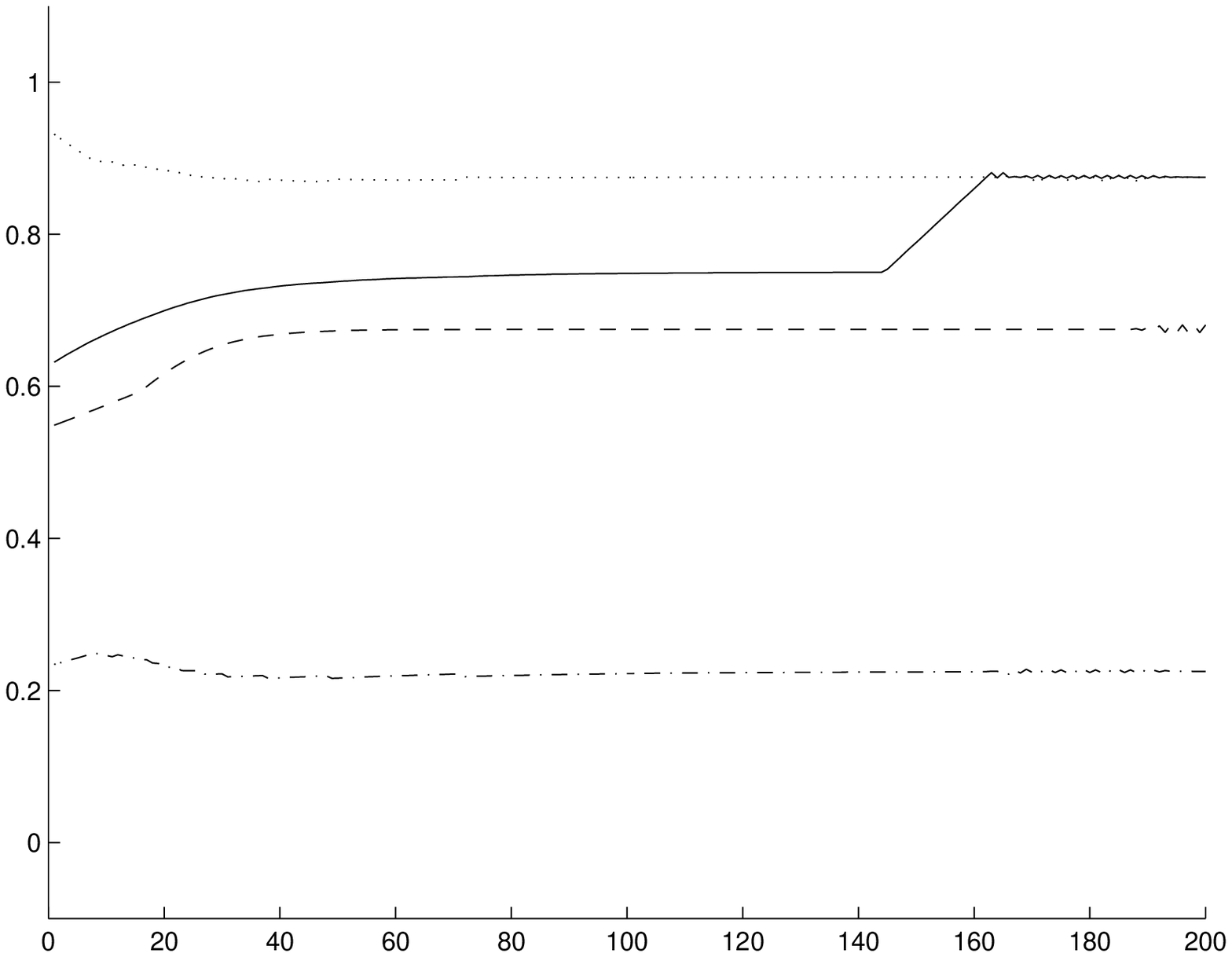}}  & \scalebox{0.5}%
{\includegraphics{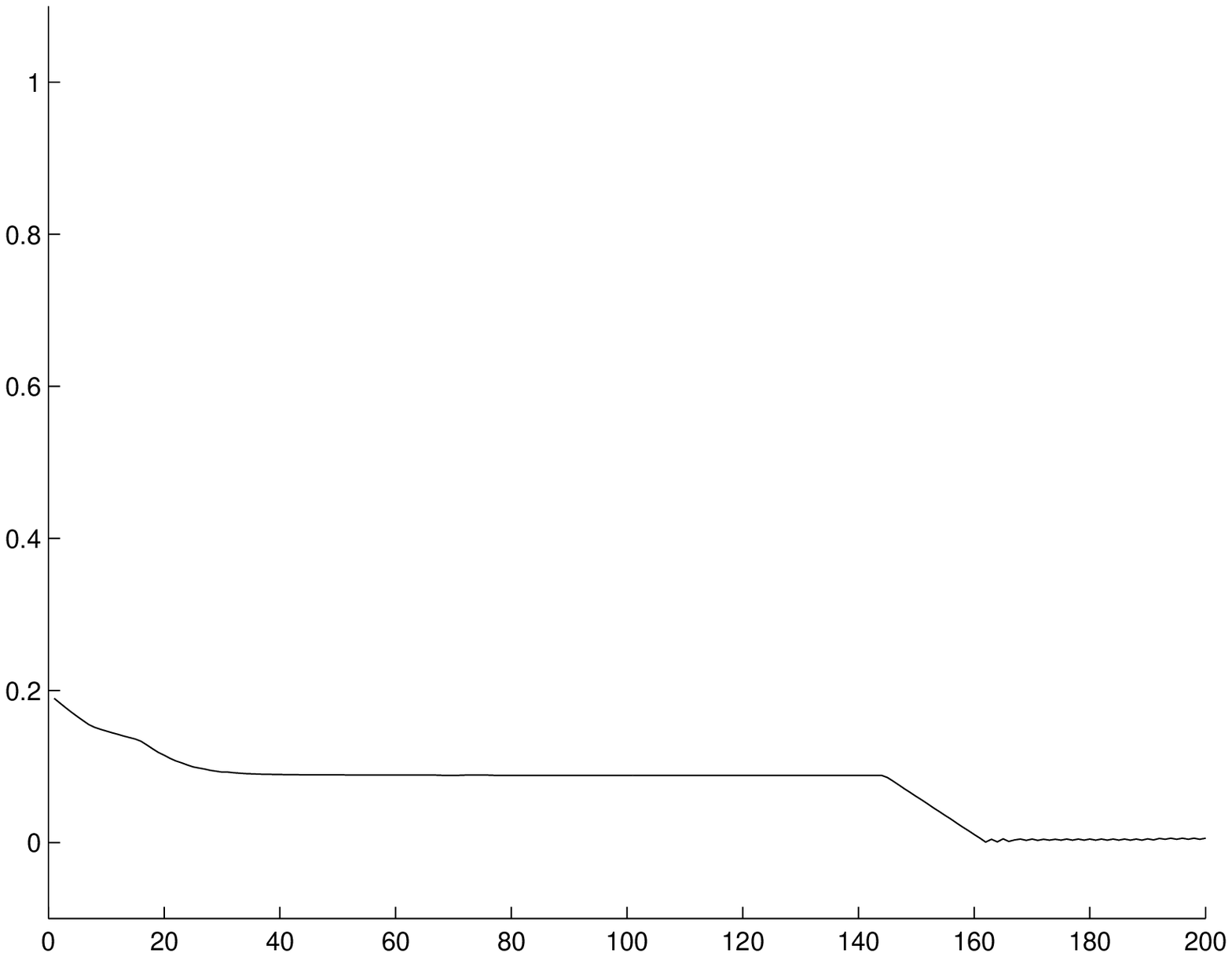}}
\end{tabular}%

\begin{tabular}
[c]{cc}%
\textbf{Fig. 25.a} & \textbf{Fig. 25.b}\\
\scalebox{0.5}{\includegraphics{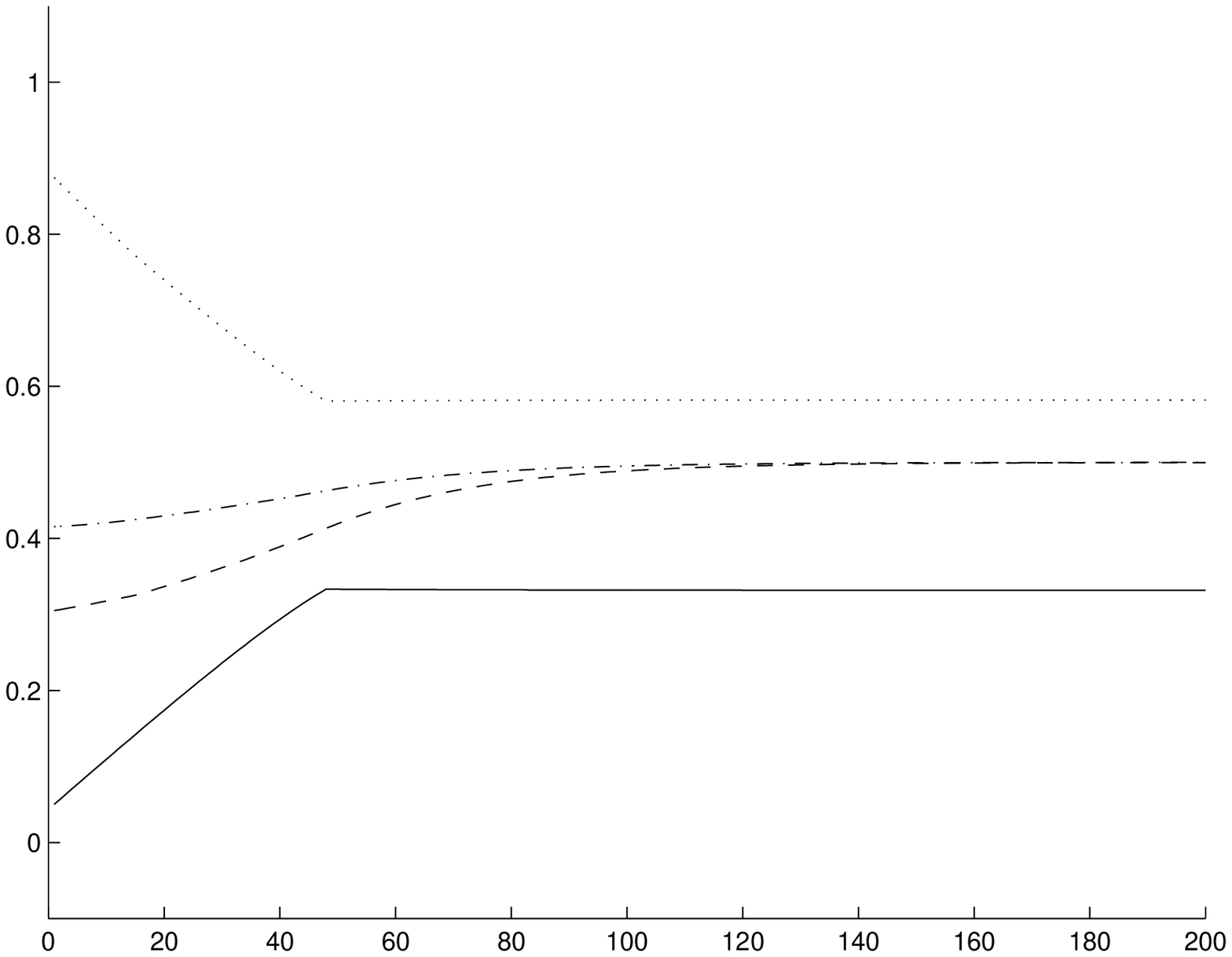}}  & \scalebox{0.5}%
{\includegraphics{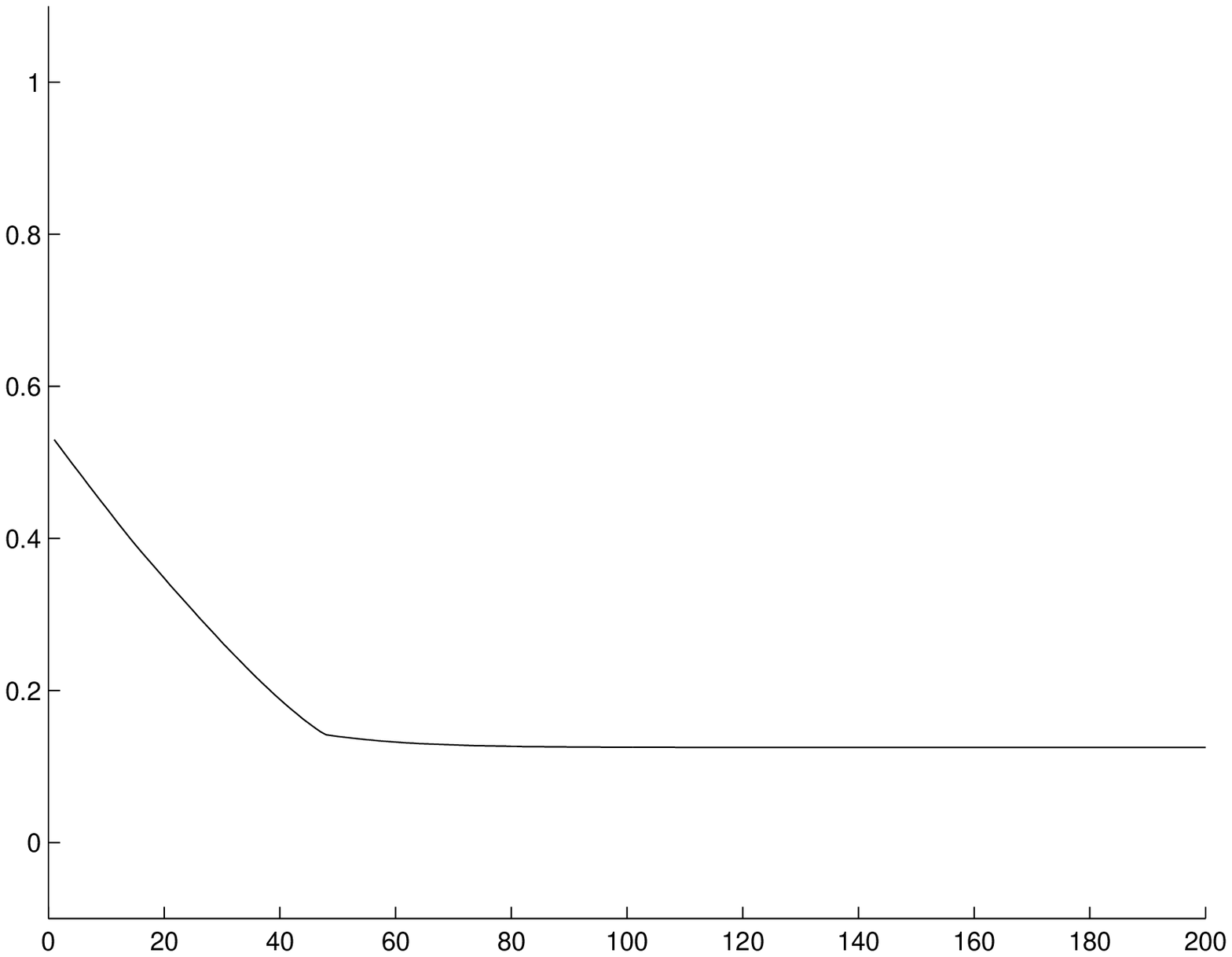}}
\end{tabular}
\end{center}

\medskip In Figures 26, 27 we present some simulation results for the control
theoretic algorithm, with various initial conditions. The algorithm always
converges to $\overline{x}$=$\left(  0.875,0.225,0.675,0.875\right)  ^{T}$
which yields zero total inconsistency. Indeed, having run a large number of
additional simulations we have noticed that the control algorithm \emph{always
}converges; furthermore the equilibrium is always the above mentioned
$\overline{x}$, which perhaps indicates that it is the unique solution of the
truth value equations (\ref{eq101})--(\ref{eq104}).

\begin{center}%
\begin{tabular}
[c]{cc}%
\textbf{Fig. 26.a} & \textbf{Fig. 26.b}\\
\scalebox{0.5}{\includegraphics{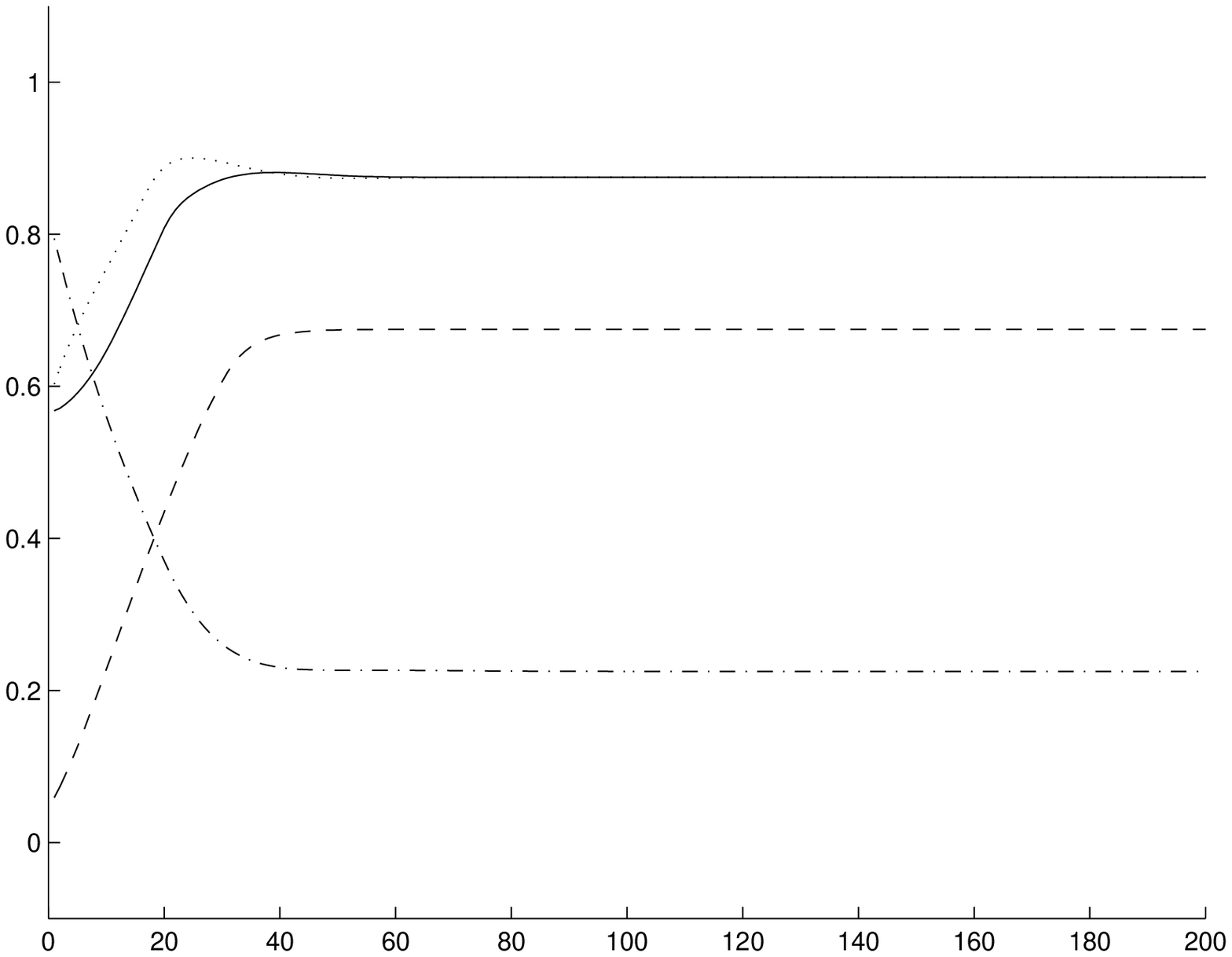}}  & \scalebox{0.5}%
{\includegraphics{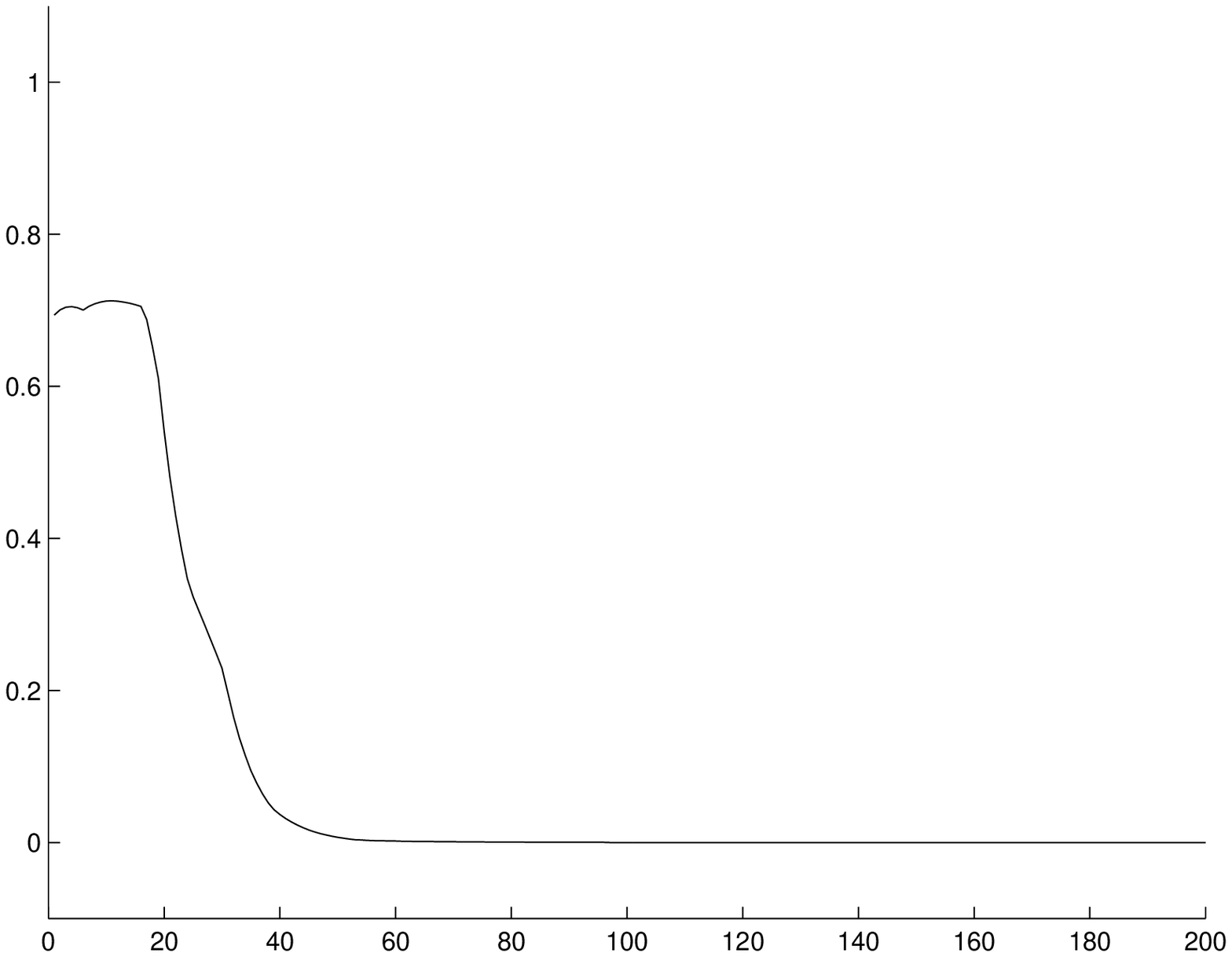}}
\end{tabular}%

\begin{tabular}
[c]{cc}%
\textbf{Fig. 27.a} & \textbf{Fig. 27.b}\\
\scalebox{0.5}{\includegraphics{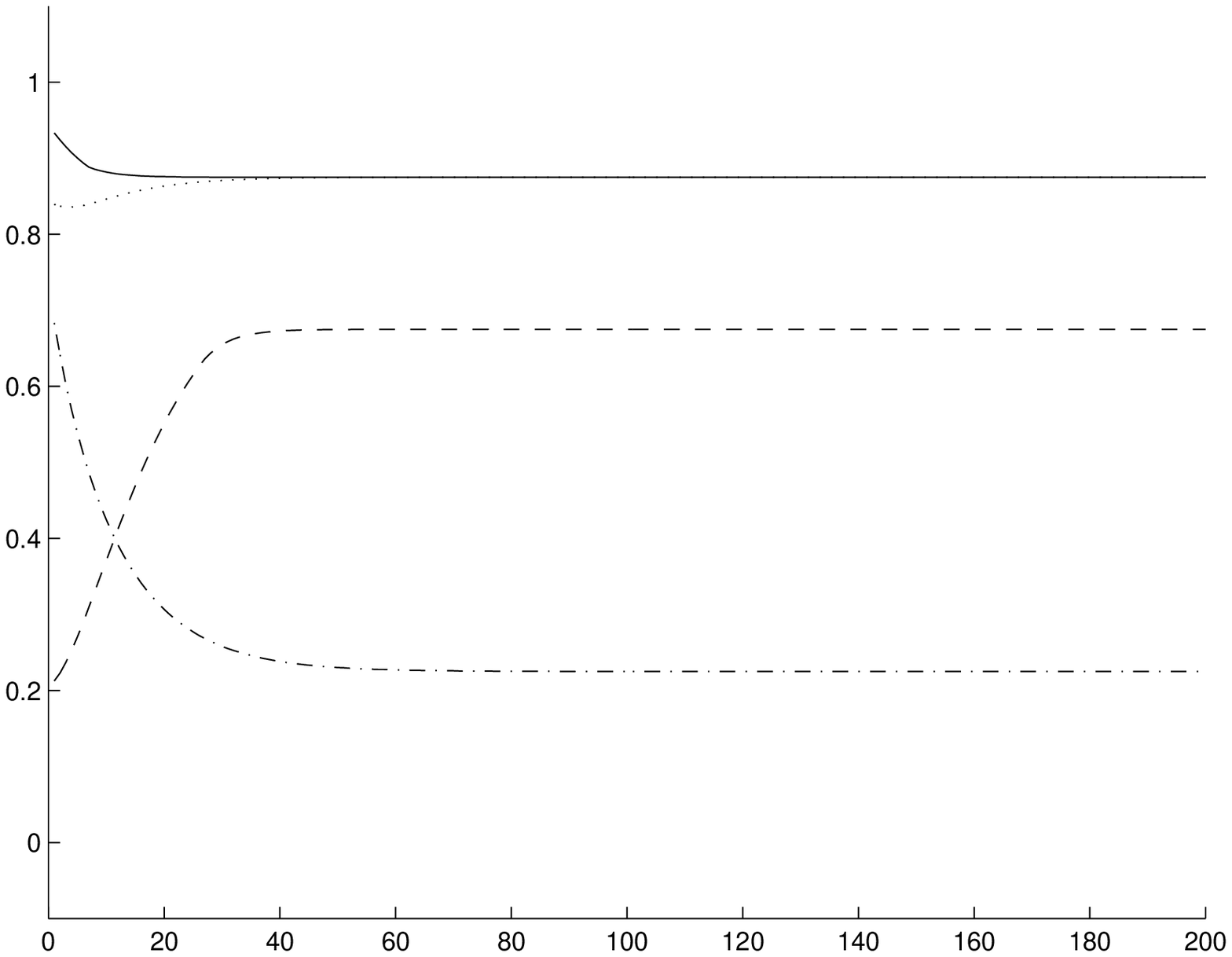}}  & \scalebox{0.5}%
{\includegraphics{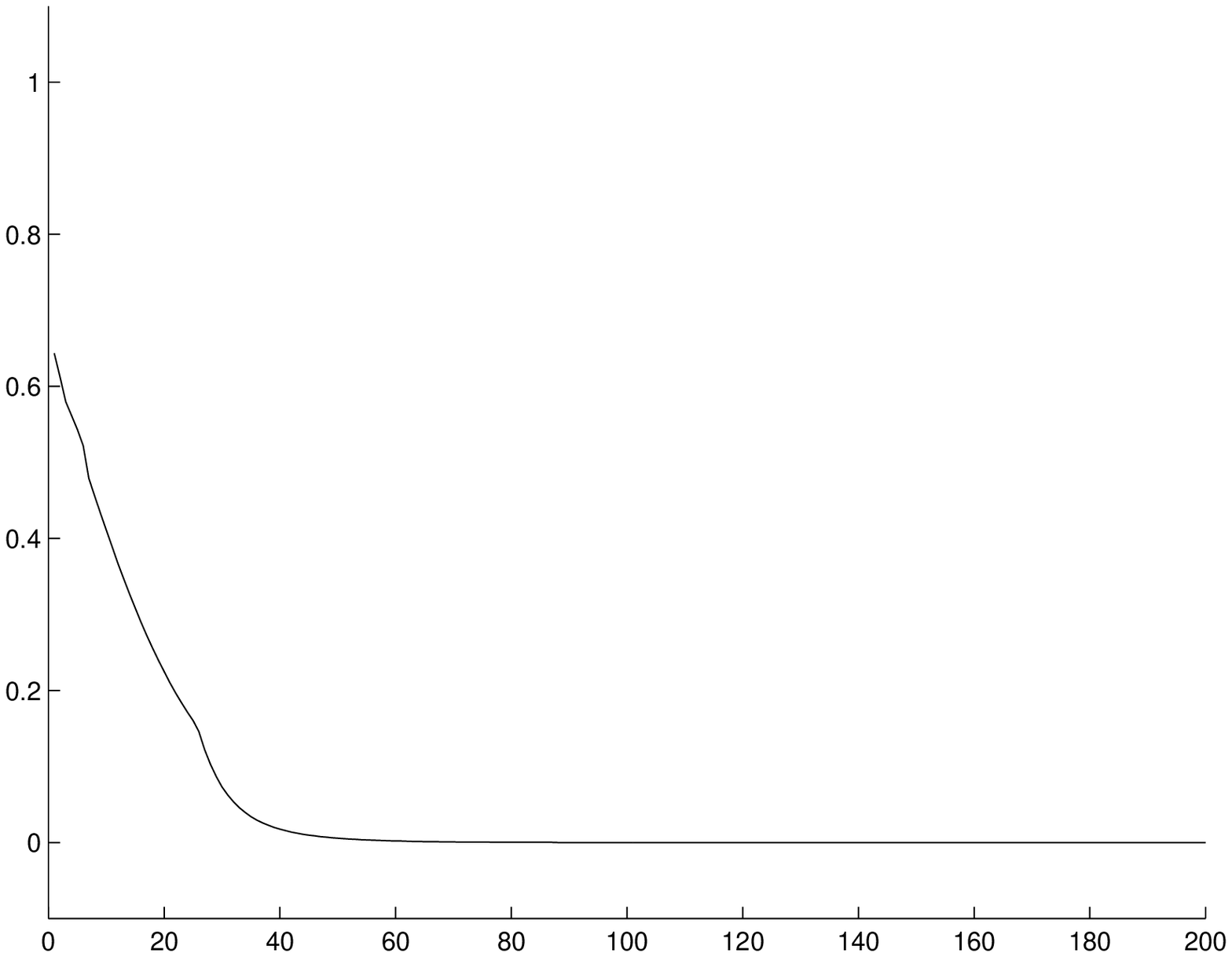}}
\end{tabular}
\end{center}

\subsubsection{$\wedge$ Implemented by Product, $\vee$ Implemented by Extended Sum}

We present the truth value equations:
\begin{align}
x_{1}  &  =\left(  1-\left|  x_{1}-0.75\right|  \right)  \cdot\left(
1-\left|  x_{2}-0.35\right|  \right)  +\left(  1-\left|  x_{4}-1.00\right|
\right) \nonumber\\
&  -\left(  1-\left|  x_{1}-0.75\right|  \right)  \cdot\left(  1-\left|
x_{2}-0.35\right|  \right)  \cdot\left(  1-\left|  x_{4}-1.00\right|  \right)
\label{eq111}\\
x_{2}  &  =\left(  1-\left|  x_{1}+x_{3}-x_{1}x_{3}-1.00\right|  \right)
\cdot\left(  1-\left|  x_{4}-0.10\right|  \right) \label{eq113}\\
x_{3}  &  =\left(  1-\left|  x_{2}-0.00\right|  \right)  \cdot\left(
1-\left|  x_{3}-0.35\right|  \right) \label{eq112}\\
x_{4}  &  =1-\left|  1-x_{1}-0.25\right|  . \label{eq116}%
\end{align}
We omit further details and directly present the results of numerical
simulation. In Figures 28 and 29 we present some simulation results for the
Newton-Raphson algorithm, with various initial conditions. The algorithm
always finds the solution $\overline{x}=\left(
0.9507,0.2942,0.5586,0.7993\right)  ^{T}$. \medskip In Figures 30 and 31 we
present some simulation results for the steepest descent algorithm, with
various initial conditions. Note that in Figure 31 the algorithm converges to
a local minimum (nonzero) of the total inconsistency.

\begin{center}%
\begin{tabular}
[c]{cc}%
\textbf{Fig. 28.a} & \textbf{Fig. 28.b}\\
\scalebox{0.5}{\includegraphics{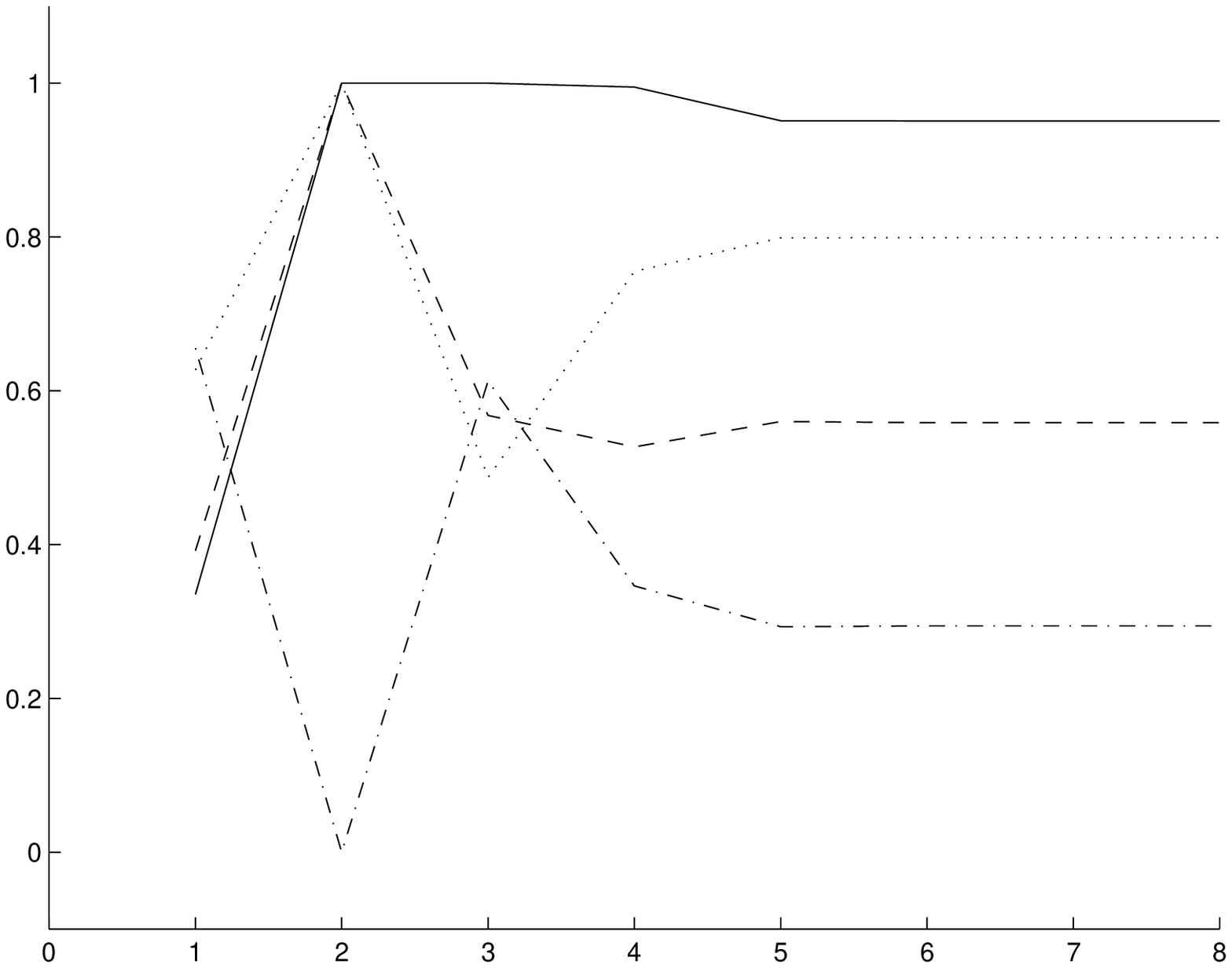}}  & \scalebox{0.5}%
{\includegraphics{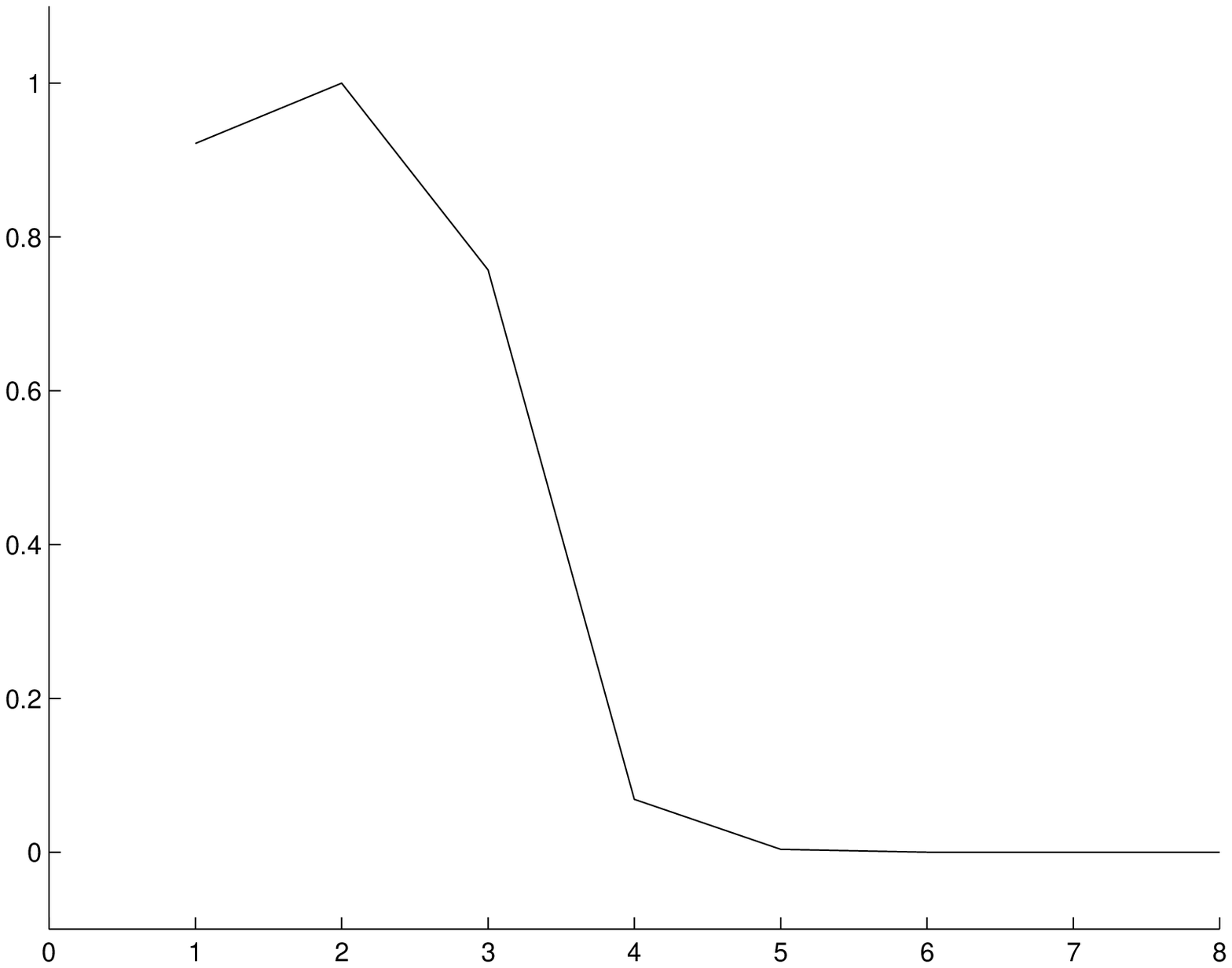}}
\end{tabular}%

\begin{tabular}
[c]{cc}%
\textbf{Fig. 29.a} & \textbf{Fig. 29.b}\\
\scalebox{0.5}{\includegraphics{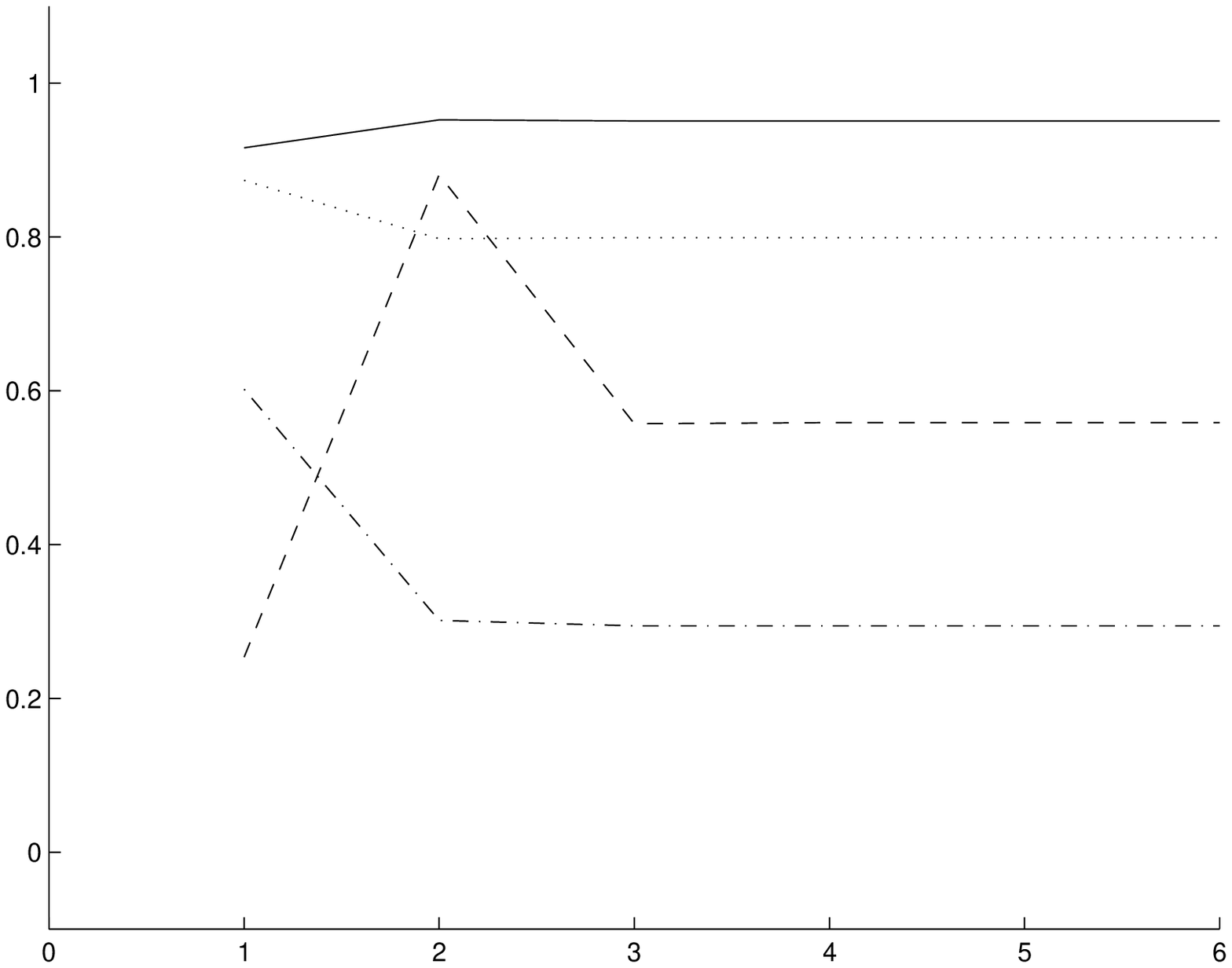}}  & \scalebox{0.5}%
{\includegraphics{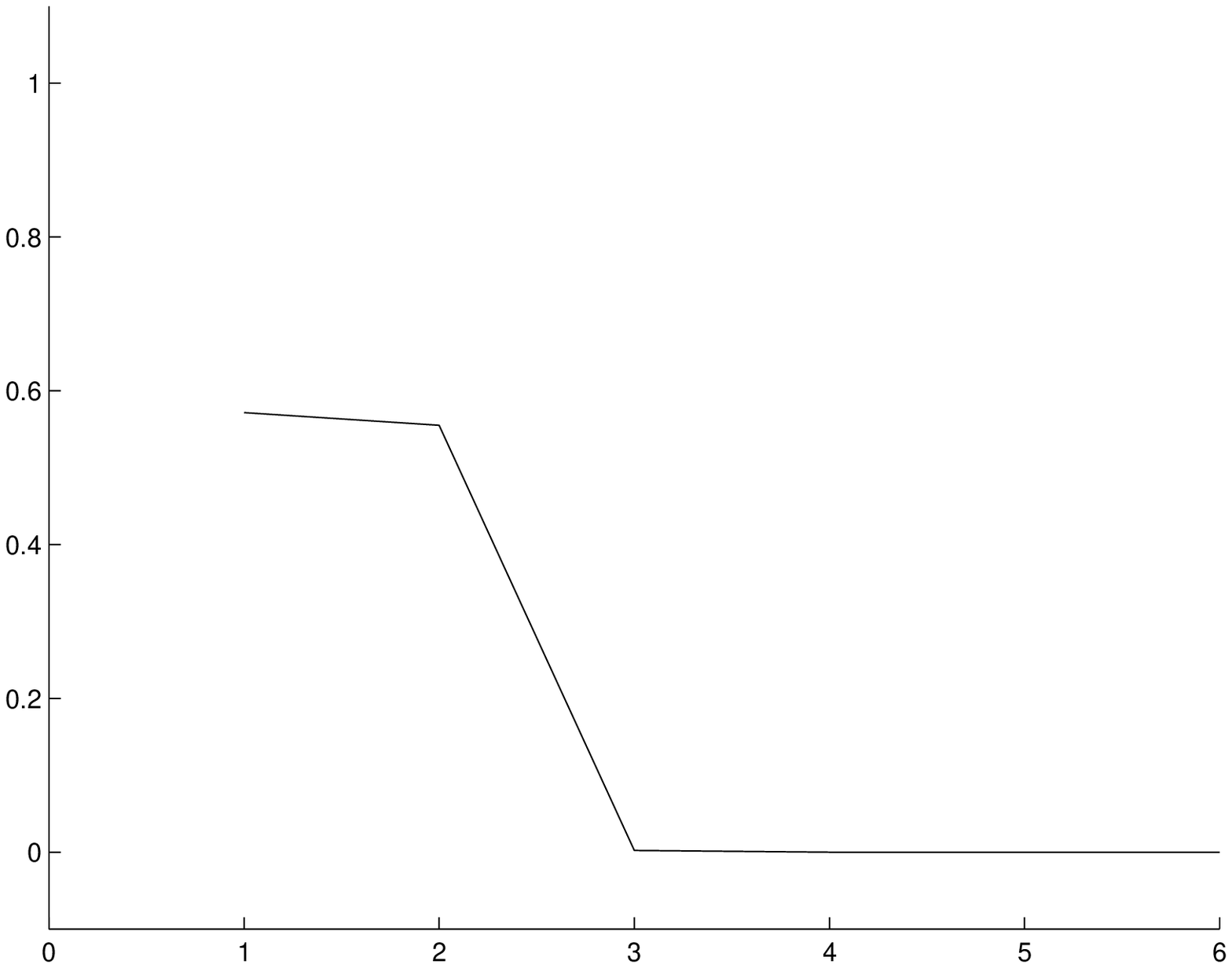}}
\end{tabular}%

\begin{tabular}
[c]{cc}%
\textbf{Fig. 30.a} & \textbf{Fig. 30.b}\\
\scalebox{0.5}{\includegraphics{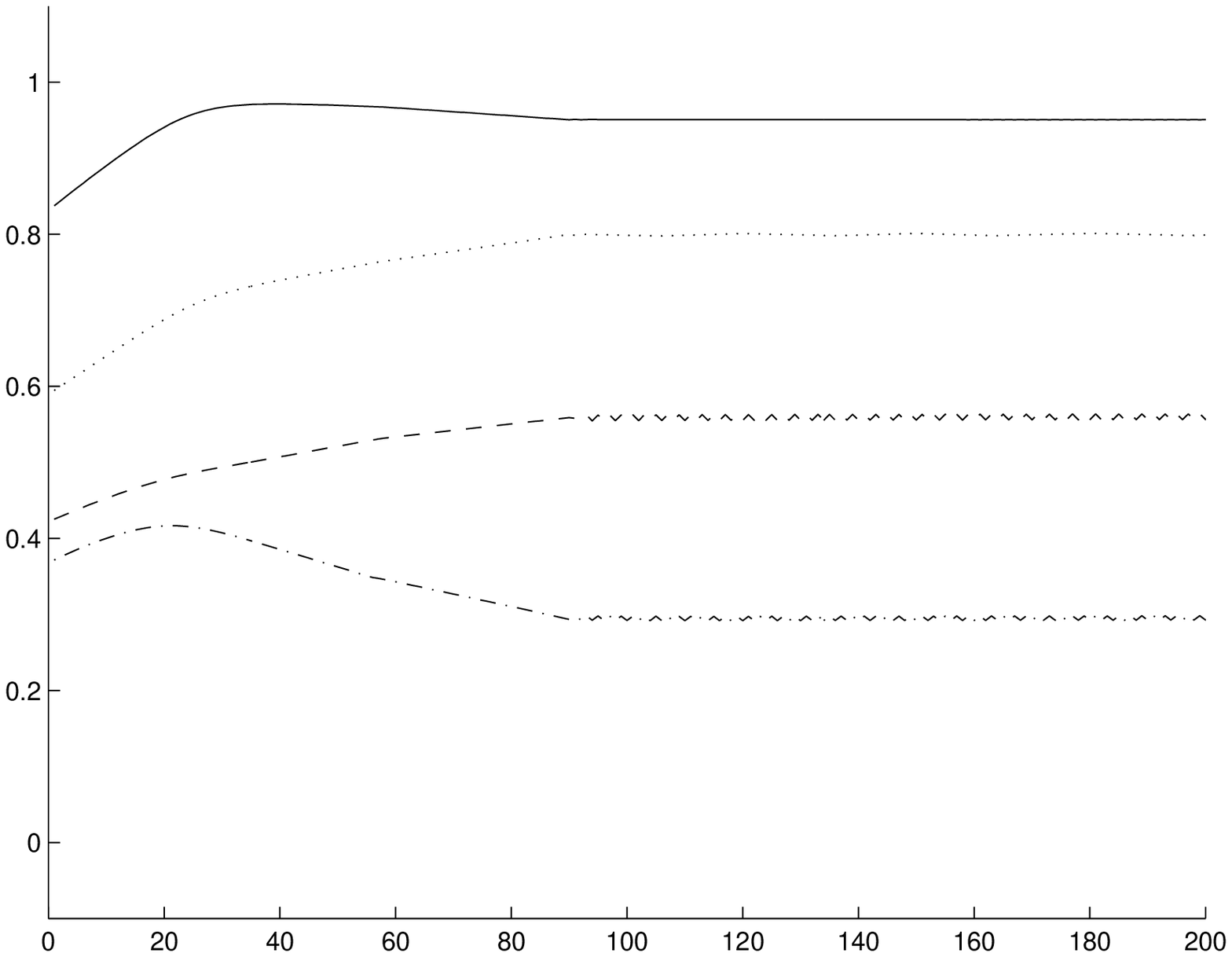}}  & \scalebox{0.5}%
{\includegraphics{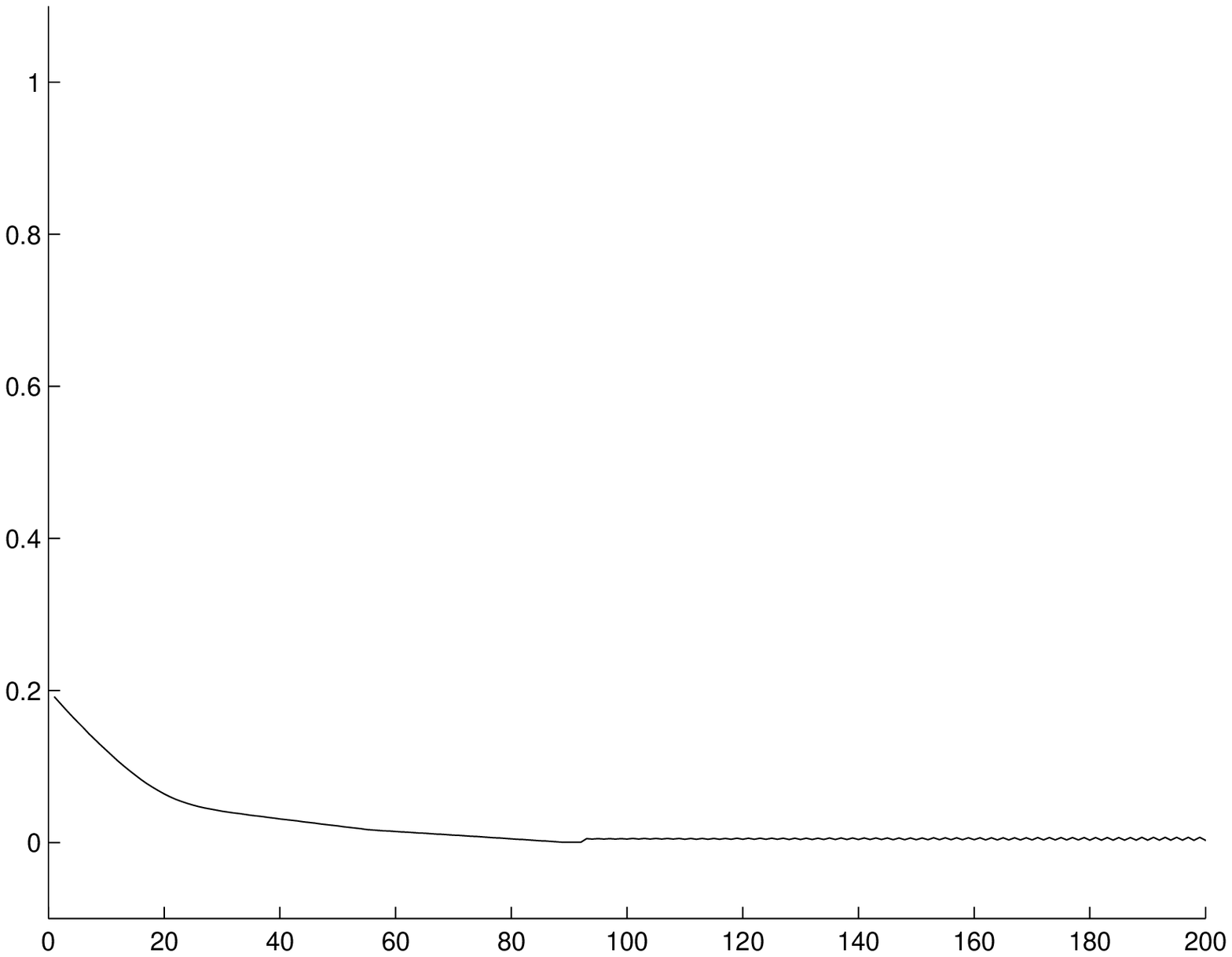}}
\end{tabular}%

\begin{tabular}
[c]{cc}%
\textbf{Fig. 31.a} & \textbf{Fig. 31.b}\\
\scalebox{0.5}{\includegraphics{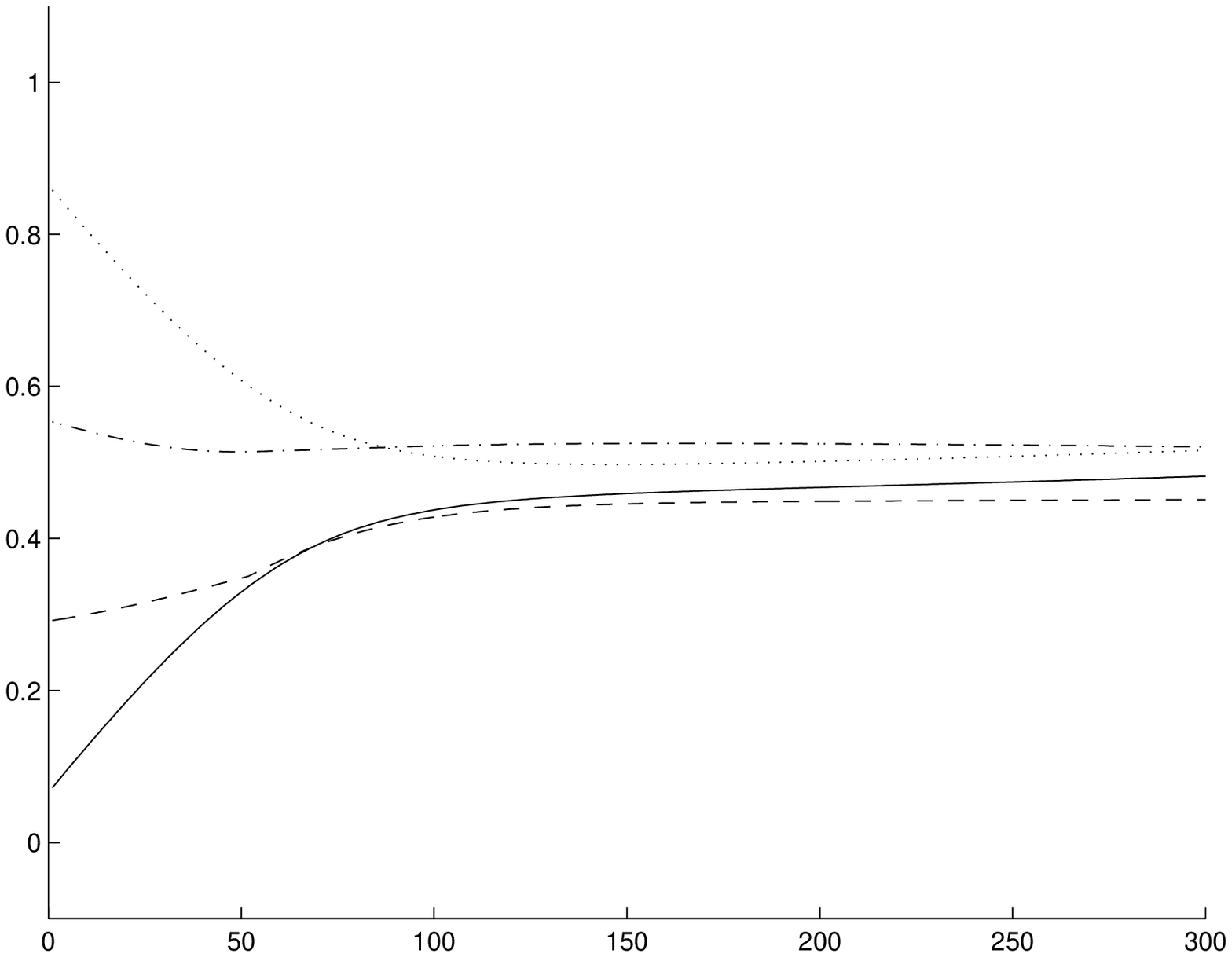}}  & \scalebox{0.5}%
{\includegraphics{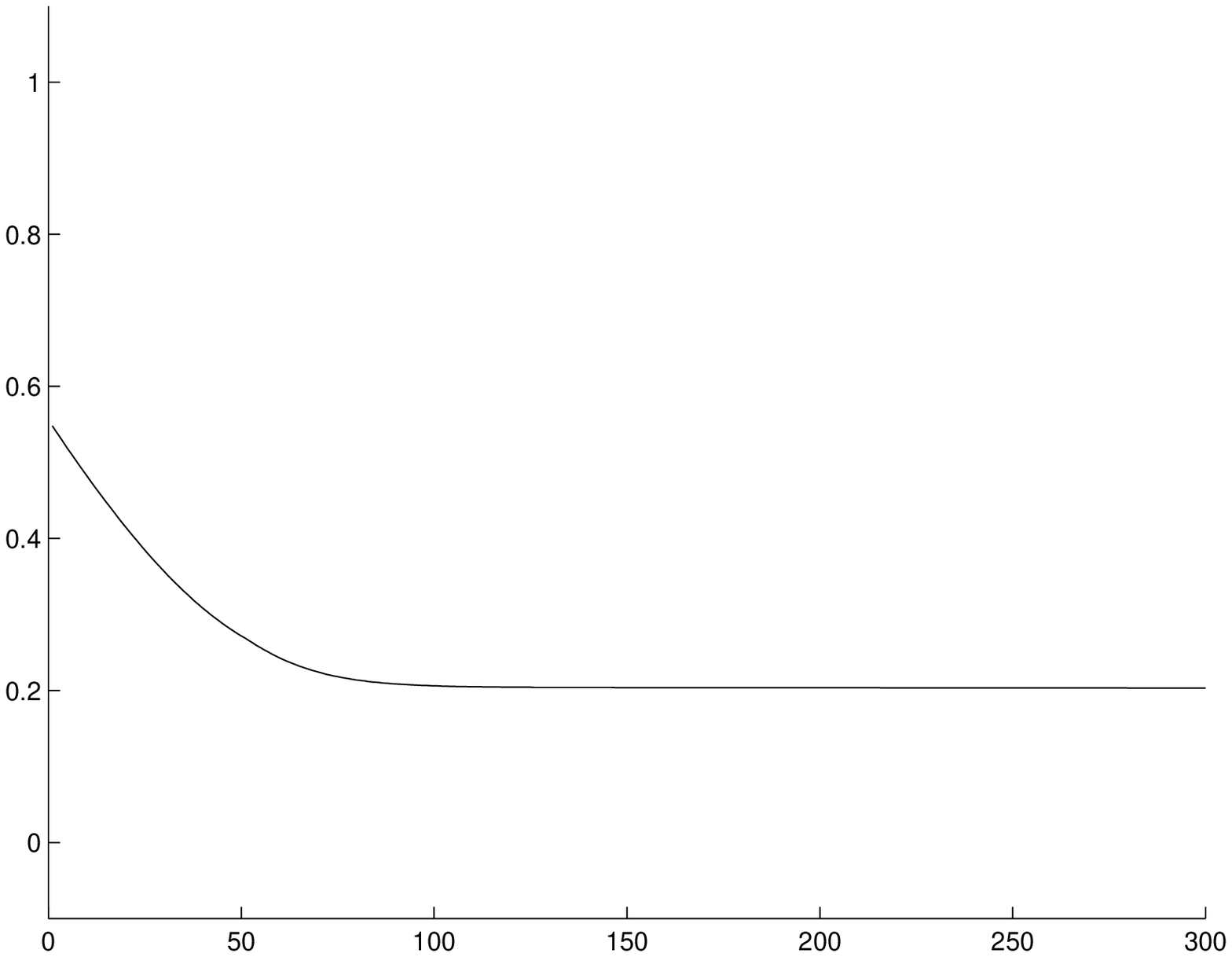}}
\end{tabular}
\end{center}

Finally, in Figures 32 and 33 we present some simulation results for the
control theoretic algorithm, with various initial conditions. The algorithm
always converges to the following solution of the truth value equations
(\ref{eq111})--(\ref{eq116}): $\overline{x}=\left(
0.9507,0.2942,0.5586,0.7993\right)  ^{T}$ , rendering likely that this is the
unique solution.

\begin{center}%
\begin{tabular}
[c]{cc}%
\textbf{Fig. 32.a} & \textbf{Fig. 32.b}\\
\scalebox{0.5}{\includegraphics{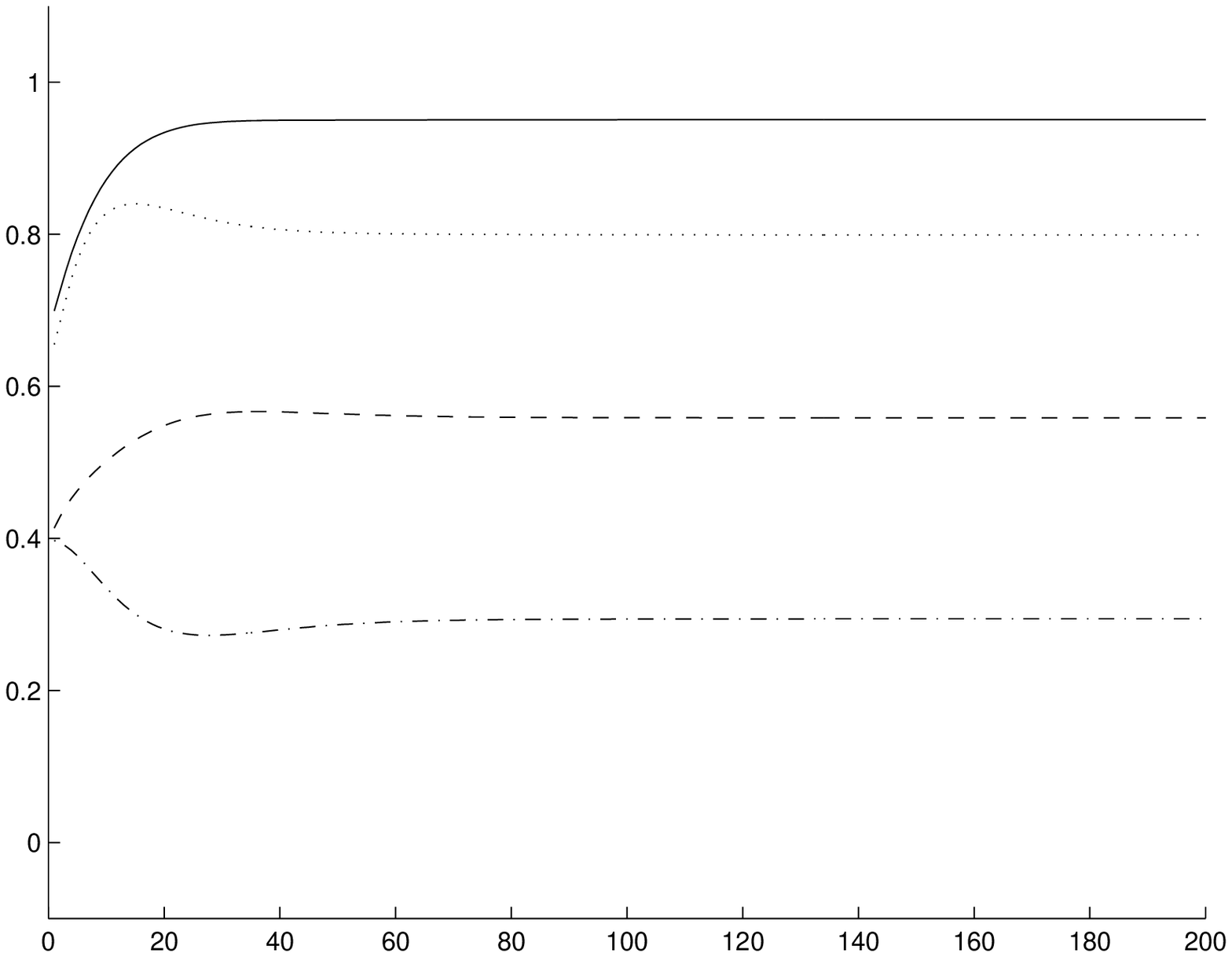}}  & \scalebox{0.5}%
{\includegraphics{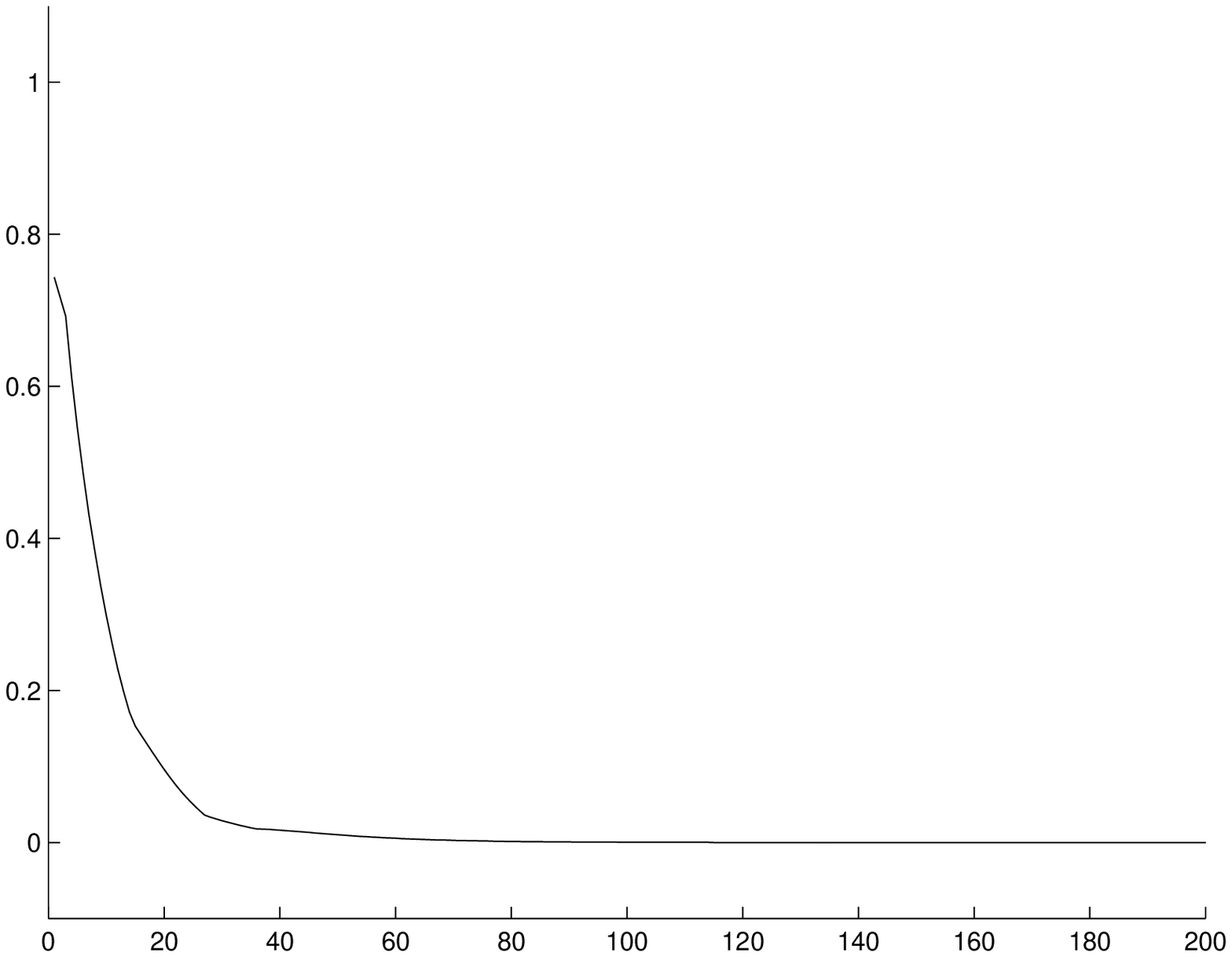}}
\end{tabular}%

\begin{tabular}
[c]{cc}%
\textbf{Fig. 33.a} & \textbf{Fig. 33.b}\\
\scalebox{0.5}{\includegraphics{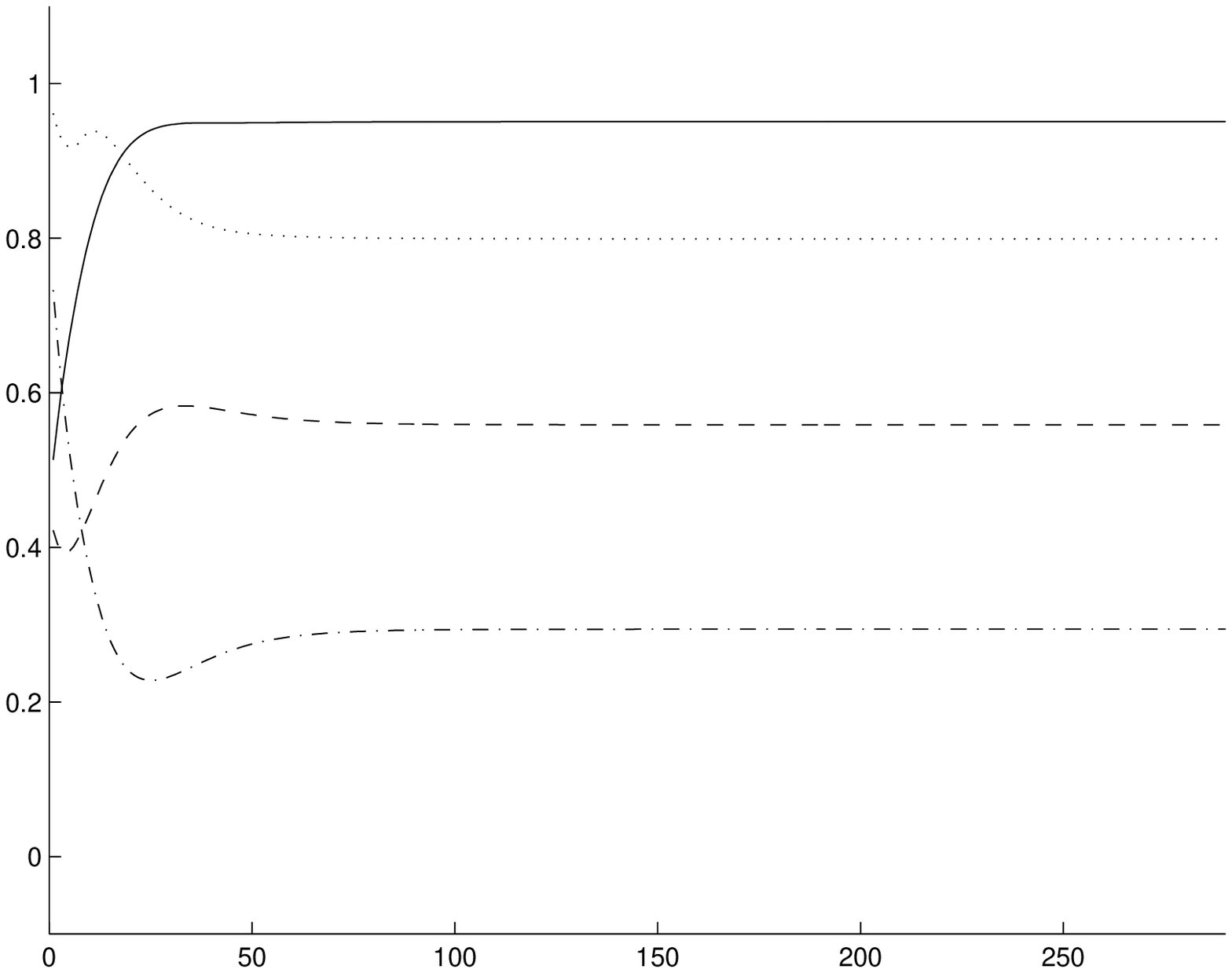}}  & \scalebox{0.5}%
{\includegraphics{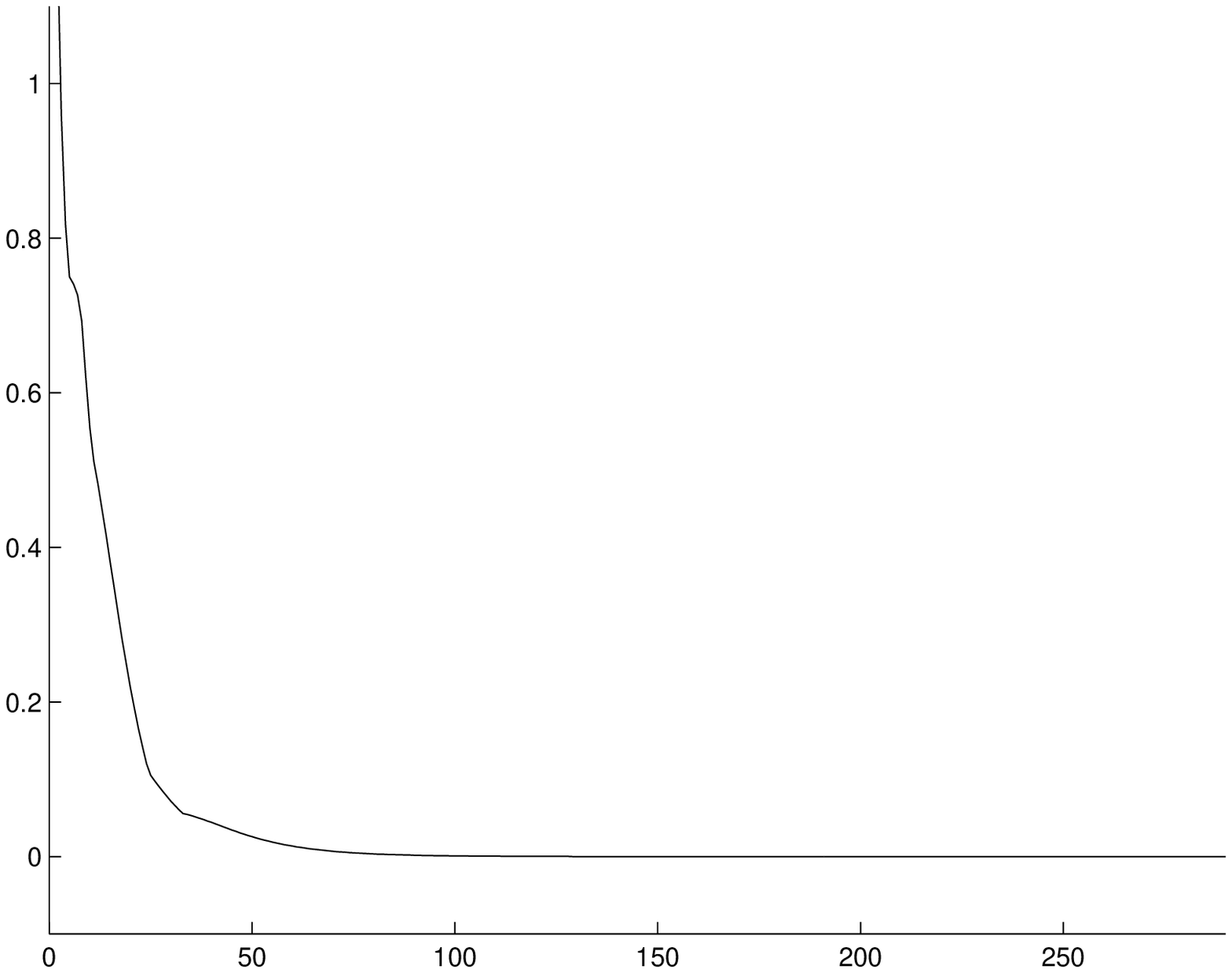}}
\end{tabular}
\end{center}

\subsection{Example 7: The Strengthened Liar}

\label{secA07}

This final example does not fall, strictly speaking, within the framework we
have presented. However it can be treated by a small, straightforward
extension of our approach.

The example is the so-called ``Strengthened Liar'', a self-referential
sentence which has been often used as a test of proposed solutions to the Liar
paradox \cite{Rieger}. The Strengthened Liar sentence is
\begin{equation}
A=\text{``}A\text{ is not true''.} \label{eq0571}%
\end{equation}

To treat this and similar sentences in the fuzzy context we must translate it
in terms of a membership function for the property of being not true. To this
end, consider the sentence
\begin{equation}
C=\text{``The truth value of }A\text{ is not }a\text{''.}\label{eq0572}%
\end{equation}
A possible truth value assignment for (\ref{eq0572}) is
\[
\text{Tr}\left(  C\right)  =\left\{
\begin{array}
[c]{cc}%
1 & \text{when Tr}\left(  A\right)  \neq a\\
0 & \text{else}%
\end{array}
\right.  .
\]
However, this is too strict. Consider the case when $a=1$ and Tr$\left(
A\right)  =0.99$. Do we really want to assign Tr$\left(  C\right)  =1$? How
about the case Tr$\left(  A\right)  =0.99999$? A more reasonable truth value
assignment is
\begin{equation}
\text{Tr}\left(  C\right)  =\left|  \text{Tr}\left(  A\right)  -a\right|
\label{eq0574}%
\end{equation}
which takes the maximum value of 1 when $\left|  \text{Tr}\left(  A\right)
-a\right|  =1$, i.e. in the cases

\begin{enumerate}
\item Tr$\left(  A\right)  =1$ and $a=0$;

\item Tr$\left(  A\right)  =0$ and $a=1$.
\end{enumerate}

Let us accept (\ref{eq0574}) and set $A=C$, i.e.
\begin{equation}
A=\text{``Tr}\left(  A\right)  \neq a\text{''.} \label{eq0575}%
\end{equation}
(\ref{eq0575}) is more general than (\ref{eq0571}); to obtain (\ref{eq0571}%
)\ we set $a=1$:
\[
A=\text{``Tr}\left(  A\right)  \neq1\text{''.}%
\]
Hence, setting $x=$Tr$\left(  A\right)  $, we must solve the truth value
equation
\[
x=\left|  x-1\right|  =1-x
\]
which has the unique solution $x=1/2$.

Obviously this approach can be extended to treat situations which involve
statements of the form
\[
\text{``The truth value of }B\text{ is not }b\text{''}%
\]
for any $B\in\mathbf{S}_{1}$. Hence our framework can be extended defining
$\mathbf{V}_{2}$, the set of 2nd level elementary sentences to include both
sentences of the form
\[
\text{Tr}\left(  B\right)  =b
\]
and
\[
\text{Tr}\left(  B\right)  \neq b.
\]
The definition of $\mathbf{S}_{2}$ remains unchanged but, since it depends on
the expanded $\mathbf{V}_{2}$, results to an expansion of $\mathbf{S}_{2}$ as
well. We leave the details to the reader

\subsection{Discussion of the Algorithms}

\label{secA08}

Let us close with a short comparison of the three numerical algorithms. We see
that the Newton-Raphson algorithm converges very quickly but not always to a
solution of the truth value equations. Regarding steepest descent, its
convergence is guaranteed but not always to point of zero inconsistency. It
appears that the control-theoretic algorithm combines the best
properties:\ practically guaranteed convergence to a point of zero
inconsistency and in a relatively small number of iterations. Of course, all
of the above remarks refer to the experiments of the previous sections;
further numerical experiments (and, perhaps, analytical work) are required to
establish their general validity.
\end{document}